\documentclass{ws-ijbc}
\usepackage{graphicx}
\usepackage{epstopdf}
\usepackage{tabularx}
\usepackage[caption=false]{subfig}
\usepackage{amsfonts}
\usepackage{algorithmic}
\usepackage{color}
\usepackage{ amssymb }
\usepackage{ amsmath }
\usepackage{amsopn}
\newcommand{\be}{\begin{equation}}
\newcommand{\ee}{\end{equation}}
\newcommand{\bes}{\begin{equation*}}
\newcommand{\ees}{\end{equation*}}

\providecommand{\ceil}[1]{\left \lceil #1 \right \rceil }
\def \eps{ { \varepsilon }}

\begin{document}

\catchline{}{}{}{}{} 

\markboth{A. L. Krause, D. Beliaev, R. A. Van Gorder, S. L. Waters}{Bifurcation analysis of lattice and continuum models of bioactive porous media}

\title{Bifurcations and dynamics emergent from lattice and continuum models of bioactive porous media}

\author{Andrew L. Krause, Dmitry Beliaev, Robert A. Van Gorder, and Sarah L. Waters\footnote{Corresponding author. email: waters@maths.ox.ac.uk}}

\address{Mathematical Institute, University of Oxford, Andrew Wiles Building, Radcliffe Observatory Quarter, Woodstock Road, Oxford, OX2 6GG, United Kingdom}

\maketitle

\begin{history}
\received{(to be inserted by publisher)}
\end{history}

\begin{abstract}
We study dynamics emergent from a two-dimensional reaction--diffusion process modelled via a finite lattice dynamical system, as well as an analogous PDE system, involving spatially nonlocal interactions. These models govern the evolution of cells in a bioactive porous medium, with evolution of the local cell density depending on a coupled quasi--static fluid flow problem. We demonstrate differences emergent from the choice of a discrete lattice or a continuum for the spatial domain of such a process.  We find long--time oscillations and steady states in cell density in both lattice and continuum models, but that the continuum model only exhibits solutions with vertical symmetry, independent of initial data, whereas the finite lattice admits asymmetric oscillations and steady states arising from symmetry-breaking bifurcations. We conjecture that it is the structure of the finite lattice which allows for more complicated asymmetric dynamics. Our analysis suggests that the origin of both types of oscillations is a nonlocal reaction-diffusion mechanism mediated by quasi-static fluid flow.
\end{abstract}


\section{Introduction}
We study a reaction--diffusion process on a two--dimensional domain modelled by a finite lattice dynamical system, which includes nonlocal coupling between nodes. This model arises as a mathematical caricature of a model developed in \citep{krause_lattice_2017} representing fluid and cell interactions in a bioactive porous medium. While lattice--dynamical systems have been studied in a variety of different settings \citep{buzano1983bifurcation,winterbottom2004mode,
wang2006spatio,kamei2009existence,sattinger1980bifurcation}, we note that processes defined on finite lattices are less well--studied, and have been shown to exhibit unique dynamics due to the influence of boundaries \citep{chow1996dynamics, gillis1997patterns, golubitsky2004some}. Additionally, our model also exhibits nonlocal coupling across the lattice, which is also of interest from the dynamical systems perspective \citep{gourley_spatio-temporal_2001, billingham_dynamics_2004, hamel_nonlocal_2014}. We also study an analogous spatially continuous system, with the chief goal of comparing the dynamics between these two paradigms, and hence elucidating the roles of the finite lattice and the nonlocal coupling in our reaction--diffusion system.

The spatial domain in our models represents a porous tissue scaffold seeded with cells and perfused with fluid in order to stimulate cell growth and the formation of viable tissue, which is a primary goal of the field of tissue engineering. A major challenge in tissue engineering is to understand and exploit processes that occur on very different spatial and temporal scales in tissues and organs \cite{vafai_porous_2010}. Such scales permit the development of mathematical models which are able to help elucidate complex physical and biological processes relevant for tissue engineering \cite{odea_continuum_2012, van_blitterswijk_tissue_2008}. In addition to providing conceptual understanding, quantitative mathematical models can be used to predict and optimize experimental operating regimes which saves on costly and time-consuming experimental trials \cite{geris_computational_2013}. While continuum models for fluid flow in porous media can be justified for pore networks with many pores, corresponding either to large porous materials (such as oil reservoirs), or to materials with densely connected pore structures (such as sponges) \cite{bear_dynamics_1972}, many tissue scaffolds used experimentally have pore sizes and scaffold geometries such that the number of pores is $O(10^3)$ or less \citep{cox_3d_2015, german_applications_2016, loh_three-dimensional_2013, vafai_porous_2010}, suggesting that lattice models may be more appropriate than continuum models, at least in some settings. 

While lattice approaches exist in related literature, such as spatial models of cell monolayers \citep{byrne2009individual}, tumour vascularization \citep{anderson1998continuous}, biofilms \citep{thullner_computational_2008}, angiogenesis \citep{mcdougall_mathematical_2002, scianna_review_2013}, and even tissue engineering \citep{barbotteau_modelling_2003}, there is still much work needed to understand the impact of finite pore networks on porous media undergoing structural changes due to cell growth; see \cite{mcdougall1997application, blunt_flow_2001, odea_continuum_2012} for a more thorough comparison between discrete and continuous models. \cite{krause_lattice_2017} introduced two models of a two-dimensional porous tissue scaffold perfused with fluid. These models incorporated interactions between a viscous fluid phase representing a culture medium, and a cell phase modelling the biomass of proliferating cells and their extracellular matrix. Cells were allowed to diffuse locally throughout the domain, but would die if the local fluid shear stress became too large. Cell proliferation was assumed to block pores, reducing the local permeability of the scaffold, and hence introducing a nonlocal interaction between cells at different parts of the domain. \cite{krause_lattice_2017} modelled the system using a continuum approach involving a system of PDEs, and a lattice approach involving a spatially embedded network of ODEs, and discussed quantitative and qualitative differences between the modelling predictions arising from these approaches. Oscillatory solutions were only present for the lattice model, and were more pronounced for smaller pore networks. 

Spatially discrete and continuous dynamical systems have been explored in a variety of related fields to model cells and tissues \citep{fraternali2014discrete, murisic2015discrete, matsiaka2018discrete}, providing further motivation to compare dynamics emergent from each modelling approach. Symmetry breaking bifurcations on lattices and other network structures have found relevance in fluid mechanics \citep{crawford1991symmetry}, optics \citep{kevrekidis2005spontaneous}, mathematical physics \citep{bendix2009exponentially, brazhnyi2011spontaneous}, chaotic systems \citep{pikovsky1991symmetry,rothos2002chaos}, and pattern formation \cite{wolfrum2012turing}, to name a few examples. For a review of symmetry breaking dynamics on lattices, see \cite{sattinger1980bifurcation}. Bifurcations due to lattice structures have been classified on a number of crystal lattices \citep{janssen1983bifurcations,silber1992symmetry}, and such results also extend to a number of other systems on lattices \cite{buzano1983bifurcation,winterbottom2004mode,wang2006spatio,kamei2009existence}, with Hopf bifurcations in particular appearing quite commonly \cite{silber1991hopf}. We remark that many of these systems are formulated on infinite lattices, yet for practical purposes, models relevant to tissue engineering will likely be posed on finite lattices. This finiteness of the lattice results in the possibility of additional broken symmetries due to boundary effects \cite{chow1996dynamics, gillis1997patterns, golubitsky2004some}. Therefore, depending on the specific functional structure of the dynamical system, and of the underlying spatial lattice, symmetry preserving or symmetry breaking bifurcations may be ubiquitous to the model dynamics. Additionally, \cite{duncan2015noise} has demonstrated multistability in discrete stochastic chemical systems, whereas the continuum limit of these systems admits only one stable equilibrium state. We will demonstrate symmetry--breaking bifurcations and multistability in our lattice system which is not present in the spatially continuous analogue.

Motivated by the aforementioned results, in the present paper we systematically compare the dynamics emergent from lattice and continuum models of bioactive porous media. We modify the models presented in \cite{krause_lattice_2017} by assuming that cells are sensitive to high pressures, rather than fluid shear stress. This is motivated by computational modelling of cell death due to hydrostatic pressure \cite{byrne2009individual, byrne2003modelling, nessler_influence_2016}, and by experimental work demonstrating cell death due to pressure \cite{manas_morphological_2004}. These modified ``fluid pressure" dependent models are more amenable to mathematical and numerical investigation, due to the simpler coupling between the fluid flow and growth processes compared with the fluid shear stress coupling used in \citep{krause_lattice_2017} (which implicitly involved gradients in pressure, rather than values of the pressure itself).

The paper is organised as follows. In Section \ref{Model}, we present our lattice and continuum models, and briefly discuss the underlying modelling assumptions and components. In Section \ref{Numerical_Overview}, we present numerical solutions of the lattice and continuum models, demonstrating typical solution behaviours. In Section \ref{Asymptotics}, we derive asymptotic solutions for the cell density appropriate for large values of the diffusion parameter, and show that these always tend to a steady spatially uniform state. We use these solutions to bound the region in parameter space where we anticipate non-equilibrium dynamics (e.g. oscillations). In Section \ref{Bifurcations}, we describe the more complicated dynamics outside of this asymptotic parameter regime. We do this by presenting numerical bifurcation diagrams over two parameters for large and small lattices alongside the continuum model. Using these diagrams we broadly characterize the steady states and long-time oscillatory states by their symmetry properties. We present continuation results showing the existence of Hopf and pitchfork bifurcations that break the vertical symmetry of steady state solutions. Additionally, we reveal that a different kind of oscillation exists, which preserves vertical symmetry and is not due to a local bifurcation from a steady state. Finally, we give an overview of these results in Section \ref{Conclusions}, and discuss implications of our results.

\section{Lattice and continuum models of a bioactive porous media}\label{Model}
\begin{figure}  
\centering
    \includegraphics[width=.46\textwidth]{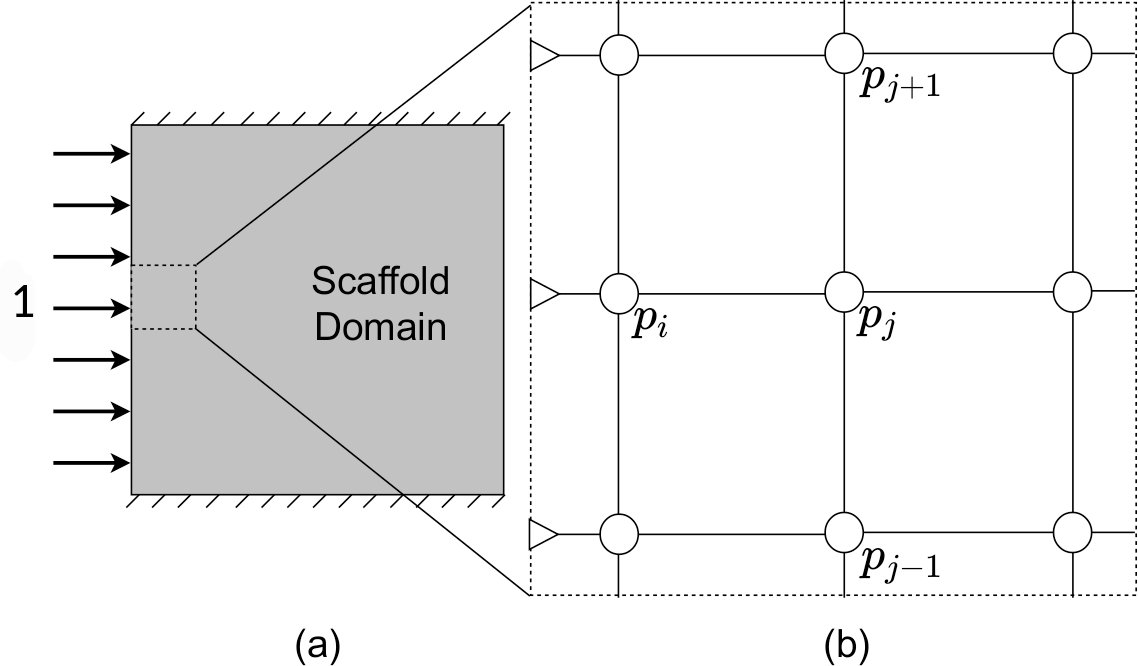}
\caption{A representation of the scaffold domain in (a), and a lattice representation of a small part of it in (b). Fluid enters the domain at unit flow rate on the left side, and leaves along the right side of the lattice, with impermeable horizontal boundaries.}
\label{lattice_diagram}
\end{figure}

We consider a square lattice of $n$ nodes per side, and at each node define a pressure $p_i$ and a cell density $N_i$ for $1 \leq i \leq n^2$. See Figure \ref{lattice_diagram} for a representation of this lattice. We label nodal variables with single indices, counting from the bottom-left upwards so that $p_1$ is the pressure at the bottom-left node, and $p_{n^2}$ is the pressure at the top-right node. We define the adjacency matrix $A$ by $A_{ij}=A_{ji}=1$ if nodes $i$ and $j$ are connected, and $A_{ij}=0$ otherwise. We write the graph Laplacian as the matrix $L$ such that $\sum_j A_{ij}(v_i-v_j) = \sum_j L_{ij}v_j$, and note that this formulation of $L$ accounts for no--flux boundary conditions at the edges of the lattice. See \cite{newman_networks:_2010} (and references therein) for a discussion of networks in general and properties of the matrices $A$ and $L$.

We assume the fluid is viscous, Newtonian, and incompressible. We model the flow rate between nodes $i$ and $j$ via Poiseuille flow in a pipe of radius $R_{ij}$, and assume that $R_{ij}$ depends on the cell density at nodes $i$ and $j$. We assume that a constant fluid flow enters the domain from the left and leaves the domain along the right side of the lattice. We assume impermeable horizontal boundaries. We prescribe a unit pressure drop across the scaffold which will have an associated fluid flow rate $Q$ into the domain, and we use a pressure rescaling procedure described in \cite{krause_lattice_2017}, based on work in \cite{shakeel_continuum_2013}, to ensure that a unit flow rate enters the domain (this is the rescaled physical boundary condition, corresponding to a constant fluid flow into the domain). We assume that there is a critical pressure $p_l$ so that cells at node $i$ grow logistically for values of the pressure $p_i < p_l$, and die for values of $p_i > p_l$. We also assume that cells can diffuse between nodes. Our system of non-dimensional equations is then
\be
 \sum_{j=1}^{n^2}A_{ij}R^4_{ij}(\tilde{p_i}-\tilde{p_j})= \begin{cases}
1-\tilde{p_i}, \quad &1 \leq i \leq n,\\
0, \quad &n < i \leq n^2-n,\\
-\tilde{p_i}, \quad &n^2-n < i \leq n^2,
\end{cases}\label{lattice_fluid_eqns}
\ee
\be
\frac{d N_{i}}{d t} = F_{1}(p_i)N_{i}(1-N_{i}) - F_{2}(p_i)N_{i} + \delta n^2 \sum_{j=1}^{n^2} A_{ij}(N_j - N_i), \quad 1 \leq i \leq n^2, \label{general_lattice_cell_eqn}
\ee
\be
R_{ij}\equiv R(N_i,N_j)=1-\frac{\rho}{2}(N_i+N_j), \label{constitutive_lawsa}
\ee
\be
 p_i = \tilde{p_i} \frac{1}{Q}, \label{constitutive_lawsb}
\ee
\be
Q = \sum_{i=1}^n 1-\tilde{p}_i, \label{constitutive_lawsc}
\ee
\be
F_1(p_i) = 1
 - \left( \frac{1}{2} \right)(\tanh[g(p_i - p_l)]+1), \label{f1}
\ee
 \be
F_2(p_i) = \left( \frac{1}{2} \right)(\tanh[g(p_i - p_l)]+1),\label{f2}
\ee
where the parameter $\delta$ represents cell diffusion, $\rho$ determines the effect of cell density on the pipe radius and hence the local fluid flow, and $g$ is a smoothing parameter.  We also prescribe the initial data,
\be
 N_i(0) = N_{i0}. \label{init_lattice_cells}
\ee
Equations \eqref{lattice_fluid_eqns} represent conservation of mass enforced at each node, with a unit pressure drop imposed across the square domain. Equations \eqref{general_lattice_cell_eqn} describe logistic cell growth and death modulated by the functions $F_1$ and $F_2$ of the pressure at that node, along with cell diffusion between nodes. Equation \eqref{constitutive_lawsa} describes how the presence of cells influences the pipe radii, and Equations \eqref{constitutive_lawsb}-\eqref{constitutive_lawsc} rescale the pressures to match the total fluid flow into the domain. Equations \eqref{f1}-\eqref{f2} describe  functions of the pressure chosen to model a `proliferation' and a `death' response of the cells to low and high values of fluid pressure respectively. The use of hyperbolic tangent functions is a modelling choice which follows \cite{shakeel_continuum_2013}, although similar choices such as step functions and mollified step functions have been used elsewhere \cite{coletti_mathematical_2006}.

The analogous dimensionless continuum model is
\be
\boldsymbol{u} = -k(N)\nabla \tilde{p}, \label{pde_fluid1}
\ee
 \be
\nabla \cdot \boldsymbol{u} = 0, \label{pde_fluid2}
\ee
\be
\frac{\partial N}{\partial t} = F_1(p)N(1-N) - F_2(p)N+\delta \nabla^2 N,\label{pde_cell_equation}
\ee
\be
Q_c = \int_0^1 -k(N(0,\hat{y},t))\frac{\partial p}{\partial x}(0,\hat{y})d\hat{y}, \label{pde_permeabilitya}
\ee
 \be
p=\frac{\tilde{p}}{Q_c}, \label{pde_permeabilityb}
\ee
 \be
k(N) = (1-\rho N)^3, \label{pde_permeabilityc}
\ee
\be
F_1(p) = 1
 - \left( \frac{1}{2} \right)(\tanh[g(p - p_c)]+1), \label{pde_f1}
\ee
\be
F_2(p) = \left( \frac{1}{2} \right)(\tanh[g(p - p_c)]+1), \label{pde_f2}
\ee
where $\delta$, $\rho$, $g$ are positive constants with the same meaning as before, $p_c$ is analogous to $p_l$, and $Q_c$ is analogous to $Q$. We prescribe the boundary and initial conditions
\be 
\boldsymbol{u}\cdot\boldsymbol{n} = 0 \enspace \text{at } y= 0,1, \enspace 0 \leq x \leq 1,\label{pde_fluid_boundarya}
\ee
\be 
\tilde{p} = 1 \enspace \text{at } x = 0,\enspace 0 \leq y \leq 1,\quad 
\tilde{p} = 0 \enspace \text{at } x = 1,\enspace 0 \leq y \leq 1,\label{pde_fluid_boundary}
\ee 
\be 
\boldsymbol{n}\cdot\nabla N = 0 \enspace \text{for } \boldsymbol{x} \in  \partial [0,1]^2,\enspace
N(x,y,0) = N_0(x,y). \label{pde_cell_inits}
\ee

Both models have no-flux (Neumann) conditions on all boundaries for the cell density, and on the horizontal boundaries for the fluid. To describe solutions with certain symmetries, we also consider spatially 1-D models without horizontal boundaries, as well as periodic conditions on both cell density and fluid flux along the horizontal boundaries. For the lattice, periodic boundary conditions are encoded in the adjacency matrix $A_{ij}$ by simply adding an edge between each node at the top and the bottom of the lattice. 

We now consider simplifications of the models available when there is a vertical symmetry - that is, when the cell density distribution (and hence the pressure) is vertically symmetric. This is motivated by the presence of this symmetry in some numerical solutions from Section \ref{Numerical_Overview}, and will provide a basis for understanding some mathematical aspects of these models.

\subsection{1-D lattice model}\label{Vertically_Symmetric}
We first simplify the lattice equations by substituting Equation \eqref{constitutive_lawsb} into Equation \eqref{lattice_fluid_eqns} to give
\be\label{simplified_lattice_eqns}
 \sum_{j=1}^{n^2}A_{ij}R^{4}_{ij}(p_i-p_j)= \begin{cases}
\frac{1}{Q}-p_i, \quad &1 \leq i \leq n,\\
0, \quad &n < i \leq n^2-n,\\
-p_i, \quad &n^2-n < i \leq n^2,
\end{cases}
\ee
where $Q$ is given in terms of $p_i$ from Equation \eqref{constitutive_lawsc} by
\bes
 Q = \frac{n}{1+\sum_{j=1}^n p_j}.
\ees
Using this expression for $Q$, our system of fluid equations \eqref{simplified_lattice_eqns} becomes
\begin{equation}\label{final_lattice_fluid}
 \sum_{j=1}^{n^2}A_{ij}R^{4}_{ij}(p_i-p_j)= \begin{cases}
\frac{1}{n}\left(1+\sum_{j=1}^n p_j\right)-p_i, \quad &1 \leq i \leq n,\\
0, \quad &n < i \leq n^2-n,\\
-p_i, \quad &n^2-n < i \leq n^2.
\end{cases}
\end{equation}

We solve the pressure equations when the solutions are vertically symmetric. We label the value at a representative node by $p_i \equiv \hat{p}_k$ where $k = \ceil{\frac{i-1}{n}}$, and $\ceil{a}$ denotes the smallest integer greater than $a$. So the first $n$ variables have the value $\hat{p}_1$, the next $n$ have the value $\hat{p}_2$, etc. We analogously denote the cell densities as $N_i = \hat{N}_k$. We rewrite Equation \eqref{final_lattice_fluid} using these variables as
\begin{equation}
 \sum_{j=1}^{n}\hat{A}_{ij}R^4(\hat{N}_i,\hat{N}_j)(\hat{p}_i-\hat{p}_j)= \begin{cases}
\frac{1}{n}, \quad &i =1,\\
0, \quad &1 < i < n,\\
-\hat{p}_n, \quad  &i =n,
\end{cases}\label{symmetric_fluid}
\end{equation}
where we have used the simplification $(1+\sum_{j=1}^n p_j)/n-p_i = (1+n \hat{p}_1)/n-\hat{p}_1 = 1/n$. The adjacency matrix $\hat{A}_{ij}$ represents the path graph on $n$ vertices, so that $\hat{A}_{ij}=1$ if $i=j\pm1$ or $\hat{A}_{ij}=0$ otherwise. Summing all $n$ equations in \eqref{symmetric_fluid} we find that $\hat{p}_n = 1/n$, where we have used the symmetry of the function $R$. For $1 \leq i<n$ we can iteratively solve these equations to find
\be\label{lattice_p_sol}
\hat{p}_i = \frac{1}{n}\left(1 + \sum_{k=i}^{n-1}R^{-4}(\hat{N}_k,\hat{N}_{k+1}) \right).
\ee
This is equivalent to saying that resistance in serial circuits is additive, where this pressure variable is playing the role of voltage and $R^{-4}(\hat{N}_k,\hat{N}_{k+1})$ is the resistance between nodes $k$ and $k+1$ \citep{oh2012design}.

Hence, in the 1-D model, we obtain a system of $n$ coupled equations for the cell density:
\be\label{1dlattice}
\frac{d \hat{N_{i}}}{d t} = F_{1}(\hat{p}_i)\hat{N_{i}}(1-\hat{N_{i}}) - F_{2}(\hat{p}_i)\hat{N_{i}} + \delta n^2 \sum_{j=1}^{n} \hat{A}_{ij}(\hat{N}_j - \hat{N}_i), \quad 1 \leq i \leq n,
\ee
where $\hat{p}_i$ given by \eqref{lattice_p_sol}.

\subsection{1-D continuum model}\label{Vertically_Symmetric_PDE}
We assume that the cell density varies only in the $x$ direction, so that the permeability $k=k(N(x))$ also only varies in this direction. We then have by Equations \eqref{pde_fluid1}-\eqref{pde_fluid2} and \eqref{pde_fluid_boundarya}-\eqref{pde_fluid_boundary} that $\tilde{p}=\tilde{p}(x)$, so the governing flow equations become
\be
\frac{\partial}{\partial_x}\left (k(x)\frac{\partial\tilde{p}}{\partial_x} \right ) = 0, \label{flow_rescale_eqn_xa}
\ee
 \be
Q_c = -k(N(0))\frac{\partial \tilde{p}}{\partial x}(0), \label{flow_rescale_eqn_xb}
\ee
 \be
p=\frac{\tilde{p}}{Q_c}. \label{flow_rescale_eqn_xc}
\ee

Integrating Equation \eqref{flow_rescale_eqn_xa} subject to the boundary conditions \eqref{pde_fluid_boundary} we find
\begin{equation}\label{pde_til_pressure}
\tilde{p}(x) = \frac{\displaystyle \int_x^1 \frac{1}{k(N(\alpha))}d\alpha}{\displaystyle \int_0^1 \frac{1}{k(N(\alpha))}d\alpha}.
\end{equation}
Substituting \eqref{pde_til_pressure} into \eqref{flow_rescale_eqn_xb} we compute $Q_c$, and from \eqref{flow_rescale_eqn_xc} we find
\begin{equation}\label{pde_sym}
p(x) = \int_x^1 \frac{1}{k(N(\alpha))}d\alpha.
\end{equation}

Substituting \eqref{pde_sym} into Equation \eqref{pde_cell_equation} we obtain a single equation for the cell density,
\be
\frac{\partial N}{\partial t} = F_1\left (\int_x^1 \frac{1}{k(N(\alpha))}d\alpha \right)N (1-N) - F_2\left (\int_x^1 \frac{1}{k(N(\alpha))}d\alpha \right)N+\delta \partial_x^2 N.\label{pde_cell_equation_sym}
\ee
Equation \eqref{pde_cell_equation_sym} is a spatially nonlocal scalar reaction-diffusion equation for the cell density $N$.

\subsection{Boundedness of dynamics}\label{bounded}
We now comment on the boundedness of solutions to both models in the full 2-D setting, which will enable us to deduce reasonable ranges of model parameters n Sections \ref{Numerical_Overview} and \ref{Bifurcations}. We first show that solutions to the Equations \eqref{lattice_fluid_eqns}-\eqref{f2} are bounded within the region $[0,1]^{n^2}$. We note that for all $p \in \mathbb{R}$, the functions in Equations \eqref{f1}-\eqref{f2} satisfy $0 \leq F_1(p) \leq 1$ and $0 \leq F_2(p) \leq 1$. Now we assume that for some $t^*$, $N_i(t^*)\in [0,1]$ for all $i$, and $N_k(t^*)=1$ for at least one $k$. We then have that
\be
\frac{d N_k}{dt} = -F_2(p_k) + \delta n^2 \sum_{j=1}^{n^2} A_{kj}(N_j - 1) \leq 0, \label{bounded_above}
\ee
Hence continuously differentiable ($C^1$) solutions to Equations \eqref{general_lattice_cell_eqn} are bounded above. 

Similarly, assume that for some $t^*$, $N_i(t^*)\in [0,1]$ for all $i$, and that $N_m(t^*)=0$ for at least one $m$. We then have that,
\be
\frac{d N_m}{dt} = \delta n^2 \sum_{j=1}^{n^2} A_{mj}(N_j) \geq 0, \label{bounded_below}
\ee
and so continuously differentiable solutions are also bounded below. Note that these bounds are independent of the pressures $p_i$. So we have that $N_i(t)\in[0,1]$ for all $i$ and for all $t \geq 0$.

While we are not aware of a formal comparison principle applicable to Equations \eqref{pde_fluid1}-\eqref{pde_cell_inits}, the above argument gives a heuristic bound on the behaviour of solutions for these equations as well. The existence of spatiotemporal oscillations, which we show in Section \ref{Numerical_Overview}, suggests that a general comparison principle does not hold for this system. Hence a rigorous treatment of Equations \eqref{pde_fluid1}-\eqref{pde_cell_inits} would be needed in order to formally extend this argument. Nevertheless, we will assume that PDE solutions are similarly bounded, and our numerical solutions confirm this.

\section{Exploration of solution behaviours}\label{Numerical_Overview}
In this section we present numerical solutions for the 2-D lattice equations \eqref{lattice_fluid_eqns}-\eqref{f2} as well as the PDE system given by \eqref{pde_fluid1}-\eqref{pde_cell_inits}. We fix $\rho = 0.9$ so that cell growth significantly affects the effective permeability of the medium. To capture sufficiently sharp behaviour in the pressure functions, we set $g = 60$. The remaining parameters are varied to demonstrate the range of solution behaviours. The PDE system was simulated using the finite element software \textsc{COMSOL} with 24,912 triangular elements, as well as a finite difference scheme to verify the results. The lattice model was solved using an explicit adaptive Runge-Kutta method, `ode45' in \textsc{Matlab}, with a maximal time step of $10^{-3}$. The restriction of the maximal time step was to ensure a good approximation of bifurcation phenomena \cite{christodoulou_discrete_2008}. Convergence checks were carried out in the time step in the numerical schemes for both models, and in the number of elements for the PDE.

From Section \ref{bounded} and the monotonicity of Equations \eqref{lattice_p_sol} and \eqref{pde_sym}, we can compute the maximum and minimum possible pressures throughout the domain in each (1-D) model. We then deduce the ranges of the pressure threshold parameters $p_l$ and $p_c$ that lie between maximum and minimum pressures in the scaffold. To compute these extremal pressures, we set the cell densities $N_i=0$ or $N_i=1$ for all $i$. Using \eqref{lattice_p_sol}, we see that if $p_l < 1/n$, then $p_i>p_l$ for all $i$, so that cells at every node will die (exponentially decay) due to high values of the pressure. Similarly, if $p_l > (1+(n-1)(1-\rho)^{-4})/n$, then $p_i<p_l$, and all nodes will approach a uniform value of $N_i=1$ logistically. In order to realize nontrivial dynamics, we vary $p_l$ as 
\be\label{lattice_range}
\frac{1}{n} < p_l  < \frac{1+(n-1)(1-\rho)^{-4}}{n} \approx 10^4,
\ee
for the lattice and using \eqref{pde_sym}, similar reasoning leads to the bounds,
\be\label{pde_range}
0 < p_c  <(1-\rho)^{-3} = 10^3,
\ee
for the PDE. Note the difference in the orders of magnitude between these two ranges, which is due to the constitutive difference between the response of the flow to the cell density. Specifically, the Poiseuille flow in the lattice given by \eqref{lattice_fluid_eqns} has a quartic dependence on cell density, whereas the cubic permeability in Darcy's law given by \eqref{pde_permeabilityc} is cubic in the cell density. See \cite{krause_lattice_2017} for a discussion of why these constitutive exponents differ between the models. Changing the exponent in the permeability, Equation \eqref{pde_permeabilityc}, from $3$ to $4$ gives quantitatively similar behaviour for the PDE and very large lattices.

For all of our simulations, we set the initial data to be a uniform state perturbed by normally distributed spatial noise with zero mean and standard deviation $10^{-4}$, so that $N_i(0) \approx 0.1$ and $N_0(x,y) \approx 0.1$ . The lattice and PDE models are both insensitive to perturbations in initial conditions, giving identical long-time behaviour across hundreds of realizations of these noisy initial data. The cell densities grow uniformly until the pressure at some point of the domain reaches the threshold $p_c$ for the PDE or $p_l$ for the lattice, where some regions then experience cell death due to high pressure. The simulations then tend to a spatial horizontal gradient in cell density with low cell density on the left side of the lattice (where the pressure is highest) and high cell density on the right side of the lattice (where the pressure is lowest). We observe one of three different long time behaviours: a stable steady state, a vertically symmetric `pulsating' oscillation, or a vertically asymmetric oscillation. The steady states we observe are either vertically symmetric or asymmetric.

\begin{figure}  
\centering
	\subfloat[]{
    \includegraphics[width=0.25\textwidth]{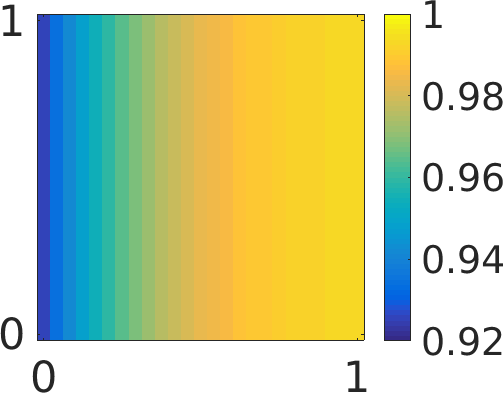}
    \label{pressure1figa}}\quad
    \subfloat[]{
    \includegraphics[width=0.25\textwidth]{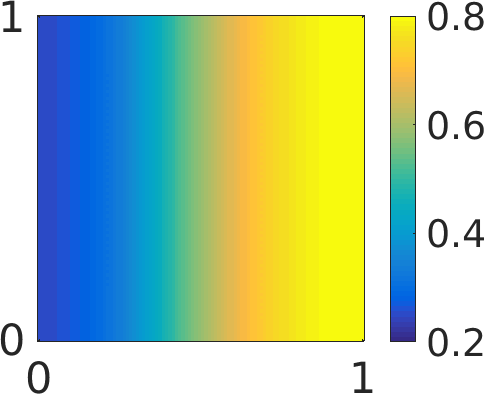}
    \label{pressure1figb}}\hfill%
    
    \subfloat[]{
    \includegraphics[width=0.25\textwidth]{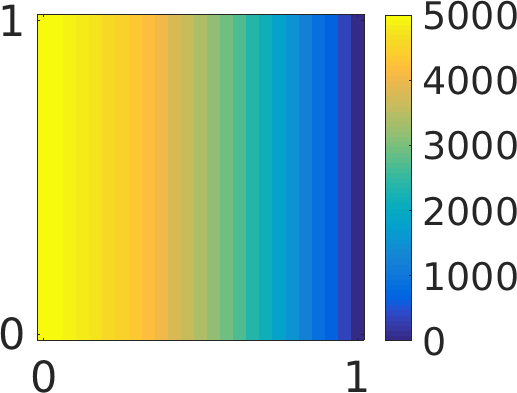}
    \label{pressure1figc}}\quad
    \subfloat[]{
    \includegraphics[width=0.25\textwidth]{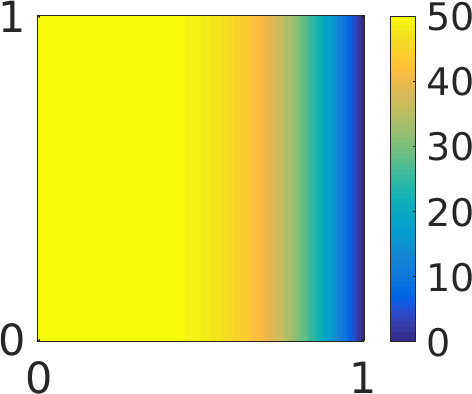}
    \label{pressure1figd}}\hfill%
    
    \subfloat[]{
    \includegraphics[width=0.35\textwidth]{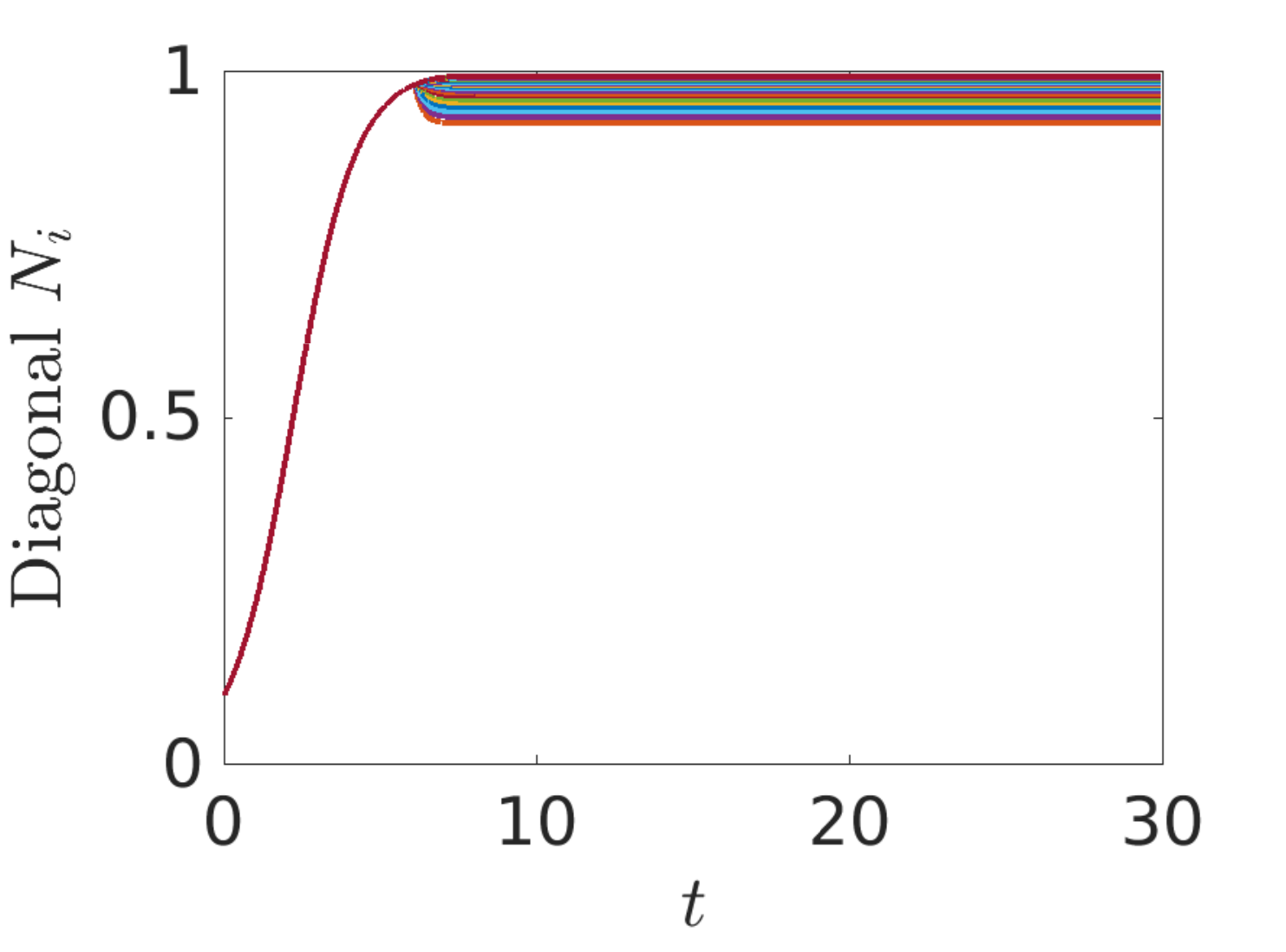}
    \label{pressure1fige}}%
    \subfloat[]{
    \includegraphics[width=0.35\textwidth]{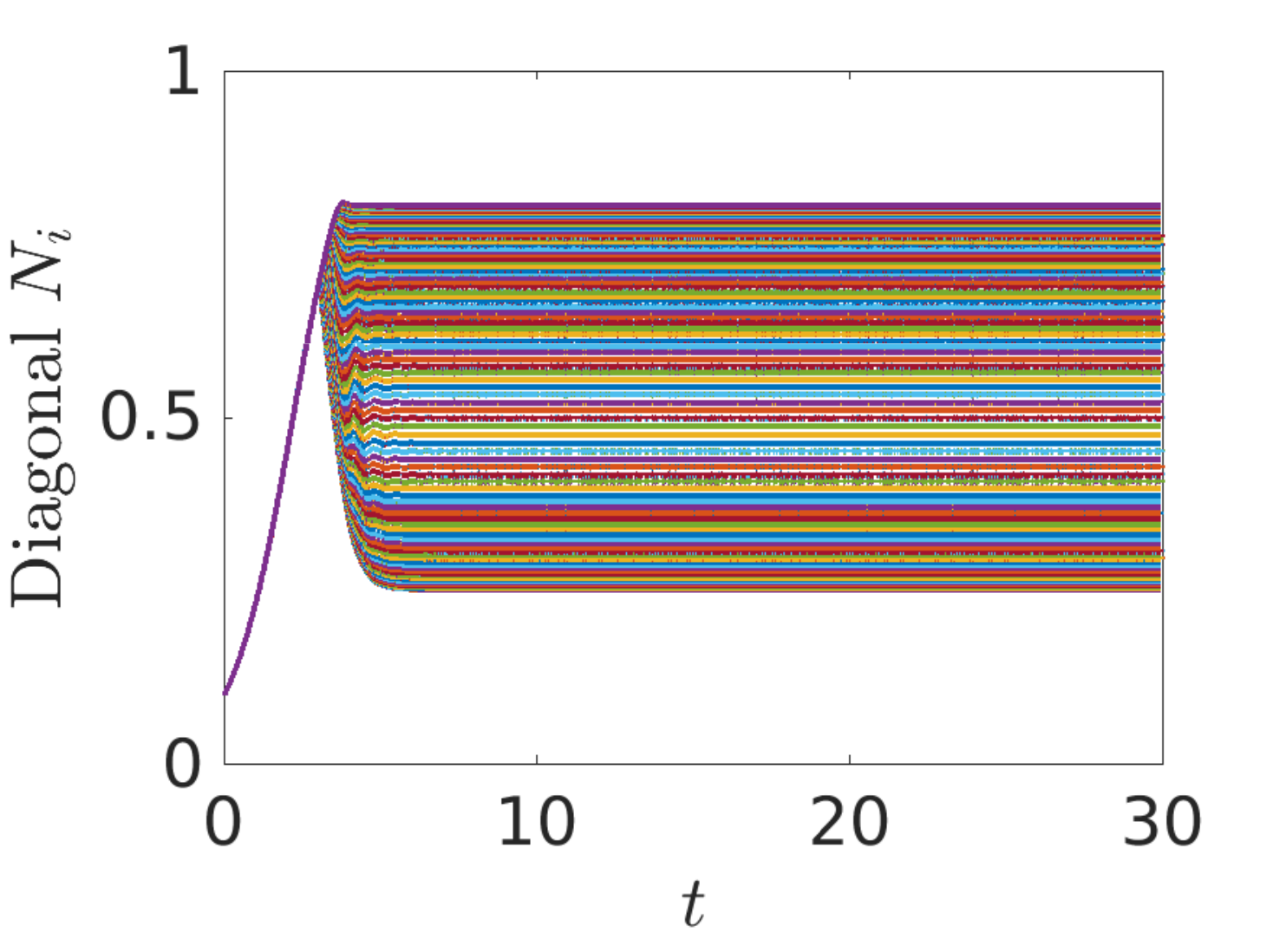}
    \label{pressure1figf}}\hfill%
\caption{Cell density plots in (a)-(b), pressure plots in (c)-(d) at $t=30$ for $\delta=10^{-1}$, and time series showing the evolution from uniform growth to these stationary patterns in (e)-(f). Plots (a), (c) and (e) correspond to simulations with $p_l=5000$, with $n=25$ and plots (b), (d), and (f) correspond to simulations with $p_l=50$ with $n=100$.}
\label{pressure1fig}
\end{figure}

At very large values of $\delta$ (e.g. $\sim O(1)$), or large values of the threshold parameters $p_l$ or $p_c$, all simulations tend to a similar steady state with a gradual horizontal gradient in cell density (not shown). In Figure \ref{pressure1fig} we plot two example cell density distributions alongside corresponding plots of pressure for $\delta = 10^{-1}$. We also plot time series of the nodal evolution along the diagonals of the scaffold, i.e. all nodes along the lines $x=y$ and $x=1-y$, in Figures \ref{pressure1fige}-\ref{pressure1figf}. For large $\delta$, there is a non-negligible cell density present in the region of high pressure despite the local cell density being governed by exponential death. The pressure does not vary significantly in this region of lower cell density near the left side of the scaffold, as the effective permeability changes little at low cell densities. In these parameter ranges, the 1-D reduction described in Section \ref{Vertically_Symmetric} is a quantitatively accurate approximation to the cell density along any row of the lattice. 

\begin{figure}  
\centering
	\subfloat[]{
    \includegraphics[width=0.35\textwidth]{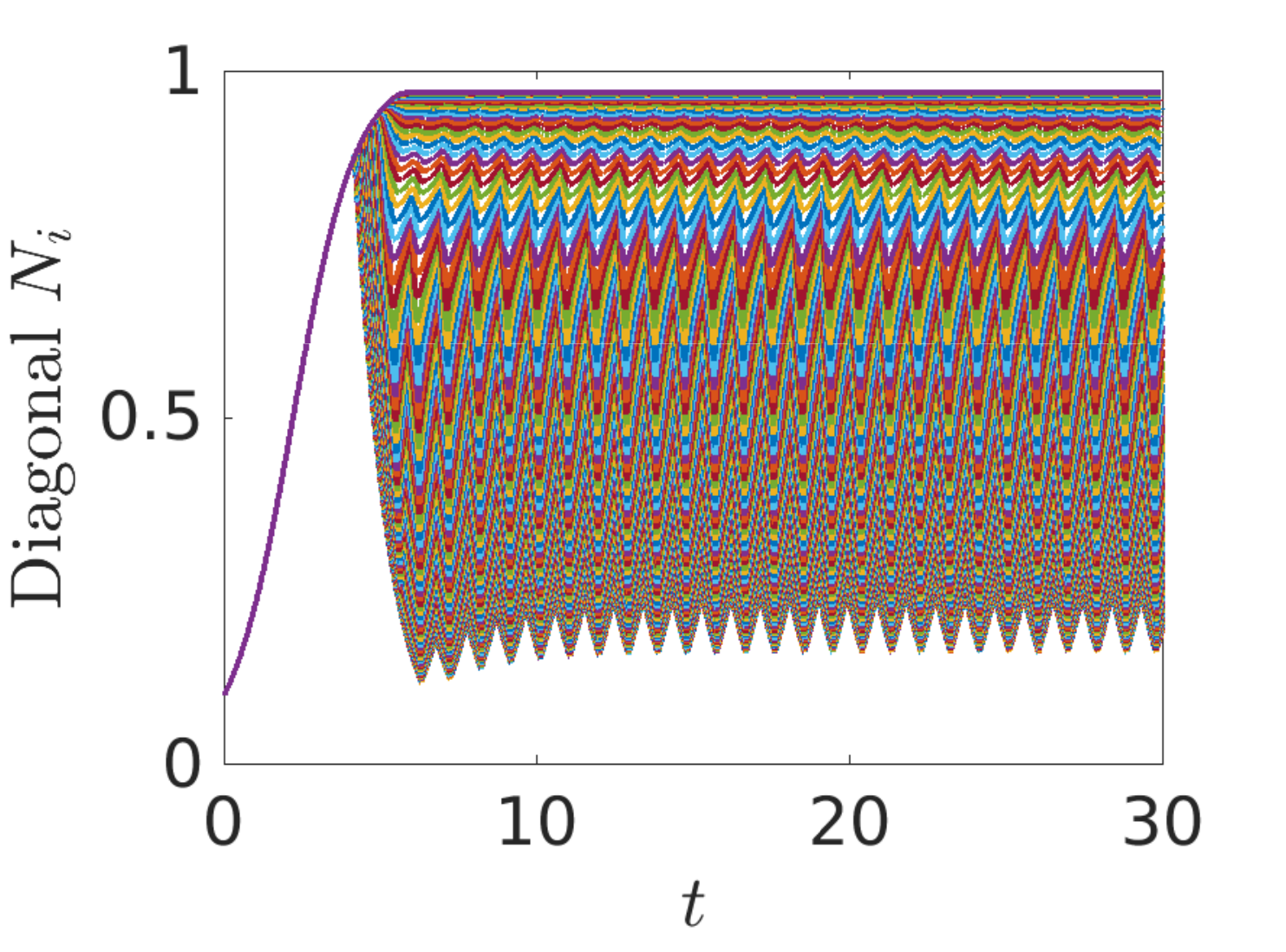}
    \label{cell1timeseriesleft}}%
    \subfloat[]{
    \includegraphics[width=0.35\textwidth]{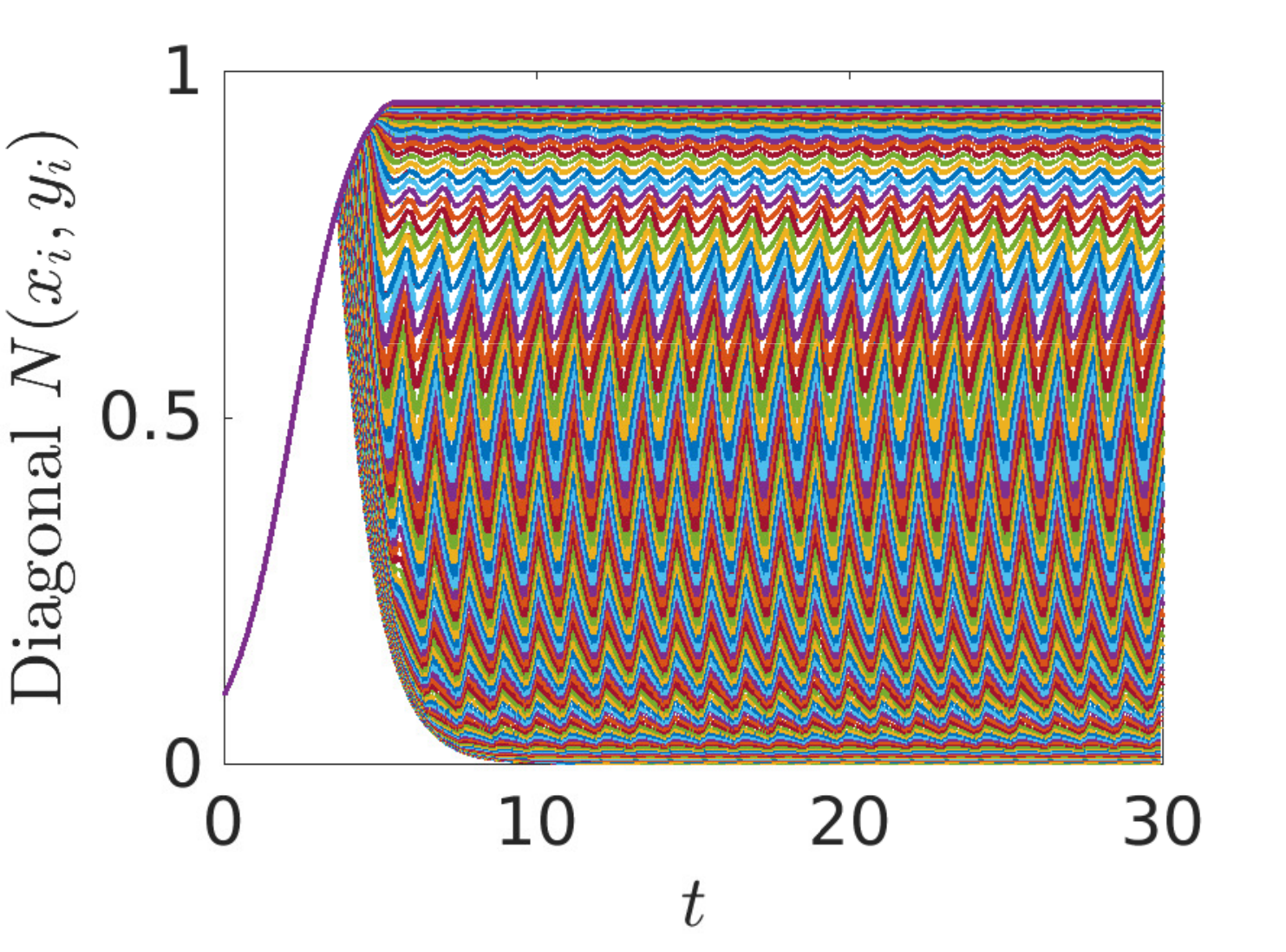}
    \label{cell1timeseriesright}}\hfill%
\caption{Time series of nodal values of the cell density for a lattice of size $n=100$, $\delta=10^{-2}$, $p_l=500$ in (a), and along 200 corresponding diagonal interpolants of the PDE system with $\delta=10^{-2}$, $p_c=50$ in (b).}
\label{cell1timeseries}
\end{figure}

\begin{figure}
\centering
	\subfloat[]{
    \includegraphics[width=0.5\textwidth]{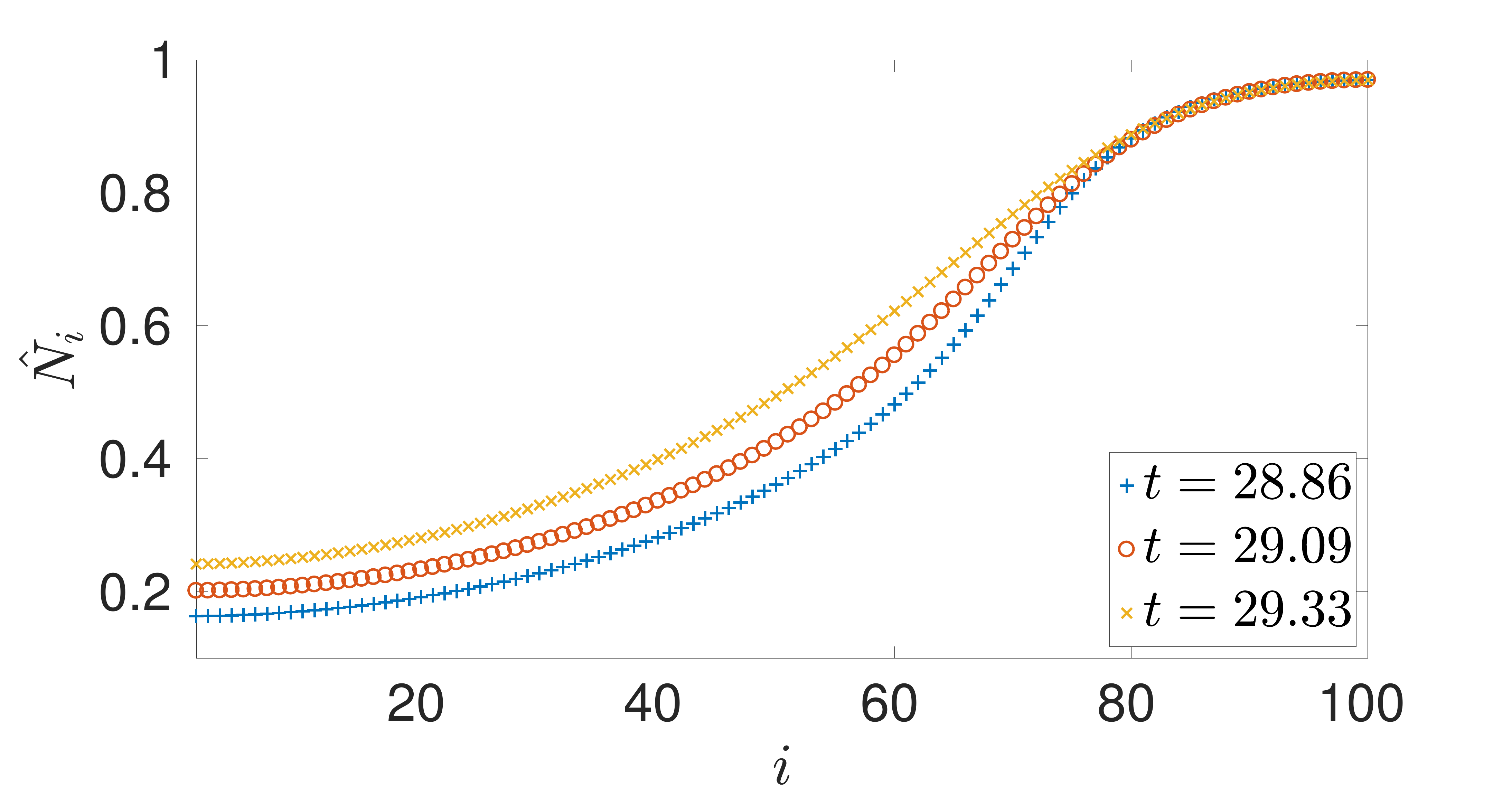}
    \label{cell1figa}}%
    \subfloat[]{
    \includegraphics[width=0.5\textwidth]{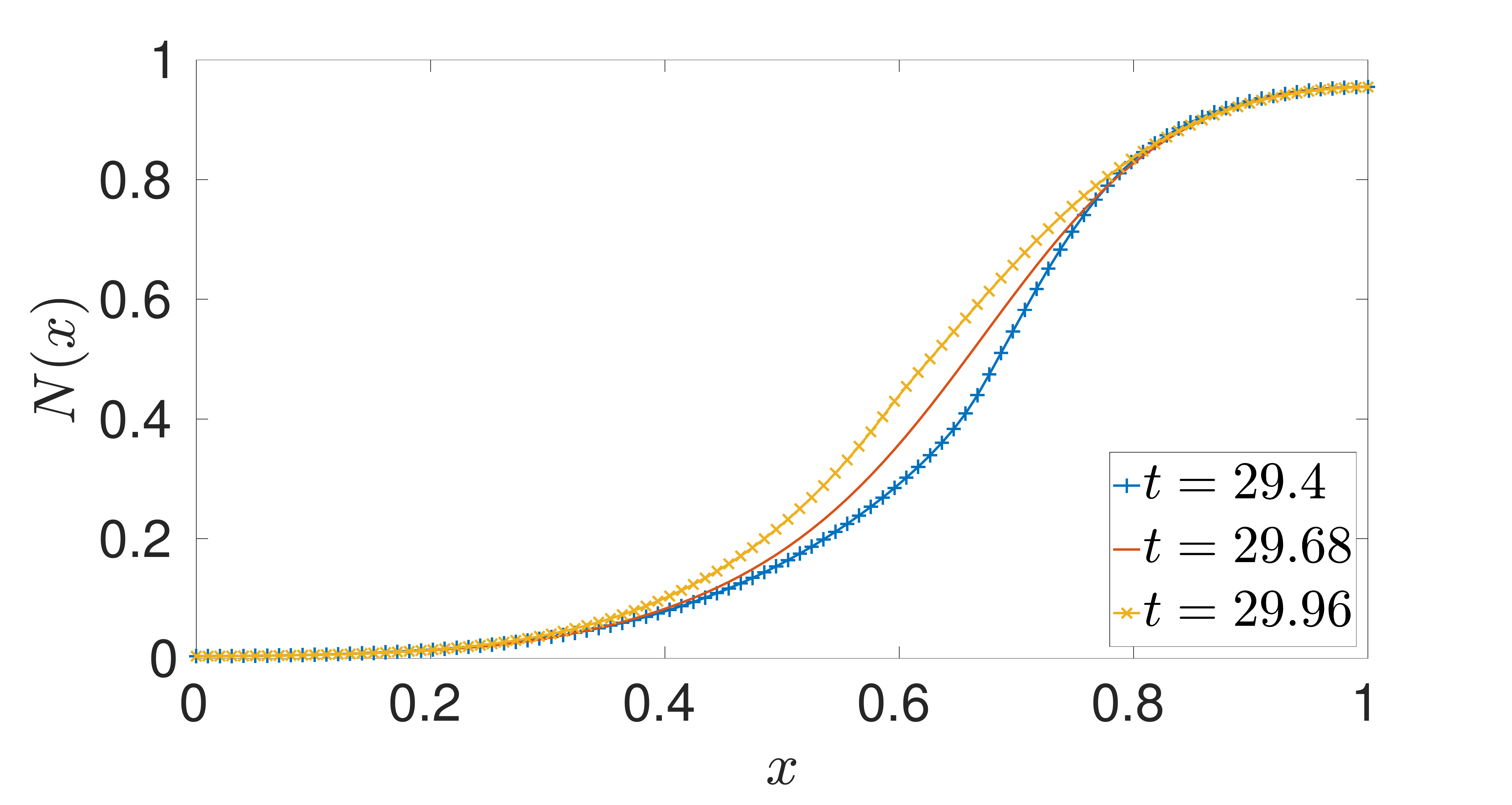}
    \label{cell1figb}}\hfill%
\caption{Plots (a) are of cell density for a lattice of size $n=100$, $\delta=10^{-2}$, $p_l=500$ at three times during one oscillation. Plots (b) are of cell density for the PDE with $\delta=10^{-2}$, $p_c=50$ at three times during one oscillation. Note that we have simulated the 2-D models but averaged over any vertical variation in cell density in both cases.}
\label{cell1fig}
\end{figure}

For smaller values of the diffusion we observe a different long-time behaviour. In some cases, the cell density throughout the scaffold oscillates with a `pulsing' between the right and the left side of the domain, maintaining a vertical symmetry throughout the oscillations. In Figure \ref{cell1timeseries} we show the cell densities across diagonal nodes for two cases: a lattice of size $n=100$ in Figure \ref{cell1timeseriesleft} and the PDE in Figure \ref{cell1timeseriesright}. For each, we observe stable oscillations throughout the domain. The lattice simulation in Figure \ref{cell1timeseriesleft} has the largest oscillations in cell density for nodes at the far left of the lattice, and these values never get close to the zero. In comparison, the PDE simulation in Figure \ref{cell1timeseriesright} has its largest oscillations in the middle of the domain, and the far leftmost part of the domain exhibits very small cell densities. In Figure \ref{cell1fig} we plot the cell density at three different points in time during one oscillation half period for the same simulations as in \ref{cell1timeseries}. In Figure \ref{cell1figa}, we plot just the 1-D values of the column-averaged cell density as these oscillations are vertically symmetric, and similarly we plot vertically-averaged cell densities for the PDE in Figure \ref{cell1figb}. The oscillations move between the yellow and blue curves with a period of $t\approx 0.94$ in Figure \ref{cell1figa} and $t \approx 1.12$ in Figure \ref{cell1figb}. These large oscillations persist in the 1-D reductions of both models, as well as when the horizontal boundary conditions for cells and fluid flow are changed from Neumann to periodic conditions. This suggests that these oscillations are inherently a 1-D phenomenon.

\begin{figure}
\centering
	\subfloat[]{
    \includegraphics[width=0.25\textwidth]{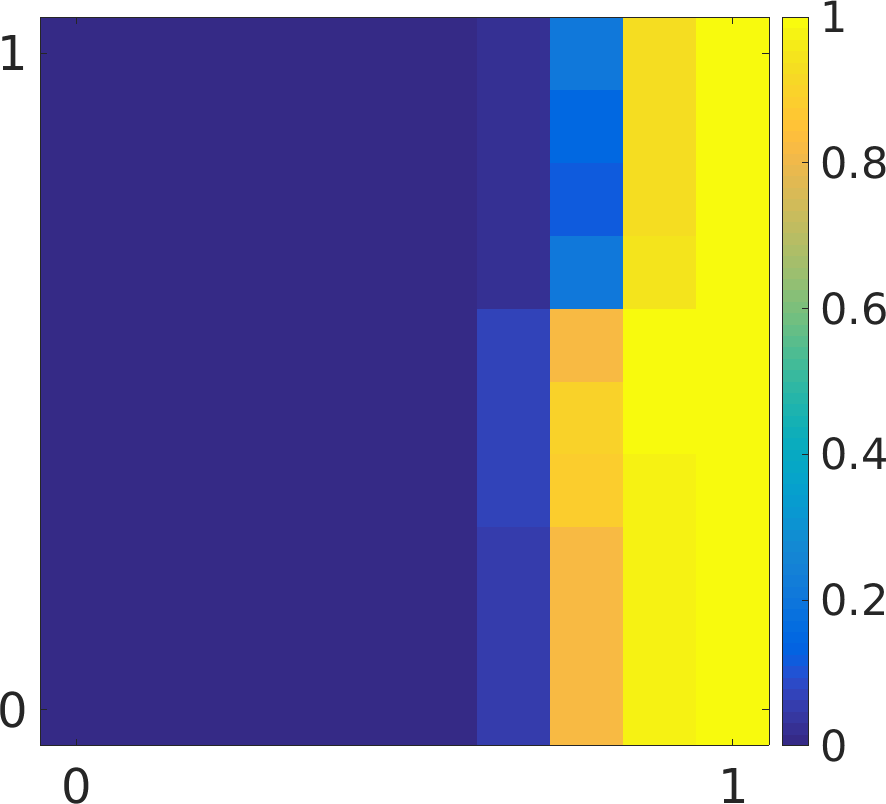}
    \label{defecta}}\hspace{1cm}%
    \subfloat[]{
    \includegraphics[width=0.25\textwidth]{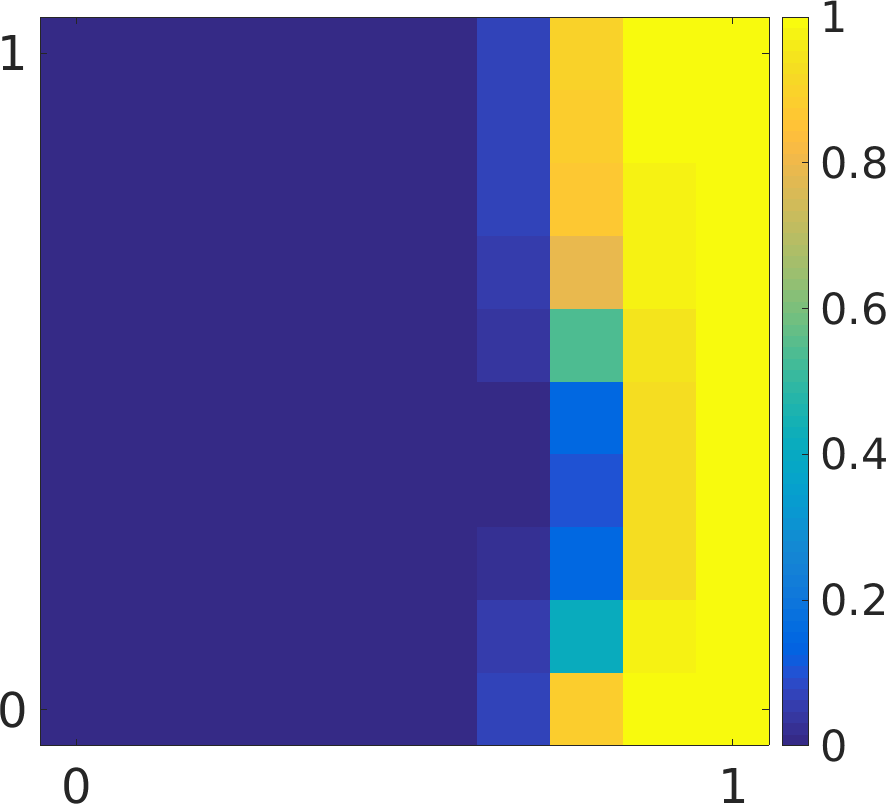}
    }\hfill%
    \\
    \subfloat[]{
    \includegraphics[width=0.25\textwidth]{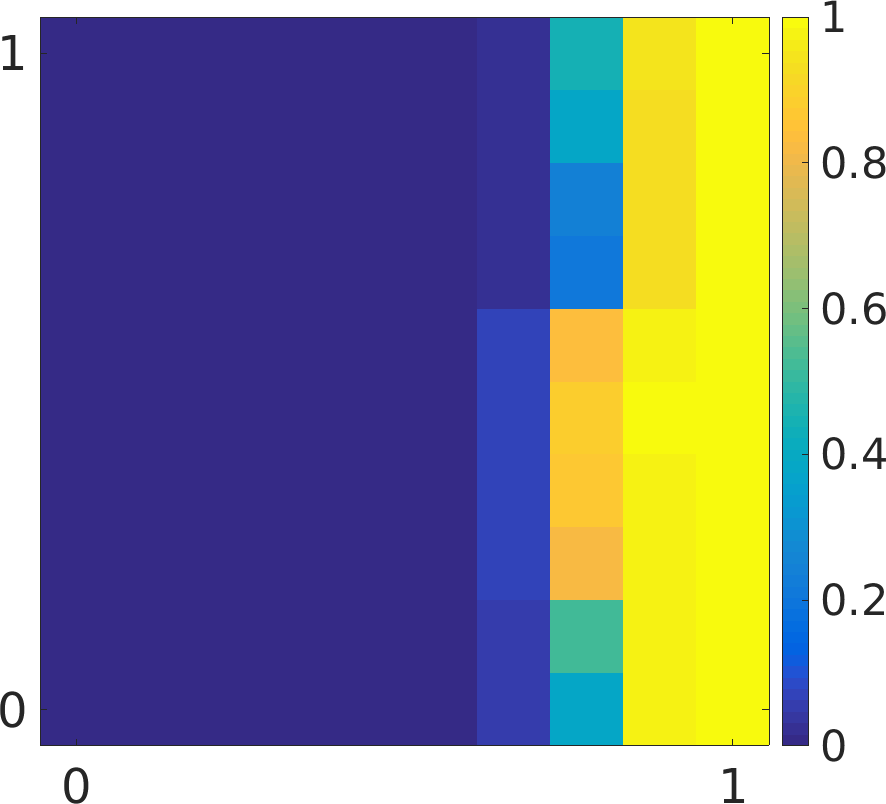}
    }\hspace{1cm}%
    \subfloat[]{
    \includegraphics[width=0.25\textwidth]{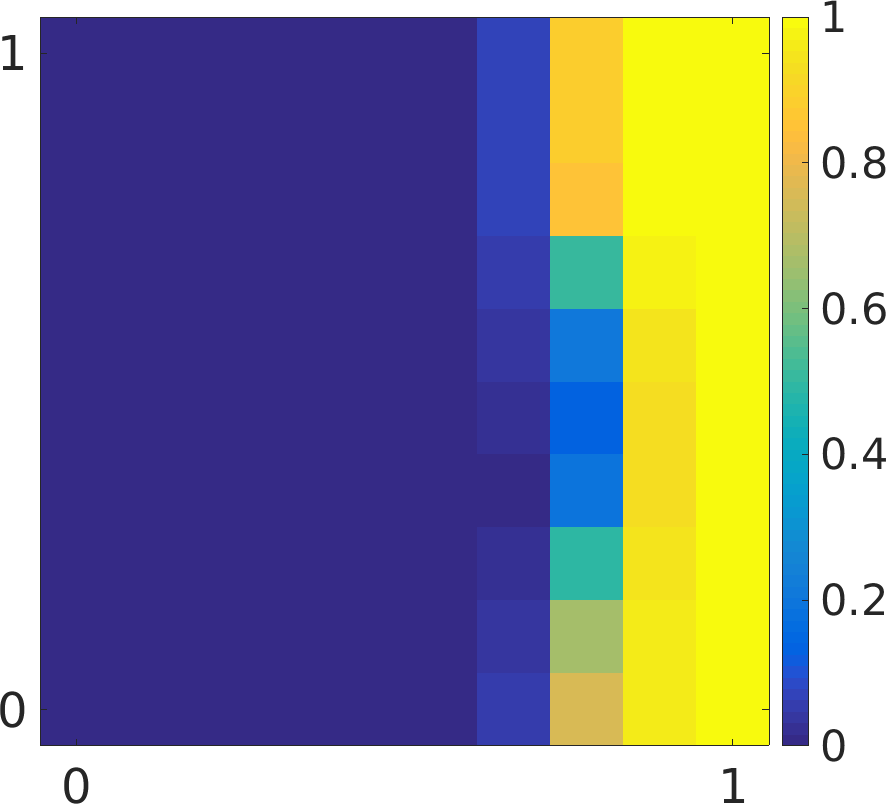}
    \label{defectd}}\hfill%
\caption{Plots of cell density for a lattice of size $n=10$, $\delta=10^{-3}$, $p_l=500$ at times $t=15$, $t=20$, $t=25$, and $t=30$ in (a)-(d) respectively. }
\label{defect_oscillation}
\end{figure}

\begin{figure}
\centering
	\subfloat[]{
    \includegraphics[width=0.32\textwidth]{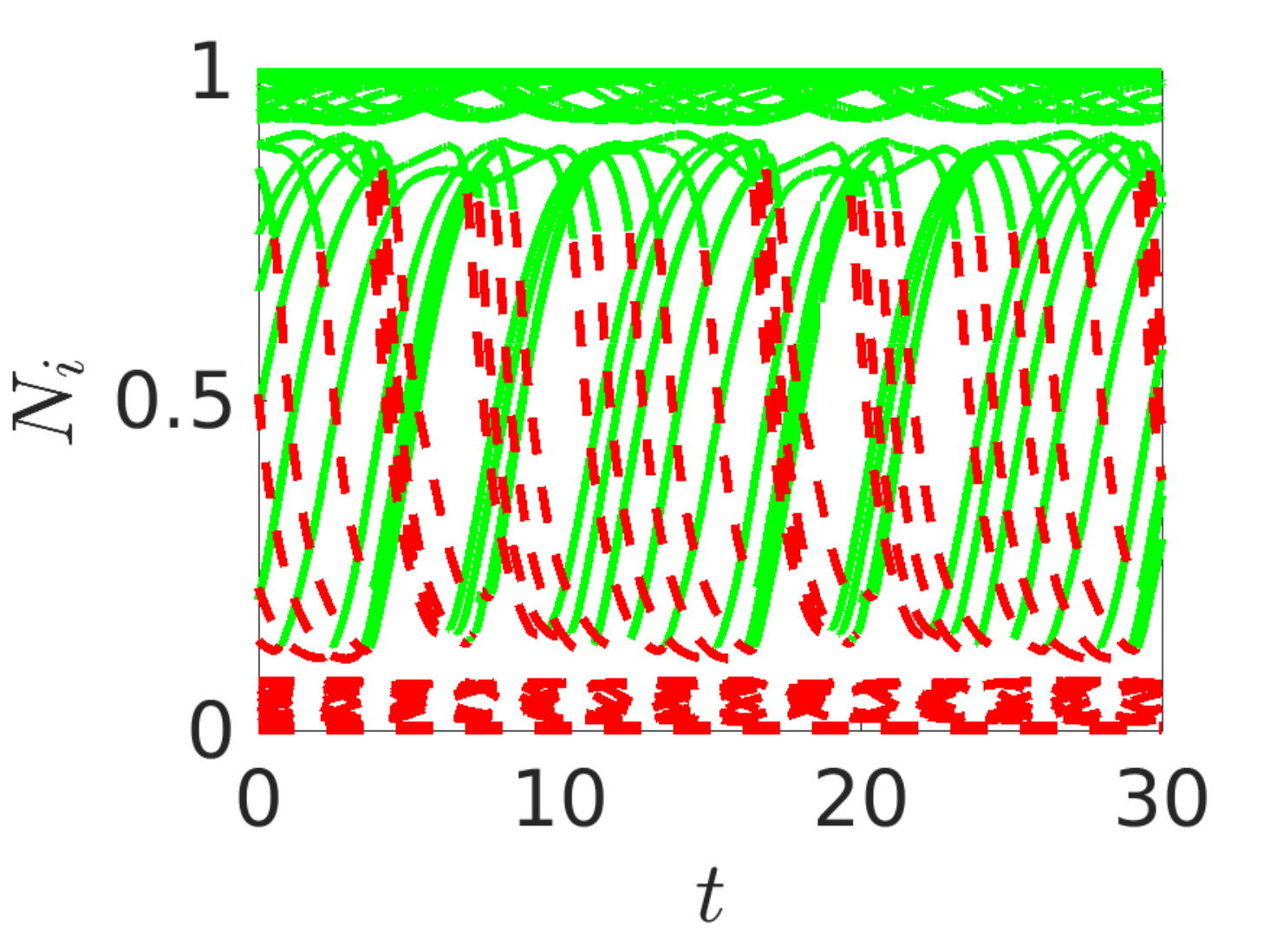}
    \label{defecte}}\hspace{0.5cm}%
    \subfloat[]{
    \includegraphics[width=0.32\textwidth]{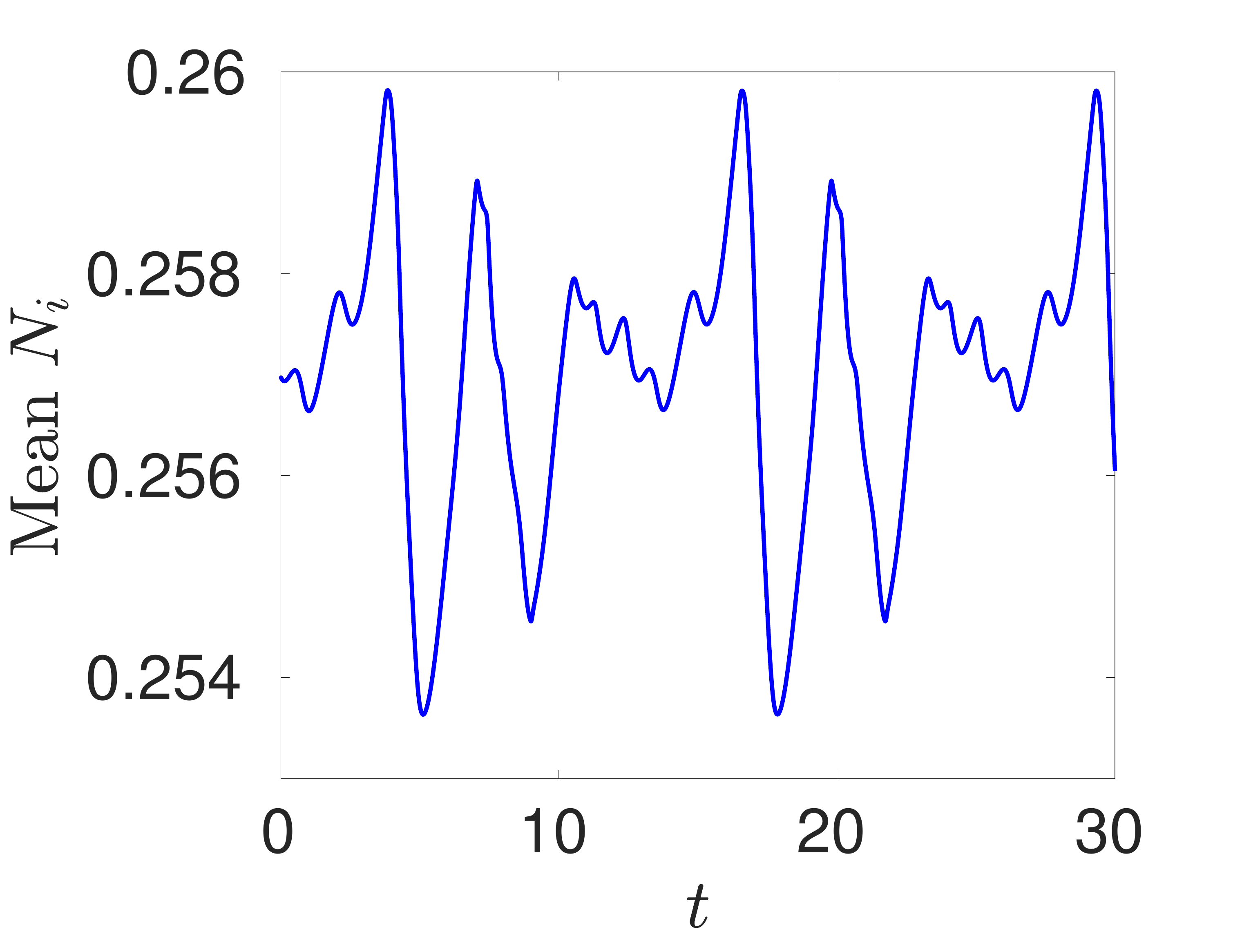}
    \label{defectf}}\hfill%
\caption{Plots of the nodal values of the cell density at every node in the lattice, corresponding to Figure \ref{defect_oscillation}, are shown in (a), where the line is coloured green if $p_i < p_l$ and red if $p_i > p_l$. In (b) we plot the spatial mean cell density during the oscillations. Note that we have only shown time series for latter time periods after the cell densities have settled into oscillations.}
\label{defect_oscillationNEW}
\end{figure}

The aforementioned behaviours - vertically symmetric steady states and oscillations - appear in the PDE and in large lattices (e.g. for $n=100$). Qualitatively different steady states and oscillations exist in the smaller lattices. In many of the smaller lattice simulations, we observe a complicated breaking of the vertical symmetry. We give an example of vertically asymmetric oscillations in Figure \ref{defect_oscillation}, where the cell density oscillates vertically between the two impermeable horizontal boundaries. We plot a corresponding time series in Figure \ref{defecte}, which demonstrates a separation of solutions into `bands,' corresponding to specific columns in the lattice. The largest of these bands that changes between a growing and a dying state (shown in dashed red and solid green in Figure \ref{defecte}) corresponds to the third column from the right boundary in the lattice. The bands which are always green (locally growing logistically) correspond to nodes to the right of this column, and those that are always red correspond to nodes to the left of this column. We note that the spatial mean cell density during an oscillation, plotted in Figure \ref{defectf}, has many turning points despite being simply periodic. For this particular parameter set, changing the horizontal boundary conditions from no-flux on the fluid and cells to periodic conditions eliminated the oscillations (and solutions tended to a steady state). Similarly, the 1-D model did not display oscillating behaviour for these parameters. 

\begin{figure}
\centering
    \subfloat[]{
    \includegraphics[width=0.3\textwidth]{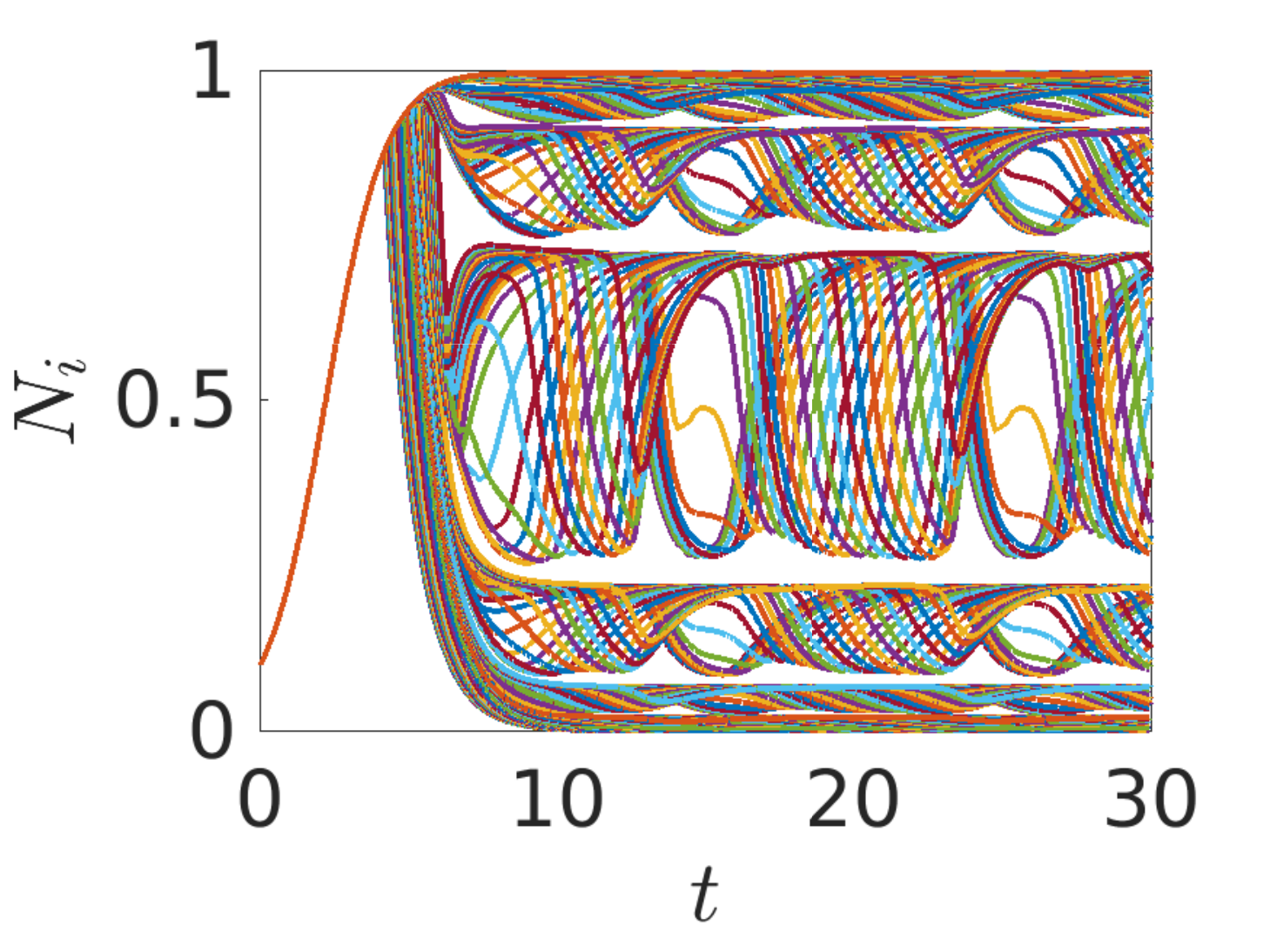}
    \label{defect_oscillation2a}}%
    \subfloat[]{
    \includegraphics[width=0.3\textwidth]{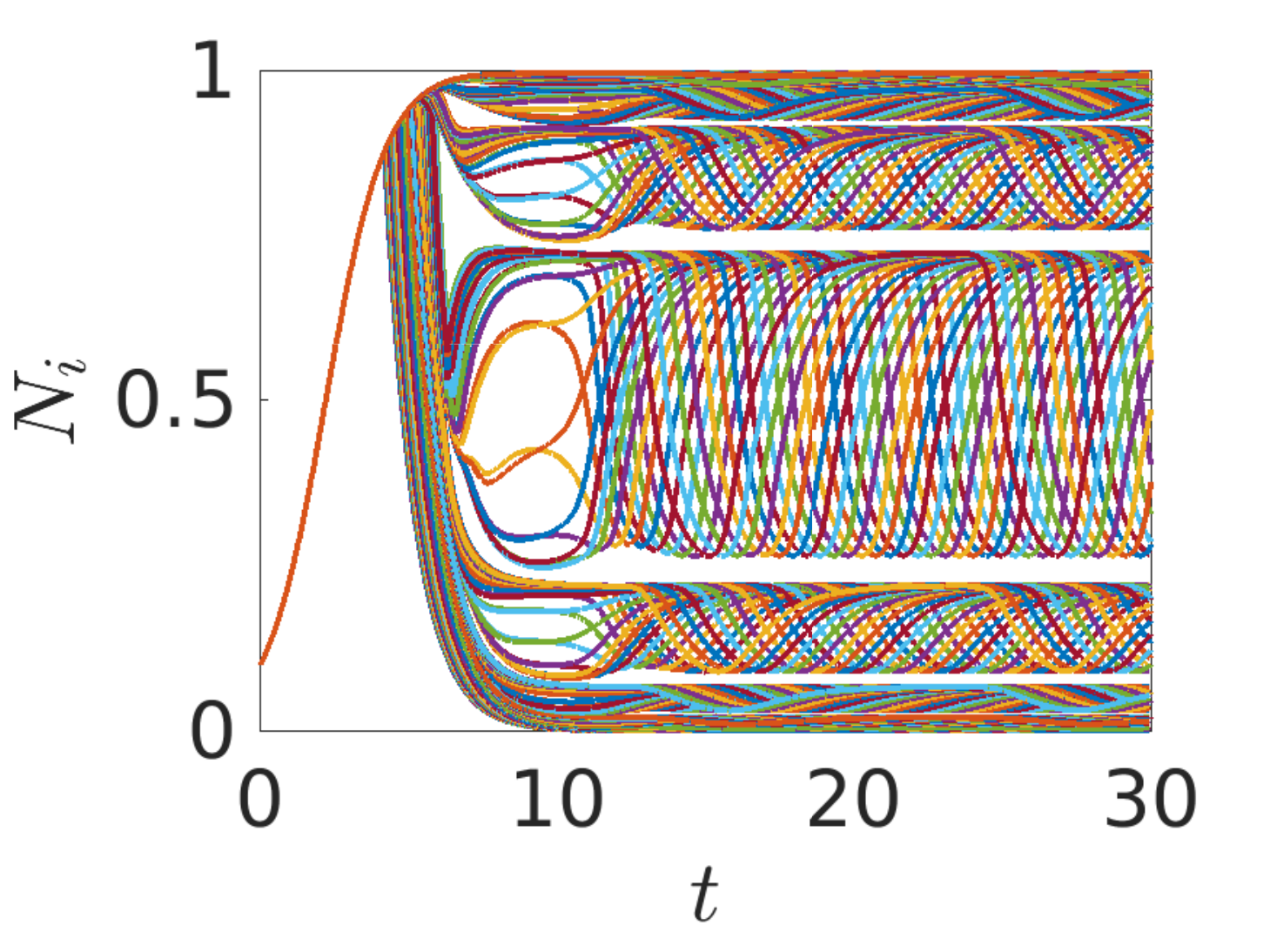}
    \label{defect_oscillation2b}}%
    \subfloat[]{
    \includegraphics[width=0.3\textwidth]{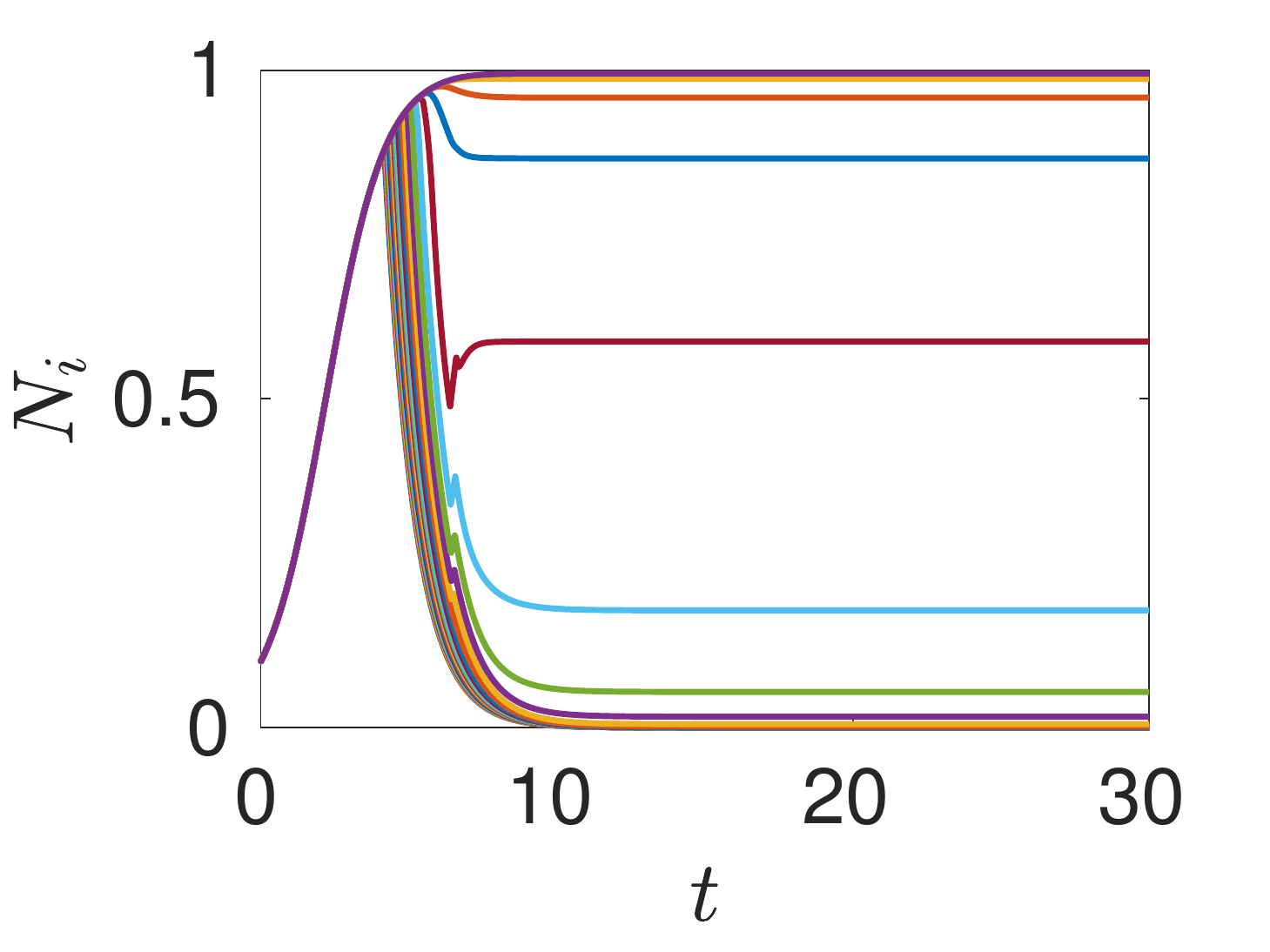}
    \label{defect_oscillation2c}}\hfill%
\caption{Nodal plots of cell densities for a lattice of size $n=25$, $\delta=10^{-3}$, $p_l=500$ with (a) no-flux conditions on the horizontal boundaries, (b) periodic conditions on these boundaries, and (c) the 1-D model.}
\label{defect_oscillation2}
\end{figure}

In Figure \ref{defect_oscillation2a} we demonstrate vertically asymmetric oscillations in an $n=25$ lattice. We plot the cell density at every node and demonstrate a similar oscillation as in Figure \ref{defect_oscillation}. We also show the effects of changing the nature of the boundary conditions, and the dimension of the domain, on the behaviour shown in Figure \ref{defect_oscillation2a}. We show plots of cell density for periodic horizontal boundary conditions for both the fluid and the cell density in Figure \ref{defect_oscillation2b}, and for the 1-D model in Figure \ref{defect_oscillation2c}. The periodic conditions in Figure \ref{defect_oscillation2b} change how the oscillation interacts with the boundary, and hence the bands do not display the same structure as in Figure \ref{defect_oscillation2a}. The 1-D model in Figure \ref{defect_oscillation2c} does not oscillate and instead reaches a steady state. The periods of the vertically asymmetric oscillations in Figures \ref{defect_oscillation} and \ref{defect_oscillation2} are much larger than the vertically symmetric oscillations in Figure \ref{cell1timeseries}. In addition to the absence of asymmetric oscillations in the 1-D models, this change in oscillation frequency is how we differentiate these two kinds of oscillatory behaviours.

\begin{figure}
\centering
    \subfloat[]{
	\includegraphics[width=0.3\textwidth]{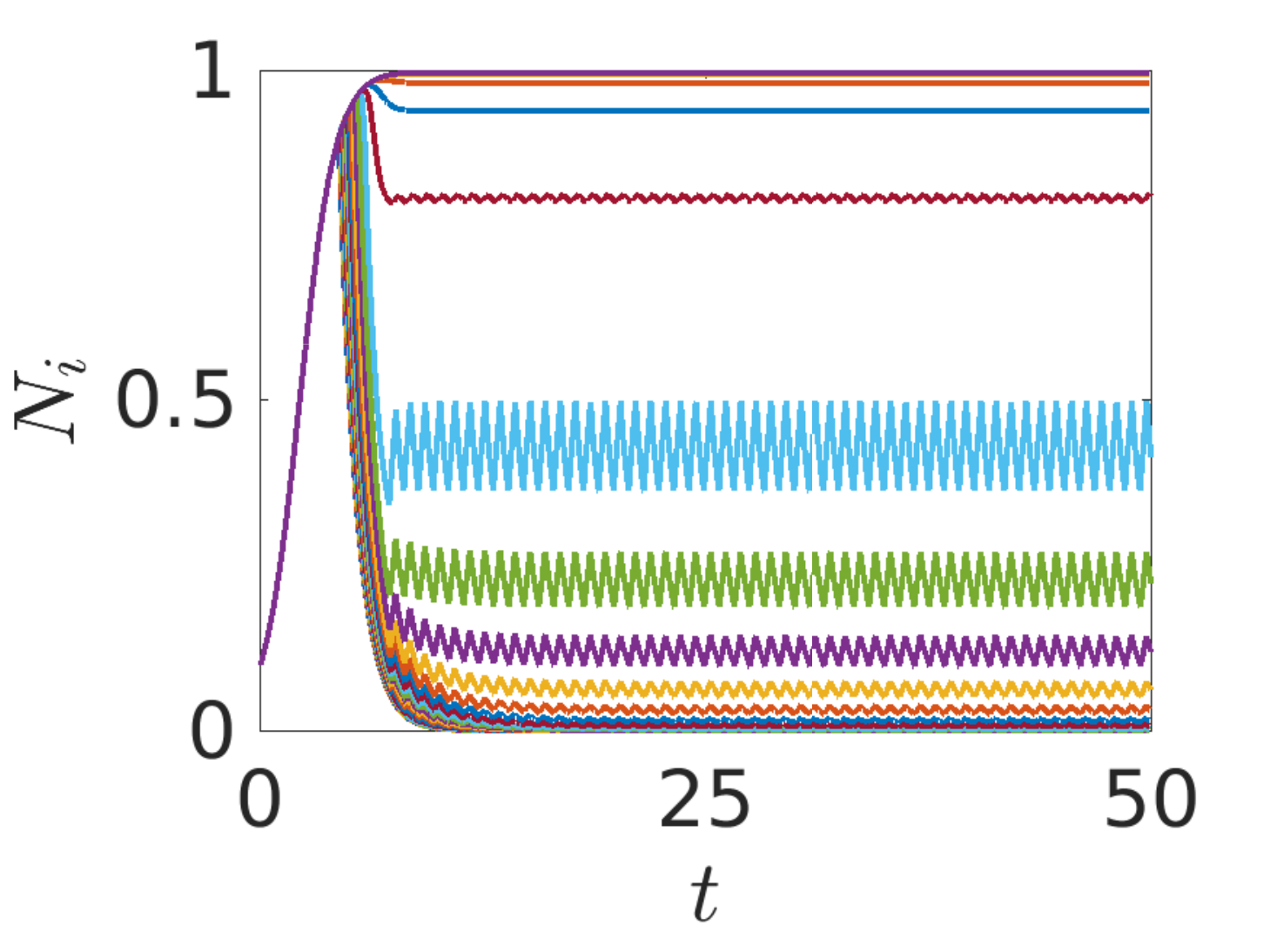}
	\label{interpa}}%
    \subfloat[]{
	\includegraphics[width=0.3\textwidth]{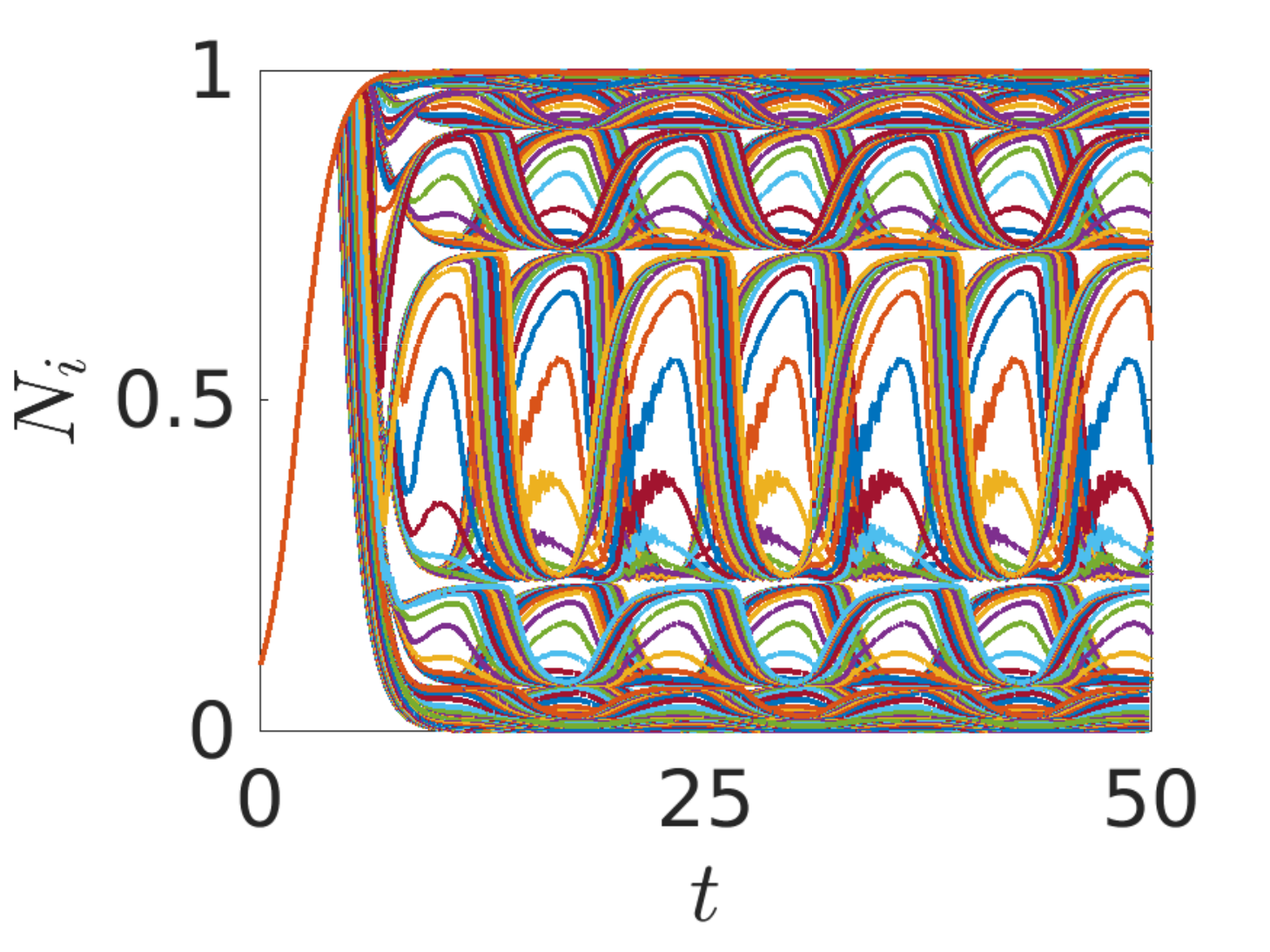}
    \label{interpb}}\hfill%
    
    \subfloat[]{
	\includegraphics[width=0.3\textwidth]{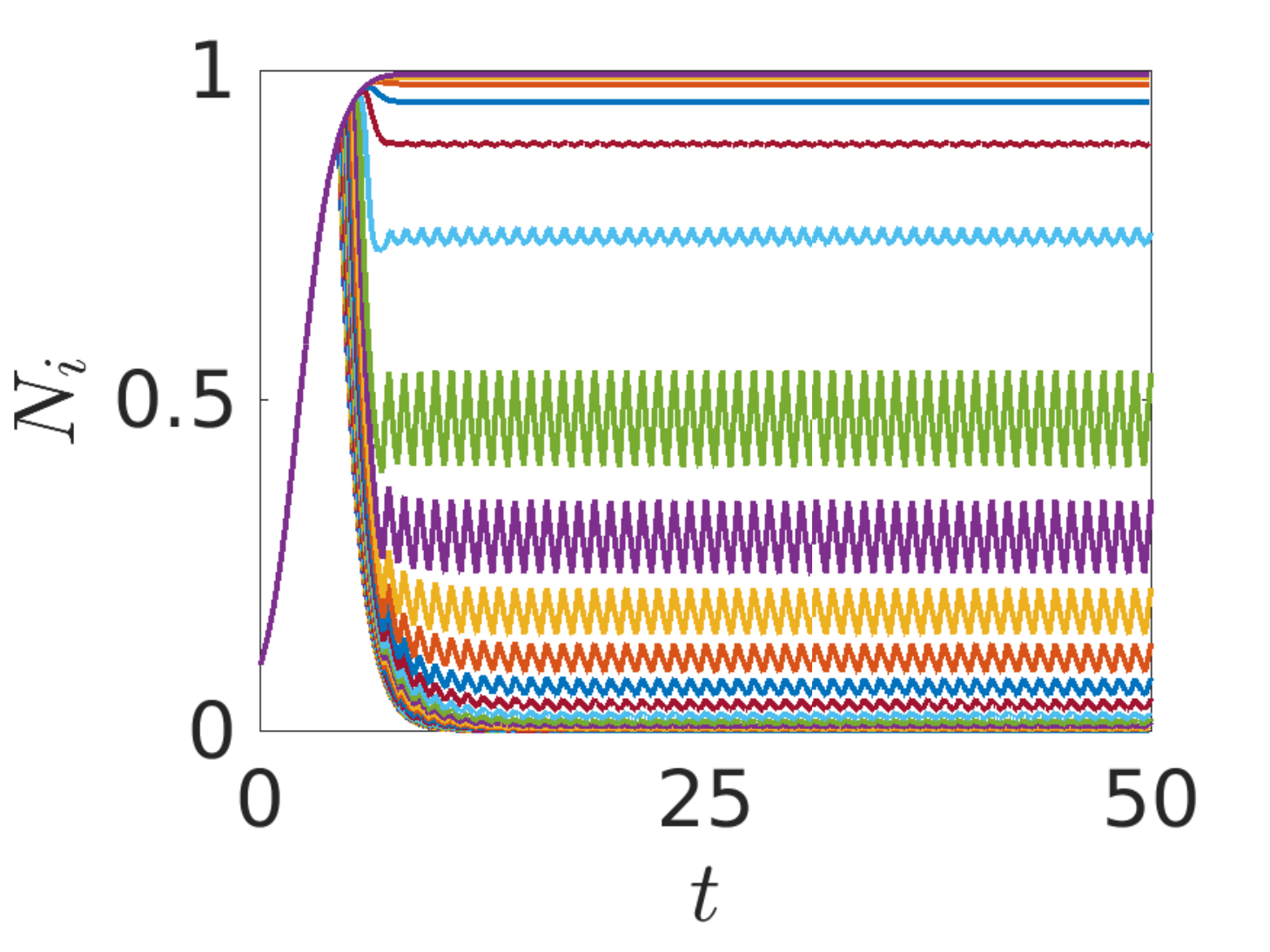}
	\label{interpc}}%
    \subfloat[]{
	\includegraphics[width=0.3\textwidth]{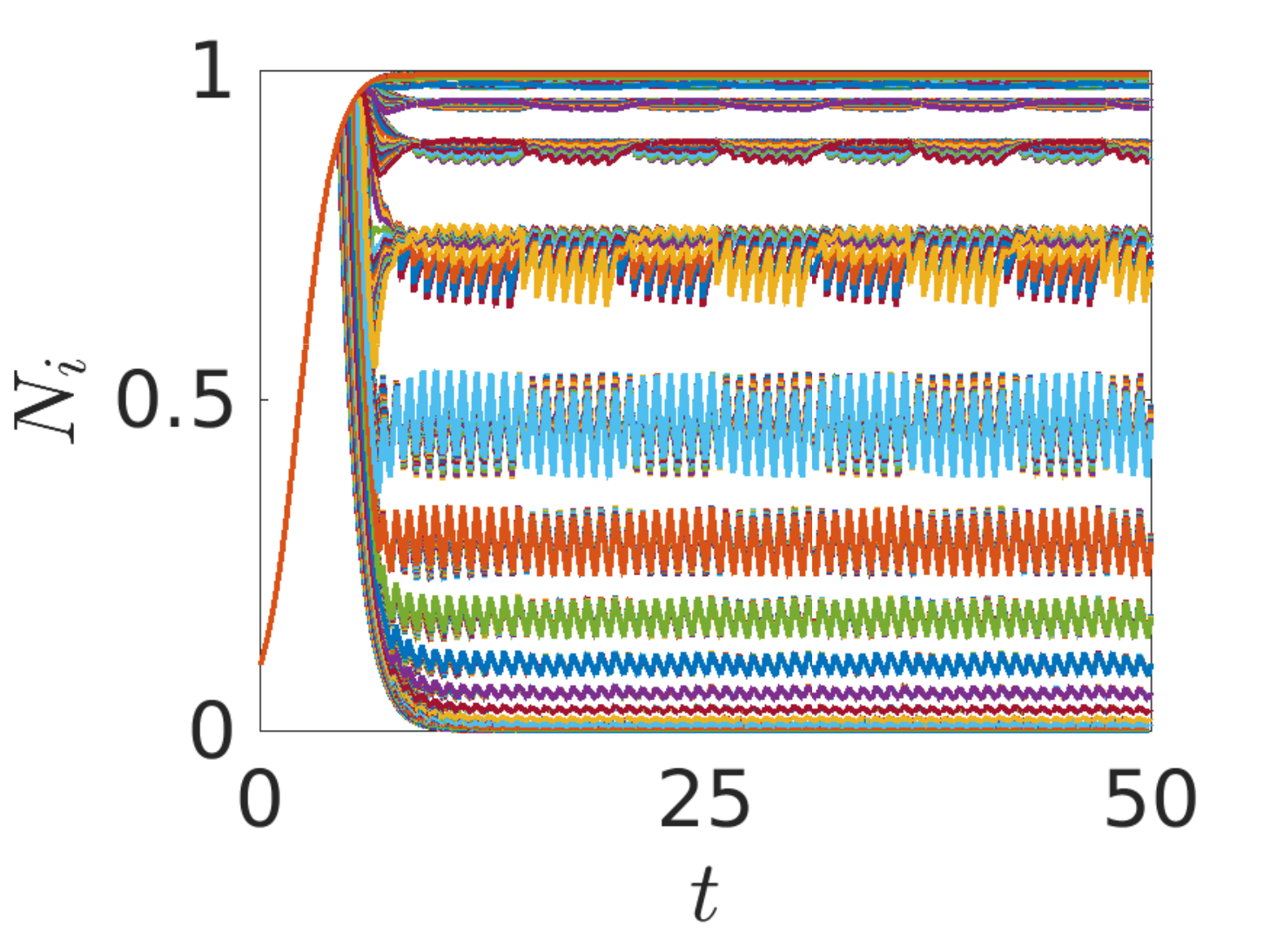}
    \label{interpd}}\hfill%
\caption{Plots of the cell density at each node for $n=25$, $p_l=750$, with $\delta=10^{-3}$ in (a) and (b) and $\delta=2 \times 10^{-3}$ in (c) and (d). The 1-D model is plotted in (a) and (c), and the 2-D in (b) and (d).}
\label{interp}
\end{figure}

We also observe combinations of both kinds of oscillations. In Figure \ref{interp} we compare 1-D and 2-D simulations of a lattice of size $n=25$ with $p_l=750$ for different values of $\delta$. In \ref{interpb} we observe asymmetric oscillations in a 2-D lattice in combination with smaller amplitude and higher frequency oscillations for some nodes. In \ref{interpd} we instead observe fast oscillations with a slow variation indicative of two kinds of oscillatory behaviours. Figures \ref{interpa} and \ref{interpc} are the 1-D versions of Figures \ref{interpb} and \ref{interpd} respectively, and indicate the strength of the underlying pulsing oscillation in the 2-D simulations. Figure \ref{interpc} shows more nodes oscillating with larger amplitudes compared to Figure \ref{interpa}, which suggests why the oscillations in Figure \ref{interpd} are qualitatively different from those in Figure \ref{interpb}. Note that the frequencies of the fast oscillations in the 2-D simulations correspond to the frequencies of the 1-D oscillations, and hence the underlying vertically symmetric behaviour. We also find parameters where a pulsing oscillation in 1-D exists along with a vertically asymmetric steady state in the 2-D model, leading to a vertically asymmetric oscillation that is at the frequency of the vertically symmetric 1-D oscillation, but it is not vertically symmetric due to the underlying cell density distribution (i.e.~the mean distribution is not vertically symmetric). In Section \ref{Bifurcations} we explore the parameter space in the 1-D and 2-D models, and demonstrate that regions with large oscillation amplitudes at some nodes in the 1-D model will correspond to pulsing oscillations in the 2-D model, so that these more exotic behaviours exist only when the 1-D model has smaller amplitude oscillations.

\begin{figure}
\centering
	\subfloat[]{
    	\includegraphics[width=0.3\textwidth]{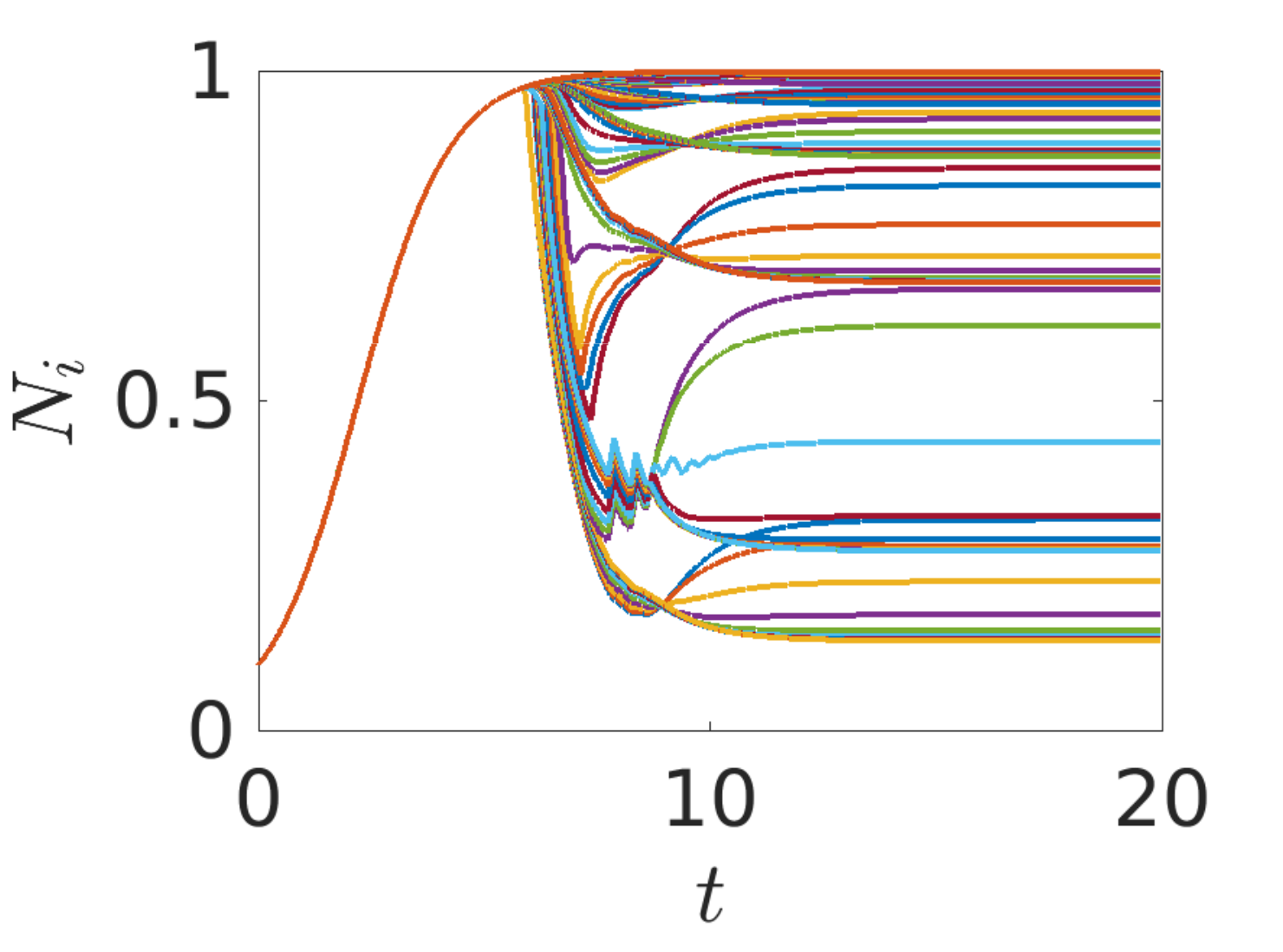}\label{excitable_behavioura}
    }
	\subfloat[]{
    	\includegraphics[width=0.3\textwidth]{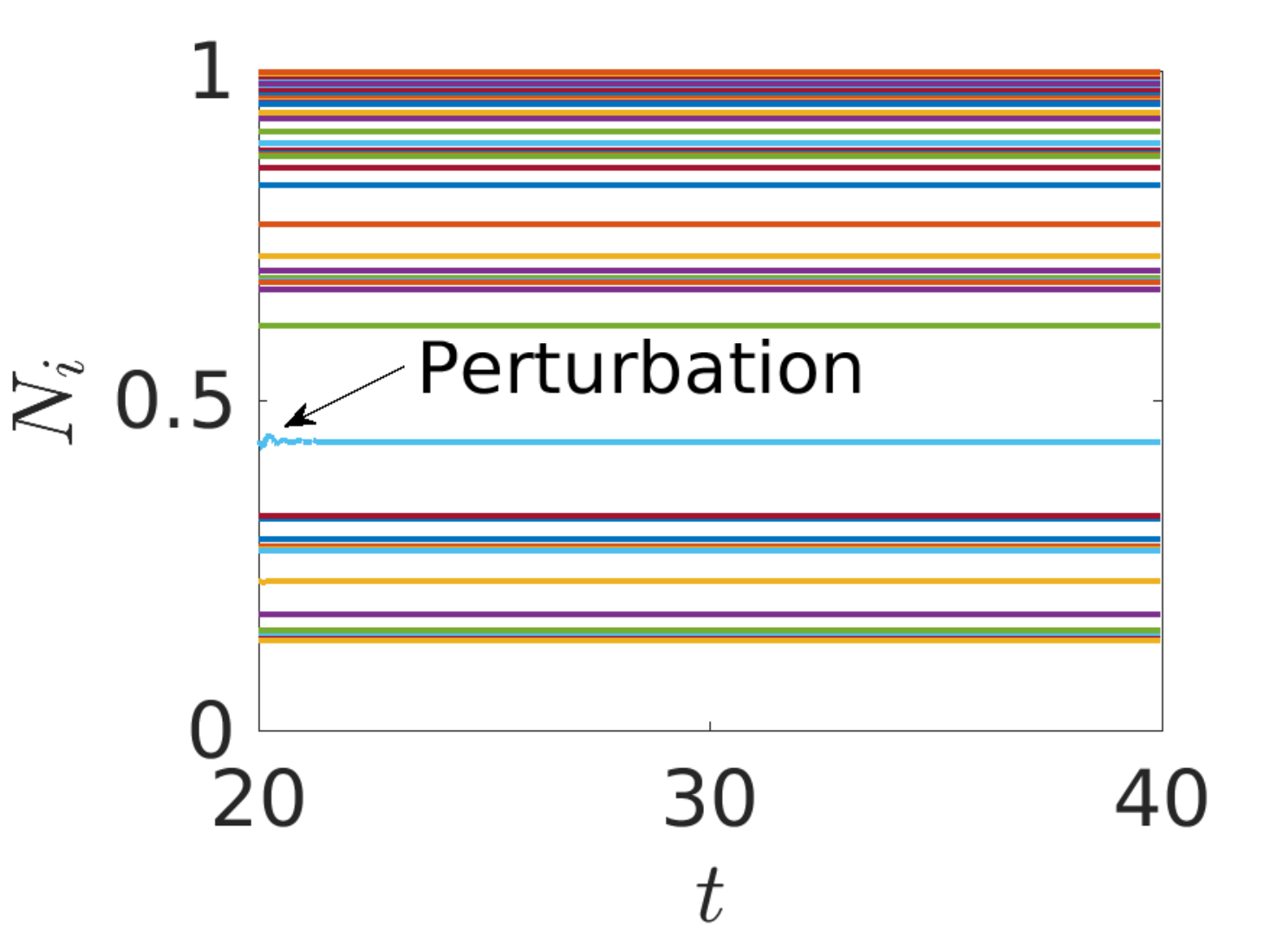}\label{excitable_behaviourb}
    }
    
    \subfloat[]{
    \includegraphics[width=0.3\textwidth]{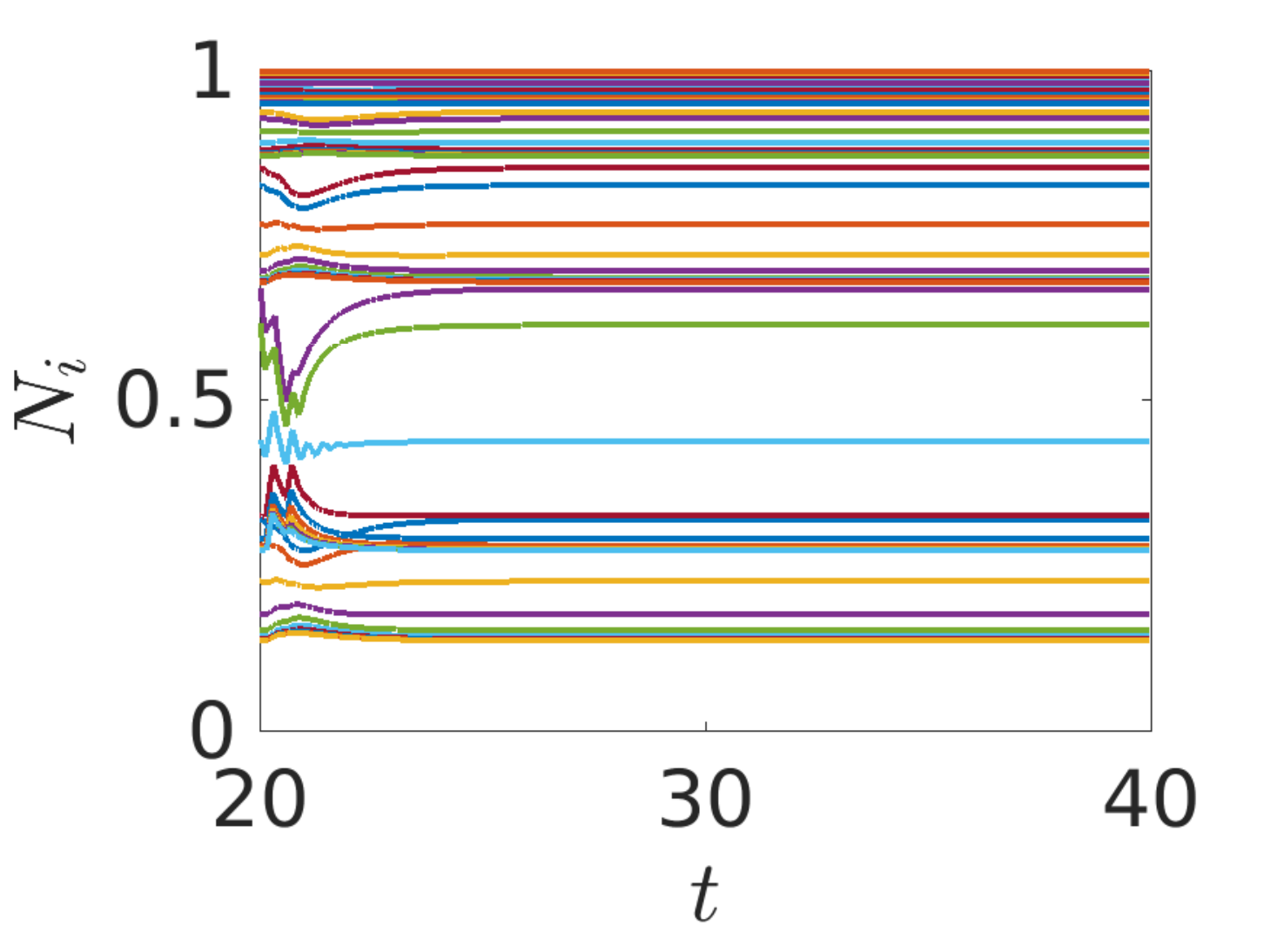} \label{excitable_behaviourc}
    }
    \subfloat[]{
    \includegraphics[width=0.3\textwidth]{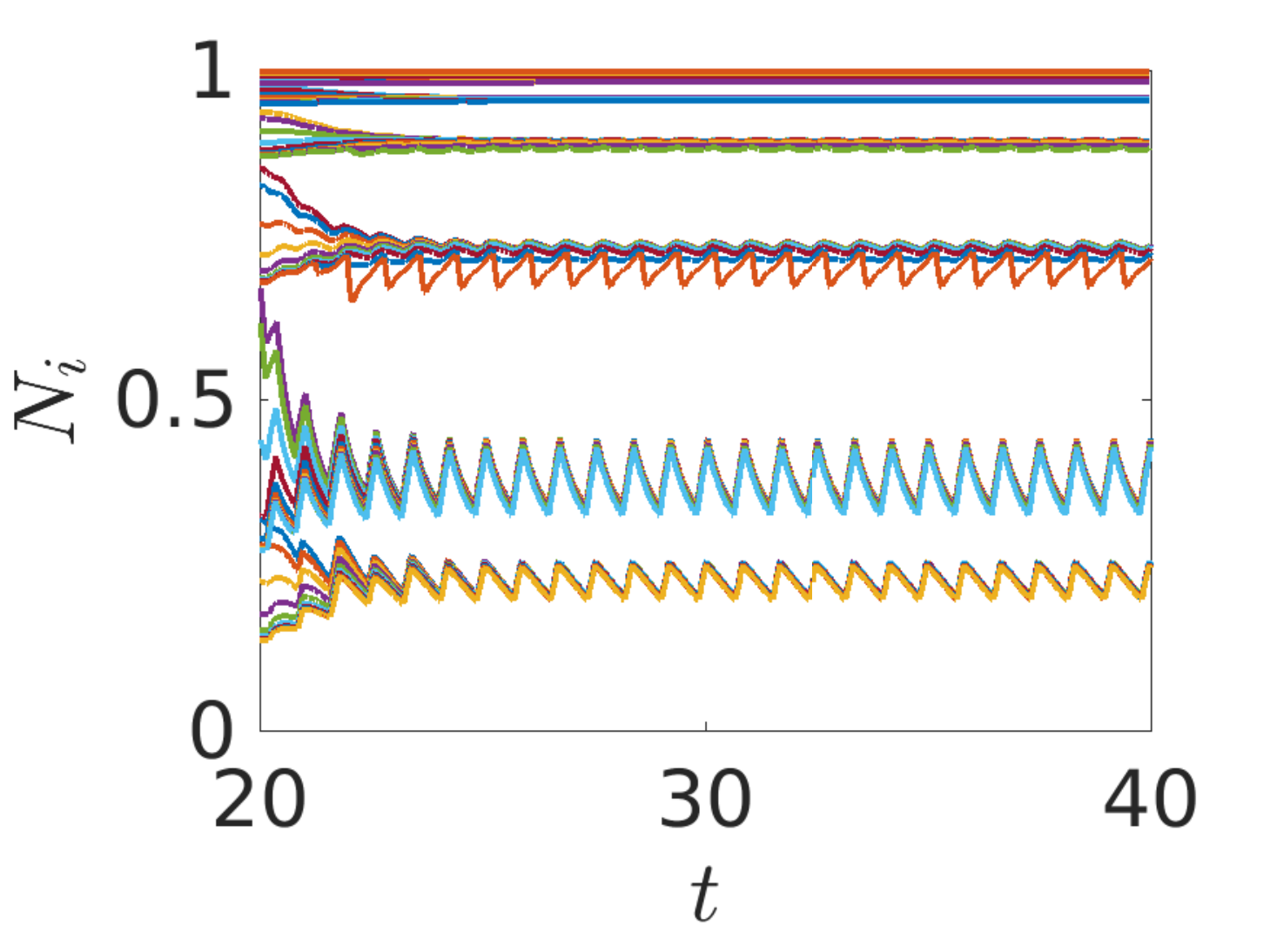}\label{excitable_behaviourd}
    }
\caption{Excitable dynamics with $n=10$, $p_l=4000$, $\delta=10^{-2}$. In (a) we show the trajectory from a uniform state to a nonuniform equilibrium at each node. In (b)-(d) we show the final state from (a) evolves after a perturbation of the cell density at a single node. These perturbations are to node $N_{91}$ at the bottom right of the lattice, and of sizes $10^{-5}$ in (b), $3\times 10^{-4}$ in (c) and $4 \times 10^{-4}$ in (d).}
\label{excitable_behaviour}
\end{figure}

Some simulations exhibit excitable dynamics. In Figure \ref{excitable_behavioura} we first show the evolution of cell density for a small ($n=10$) lattice to steady state. Then in Figures \ref{excitable_behaviourb} through \ref{excitable_behaviourd} we add increasing perturbations to this steady state cell density at a single node at the top-right of the lattice. We observe small excursions from the equilibrium for very small perturbations (Figure \ref{excitable_behaviourb}), larger excursions from equilibrium for larger perturbations (Figure \ref{excitable_behaviourc}), and finally a sufficiently large perturbation induces a pulsing oscillation in Figure \ref{excitable_behaviourd}. This excitability is due to the nonlocal nature of the quasi-static fluid coupling, in that an increase in cell density at one node in the lattice affects the overall flow through the scaffold instantaneously and hence the pressure at every other node. This is particularly true for nodes that have high cell density as the nonlinearity in Equations \eqref{lattice_fluid_eqns} and \eqref{constitutive_lawsa} amplifies the effect of large cell densities. Note in particular that we only perturb one node in Figures \ref{excitable_behaviourb}-\ref{excitable_behaviourd}, yet several nodes grow or die rapidly at $t=20$ due to this nonlocal effect. In Section \ref{Bifurcations}, we use numerical continuation to show the coexistence of locally stable steady states and pulsing oscillations for large regions of parameter space. A sufficiently large perturbation can move the state of the system between basins of attraction of these different long-time behaviours.

The smoothing parameter $g$ plays only a marginal role in these oscillations. The behaviours found in all of the simulations we performed are qualitatively and quantitatively the same in the limit of $g \to \infty$ (we replaced the functions $F_1$ and $F_2$ in Equations \eqref{f1}-\eqref{f2} with Heaviside step-functions, and obtained quantitatively comparable results). In the other direction, we set $g=1$ and repeated the simulation in Figure \ref{cell1timeseriesleft}, and the results displayed only small differences in the initial transient leading up to the oscillation, and otherwise the same qualitative behaviour. In contrast, the oscillations in Figures \ref{defect_oscillation} and \ref{defect_oscillation2} were not present if $g=1$ for any parameters simulated, and instead steady state behaviour was obtained. This indicates that these vertically asymmetric oscillations are related to sharp transitions between cell growth and death due to pressure. 

Aside from differences in the constitutive relations, the behaviour of the PDE model and larger lattices is similar, whereas the asymmetric solutions in smaller lattices do not have analogues in the PDE model. This is expected due to the similarities between the lattice equations and a method-of-lines discretization of the PDE. In this Section we have illustrated typical solution behaviours displayed by lattice and PDE models. In Section \ref{Classification} we use simulations from a large subset of the feasible parameter space ($n$, $\delta$, $p_l$, and $p_c$) to understand where in parameter space these behaviours arise. In particular, we provide evidence that smaller lattices exhibit more complicated dynamics. Before demonstrating these results, we describe some asymptotic approaches that allow us to restrict our numerical efforts to specific parameter regimes.

\section{Asymptotic solutions in the large diffusion limit}\label{Asymptotics}
For larger values of the diffusion parameter $\delta$, and some values of $p_l$, the numerical solutions approach an equilibrium with little variation between nodal values (see Figures \ref{pressure1figa} and \ref{pressure1fige} where $\delta = 10^{-1}$ is sufficient to observe very little spatial variation). We pursue an asymptotic solution for both models in the limit where $1/\delta \ll 1$. This asymptotic solution allows us to quantitatively predict the solution structure for large values of $\delta$. In particular, large-$\delta$ values give rise to simple steady state solutions, so that when we exhaustively explore a subset of the parameter space in Section \ref{Classification}, we only consider small values of $\delta$, namely $\delta \leq 10^{-1}$, in order to capture oscillations.

To make analytical progress we simplify the pressure forcing terms throughout the rest of this Section. Letting $g\to \infty$ our pressure functions become $F_1(p_i) = H(p_l-p_i)$ and $F_2(p_i)=H(p_i-p_l)$, where $H$ is the Heaviside step function. For convenience, we take $H$ to be continuous on the right, so that $\lim_{\eps \to 0^{+}}H(\eps) = H(0) \equiv 1$.

\subsection{Lattice asymptotics} \label{lattice_d_asymptotics}
We first consider solutions in the large diffusion limit for the lattice. We rewrite Equation \eqref{general_lattice_cell_eqn} as,
\begin{equation} \label{Cell_Expansion}
\delta^{-1}\frac{d N_{i}}{d t} = \delta^{-1}(H(p_l-p_i)N_{i}(1-N_{i}) - H(p_i-p_l)N_{i}) - n^2\sum_{j=1}^{n^2} L_{ij}N_j, \quad 1 \leq i \leq n^2.
\end{equation}
We expand our variables as $N_i = N_{0,i} + \delta^{-1} N_{1,i} + O(\delta^{-2})$ and $p_i = p_{0,i} + \delta^{-1} p_{1,i} + O(\delta^{-2})$. Substituting these expansions into Equation \eqref{Cell_Expansion} and formally equating powers of $\delta$, the leading-order equation for the cell density, $N_{0,i}$, is
\begin{equation}\label{Cells_Delta}
0 = -n^2\sum_{j=1}^{n^2} L_{ij}N_{0,j}, \quad 1 \leq i \leq n^2.
\end{equation}
The graph Laplacian $L$ has one zero eigenvalue with the corresponding eigenvector $v=(1,1,,\dots,1)^T$ \cite{newman_networks:_2010}. Therefore the leading-order cell densities are all equal; that is $N_{0,i} = N_0$ for some $N_0$, for all $i$. 

To compute $N_0$ we consider the next-order problem,
\begin{equation}\label{Cell_Expansion_2}
\frac{d N_{0}}{d t} = (H(p_l-p_{0,i})N_{0}(1-N_{0}) - H(p_{0,i}-p_l)N_{0}) - n^2\sum_{j=1}^{n^2} L_{ij}N_{1,j},\quad 1 \leq i \leq n^2.
\end{equation}
Equations \eqref{lattice_fluid_eqns} do not explicitly depend on $\delta$ and so the pressures $p_{0,i}$ satisfy Equations \eqref{lattice_fluid_eqns} with the uniform cell density $N_{0}$. As a spatially-constant cell density $N_0$ is vertically symmetric, we can use Equation \eqref{lattice_p_sol} to determine the pressures $p_{0,i}$. In order to eliminate the $N_{1,j}$ terms, we sum Equation \eqref{Cell_Expansion_2} over $i$ and note that $\sum_{i=1}^{n^2}\sum_{i=1}^{n^2} L_{ij}N_{1,j}=0$, to find 
\begin{equation}\label{Cell_Expansion_2_sum}
n\frac{d N_{0}}{d t} = \sum_{i=1}^{n}(H(p_l-\hat{p}_{0,i})N_{0}(1-N_{0}) - H(\hat{p}_{0,i}-p_l)N_{0}),
\end{equation}
where we have relabelled the pressures to reflect the 1-D symmetry, and divided both sides by $n$.

The pressures $\hat{p}_{0,i}$ are monotonically decreasing in $i$ with $\hat{p}_{0,n} = 1/n$. Excluding cases corresponding to $p_l$ outside of the ranges given by \eqref{lattice_range}, there exists an $m=m(N_0)$ such that for all $j \leq m$ and $k > m$, $\hat{p}_{0,j} \geq p_l$ and $\hat{p}_{0,k} < p_l$. Thinking of $m$ as an integer, we have that solutions of Equation \eqref{Cell_Expansion_2_sum} satisfy,
\begin{equation}\label{Cell_Expansion_2_unsteady}
n\frac{d N_{0}}{d t} = (n-m(N_0))N_0(1-N_{0}) - m(N_0)N_0.
\end{equation}

We now approximate $m(N_0)$ as a real variable, rather than an integer, to make analytical progress and avoid many tedious details. We will justify this approximation later by observing how well the solution matches numerical solutions of the full problem. By setting the left hand side of Equation \eqref{lattice_p_sol} equal to $p_l$, we find $m(N_0)$ to be
\be\label{m_eq}
m(N_0) = [n-R^4(N_0,N_0)(p_l n-1)]^+ = [n-(1-\rho N_0)^4(p_l n-1)]^+,
\ee
where $[f]^+=\max(f,0)$ is the positive part of the function $f$. It is necessary to define $m$ using the notation                                                                                                                                                                                                                                                                                                                                                                                                                     $[f]^+$ as for small $N_0$, corresponding to $p_i < p_l$ for all $i$, the value of $m(N_0)$ is negative.  Equations \eqref{Cell_Expansion_2_unsteady} and \eqref{m_eq} represent a single scalar first-order ordinary differential equation, compared with the original $n^2$ differential equations \eqref{general_lattice_cell_eqn} and $n^2$ algebraic equations \eqref{lattice_fluid_eqns}, although our scalar equation is not smooth (e.g. not differentiable) due to the function $m(N_0)$. 

We now look for steady state solutions, $N_0^*\geq 0$, to Equation \eqref{Cell_Expansion_2_unsteady}. If $m(N_0^*)=0$, then $N_0^*=0$ or $N_0^*=1$ are the only solutions to Equation \eqref{Cell_Expansion_2_unsteady}, and we will show that they cannot be stable steady states within the parameters we consider. Assuming that $m(N_0^*) >0$, we substitute \eqref{m_eq} into Equation \eqref{Cell_Expansion_2_unsteady}, and find that  $N_0^*$ satisfies the quintic polynomial
\begin{equation}\label{lattice_d_ss}
0 = R^4(N_0^*,N_0^*)(p_l n-1)(2-N_{0}^*) - n = (1-\rho N_0^*)^4(p_l n-1)(2-N_{0}^*) - n.
\end{equation}
We will first argue that any root of Equation \eqref{lattice_d_ss} must be unique within the interval $(0,1)$, and then show the existence and global stability of $N_0^* \in (0,1)$.

Let $\tilde{N} = R(N_0^*,N_0^*) = 1-\rho N_0^*$. Since $0 < \rho < 1$, all roots $N_0^*\in (0,1)$ of Equation \eqref{lattice_d_ss} must be associated to roots $\tilde{N}\in (0,1-\rho)$ of
\begin{equation}\label{lattice_d_sss}
0= \tilde{N}^4(p_l n-1)\left(2-\frac{1}{\rho} + \frac{\tilde{N}}{\rho}\right) - n = \tilde{N}^5\frac{(p_l n-1)}{\rho} + \tilde{N}^4(p_l n-1)\left(2-\frac{1}{\rho}\right) - n,
\end{equation}
and vice versa. Equation \eqref{lattice_d_sss} has only one sign change in its coefficients (as $p_l > 1/n$ by \eqref{lattice_range}). By Descartes' rule of signs, Equation \eqref{lattice_d_sss} has at most one positive root, implying  that there cannot be more than one root in the interval $0 \leq \tilde{N} \leq 1-\rho$. Hence any root of equation \eqref{lattice_d_ss} in $(0,1)$ must be unique. 

We now argue by monotonicity that this root exists and is globally attractive for initial data in the full interval $[0,1]$. The function $m(N_0)$ given by \eqref{m_eq} is monotonically increasing in $N_0$, and for $1+1/n < p_l< (1+(n-1)(1-\rho)^{-4})/n$ we have that $m(0)=0$ and $m(1) > 0$. 
So there is a unique value $\hat{N_0}$ such that $m(\hat{N_0})=0$ and for all $N > \hat{N_0}$, $m(N) > 0$. From Equation \eqref{m_eq} we compute that
\be\label{Nhat}
\hat{N_0} = \frac{\displaystyle 1-\left (\frac{n}{p_l n - 1} \right)^\frac{1}{4}}{\rho},
\ee
which is within the interval $[0,1]$. 

By monotonicity, we have that $m(N_0) \geq 0$ for all $N_0 \in [\hat{N_0},1]$. Since $m(0)=0$ and $m(1)>1$, we have that $N_0=0$ is an unstable steady state and $N_0=1$ is not a steady state solution to Equation \eqref{Cell_Expansion_2_unsteady} for these parameters. From Equation \eqref{Cell_Expansion_2_unsteady} we have that $N_0'(t) < 0$ when $N_0=1$ and $N_0'(t) > 0$ when $N_0=0$. This implies that continuous solutions to Equation \eqref{Cell_Expansion_2_unsteady} do not leave the interval $[0,1]$, which is consistent with the boundedness of solutions to the full system shown in Section \ref{bounded}. We note that for $m(N_0)=0$, the right side of \eqref{Cell_Expansion_2_unsteady} is positive, so $N_0(t)$ grows in time and approaches $\hat{N_0}$. Similarly, for $N_0 = \hat{N_0}$, the right side of \eqref{Cell_Expansion_2_unsteady} is positive as $1 -\hat{N_0}>0$. So the region $(\hat{N_0},1)$ absorbs solutions with initial data $N_0(0) \in (0,1]$, and hence we must have that there exists a steady state solution $N_0^* \in (\hat{N_0},1)$ to \eqref{Cell_Expansion_2_unsteady}. Both existence and global attractivity of this steady state follows from the fact that Equation \eqref{Cell_Expansion_2_unsteady} is just a scalar first-order ODE within the region $N_0^* \in (\hat{N_0},1)$.

We can solve equation \eqref{Cell_Expansion_2_unsteady} by considering the cases when $N_0(0) < \hat{N_0}$ and $N_0(0) > \hat{N_0}$ separately. We first assume that $N_0(0) \in (0,\hat{N_0})$. By the monotonicity of $m$, and the fact that $N'(t) > 0$ as long as $m(N) = 0$, there exists some $t_s$ such that $N_0(t_s)=\hat{N_0}$. So substituting \eqref{m_eq} into \eqref{Cell_Expansion_2_unsteady} we have
\begin{equation}\label{Cell_Expansion_3}
\frac{d N_{0}}{d t} = \begin{cases}
N_0(1-N_{0}), \quad &t \leq t_s,\\
\displaystyle (1-\rho N_0)^4\left(p_l-\frac{1}{n}\right)N_0(2-N_{0}) - N_0, \quad &t > t_s.
\end{cases}
\end{equation}

The solution to equation \eqref{Cell_Expansion_3} for $t \leq t_s$ is a logistic function, 
\be\label{Logistic}
N_0(t) = \frac{N_0(0)e^t}{N_0(0)(e^t-1)+1}, \quad t \leq t_s,
\ee
which we use to compute $t_s$. We set $N_0(t_s) = \hat{N_0}$ and use equations \eqref{Nhat} and \eqref{Logistic} to find
\be\label{ts_one}
t_s = \ln\left(\frac{\displaystyle (1-N_0(0))\left (1-\left (\frac{n}{n p_l-1}\right)^\frac{1}{4}\right)}{\displaystyle N_0(0)\left(\left(\frac{n}{n p_l-1}\right)^\frac{1}{4}+\rho-1\right)}\right).
\ee

We consider equation \eqref{Cell_Expansion_3} for $t > t_s$. This equation has a sixth-order polynomial nonlinearity, and we are not aware of a method to explicitly integrate it. We know by the boundedness of $N_0(t)$, however, that $N_0(t_s)$ is nearby in a nondimensional sense to our equilibrium value $N_0^*$ for any initial mean cell density $N_0(0)$. So we expect approximately exponential convergence to $N_0^*$ for $t > t_s$. We substitute the ansatz
\be\label{ansatz}
N_0(t) = N_0^* + Ae^{-r(t-t_s)}, \quad t > t_s,
\ee
for $\left|A\right| \ll 1$ into equation \eqref{Cell_Expansion_3}, and take the Taylor series of the right hand side about $A=0$ to find
\begin{multline}\label{TaylorSubs}
-r Ae^{-r(t-t_s)} =
(1-\rho N_0^*)^4\left(p_l-\frac{1}{n}\right)N_0^*(2-N_{0}^*) - N_0^*\\
+ \left\{\left(p_l-\frac{1}{n}\right)\left(2(1- \rho N_0^*)^4(1-1N_0^*)-4(1-\rho N_0^* )^3 N_0^* (2-N_0^*) \rho\right)-1 \right\}Ae^{-r(t-t_s)} + O(A^2).
\end{multline}

The $O(1)$ term on the right hand side of equation \eqref{TaylorSubs} is zero, as $N_0^*$ is the solution to equation \eqref{lattice_d_ss}, which is obtained numerically. Comparing $O(A)$ terms we have
\be\label{r}
r \sim \left(p_l-\frac{1}{n}\right)\left(4(1-\rho N_0^* )^3 N_0^* (2-N_0^*) \rho-2(1- \rho N_0^*)^4(1-N_0^*)\right)+1.
\ee
In order to make the solution \eqref{ansatz} continuous with our solution for $t \leq t_s$, we take $A=N_0(t_s)-N_0^*$ (which for our dynamics satisfies $|N_0(t_s)-N_0^*| \leq 1$) where $N_0(t_s)$ is the solution from equation \eqref{Logistic} evaluated at $t=t_s$. Our asymptotic solution can then be written as
\be\label{Asym_Sol}
N_0(t) = \begin{cases}
\displaystyle \frac{N_0(0)e^t}{N_0(0)(e^t-1)+1}, \quad &t \leq t_s,\\
\displaystyle N_0^*+\left(\frac{N_0(0)e^{t_s}}{N_0(0)(e^{t_s}-1)+1}-N_0^*\right)e^{-r(t-t_s)}, \quad &t > t_s,
\end{cases}
\ee
where $t_s$ is defined by equation \eqref{ts_one} and $r$ by equation \eqref{r}. 

Next we consider the case that $N_0(0)> \hat{N_0}$. We have that $m(N_0(0))>0$, and so we can use the method above to find an exponential solution of the form given by equation \eqref{ansatz} as before. The value of $A$ is again determined by forcing the solution to continuously approach the initial data, which gives that $A = N_0(0)-N_0^*$. So the solution in this case is 
\be\label{Asym_Sol2}
N_0(t) = N_0^* + (N_0(0)-N_0^*)e^{-rt},
\ee
with $r$ given by equation \eqref{r}.

\begin{figure}
\centering
\includegraphics[width=0.7\textwidth]{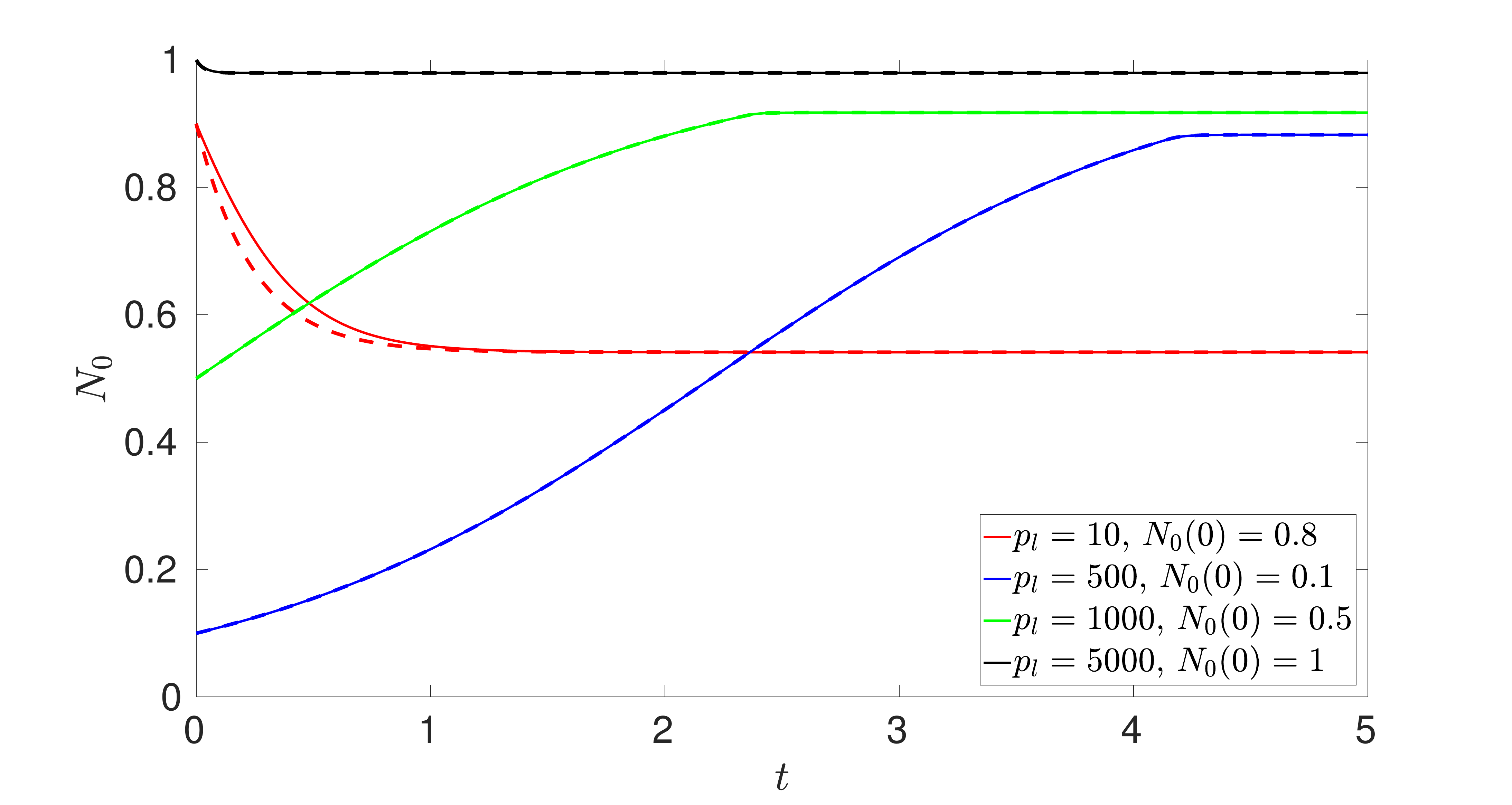}
\caption{Plots of $N_0(t)$ computed from simulations of equations \eqref{Cell_Expansion_2_unsteady}-\eqref{m_eq} as solid lines, as well as the analytical solutions \eqref{Asym_Sol}-\eqref{Asym_Sol2} as dashed lines, for $n=10$ and different initial mean cell densities $N_0(0)$ and thresholds $p_l$.}\label{temporal_asymptotics}
\end{figure}

The solution given by equation \eqref{Asym_Sol} describes a period of logistic cell growth, followed by an exponentially fast convergence to a steady state, and depends on the initial mean cell density $N_0(0)$, $p_l, \rho$ and $n$. If the initial mean cell density is high enough to immediately cause cells to die due to pressure, there is no logistic growth phase and Equation \eqref{Asym_Sol2} describes the exponential convergence to a steady state. These analytical solutions also match numerical solutions of \eqref{Cell_Expansion_2_unsteady} extremely well; see Figure \ref{temporal_asymptotics}. In the worst case of $N_0(0)=0.9$ and $p_l=10$, where there is no logistic growth, we note that the analytical solution is still quite close to the numerical solution despite the value of $A=0.45$, which is not asymptotically small. We suspect that this close agreement is due in part to solutions being bounded in $[0,1]$, so that we always have $\left |A \right | < 1$, along with the existence of a unique stable steady state to equation \eqref{Cell_Expansion_2_unsteady}.

\begin{figure}
\centering
\includegraphics[width=0.7\textwidth]{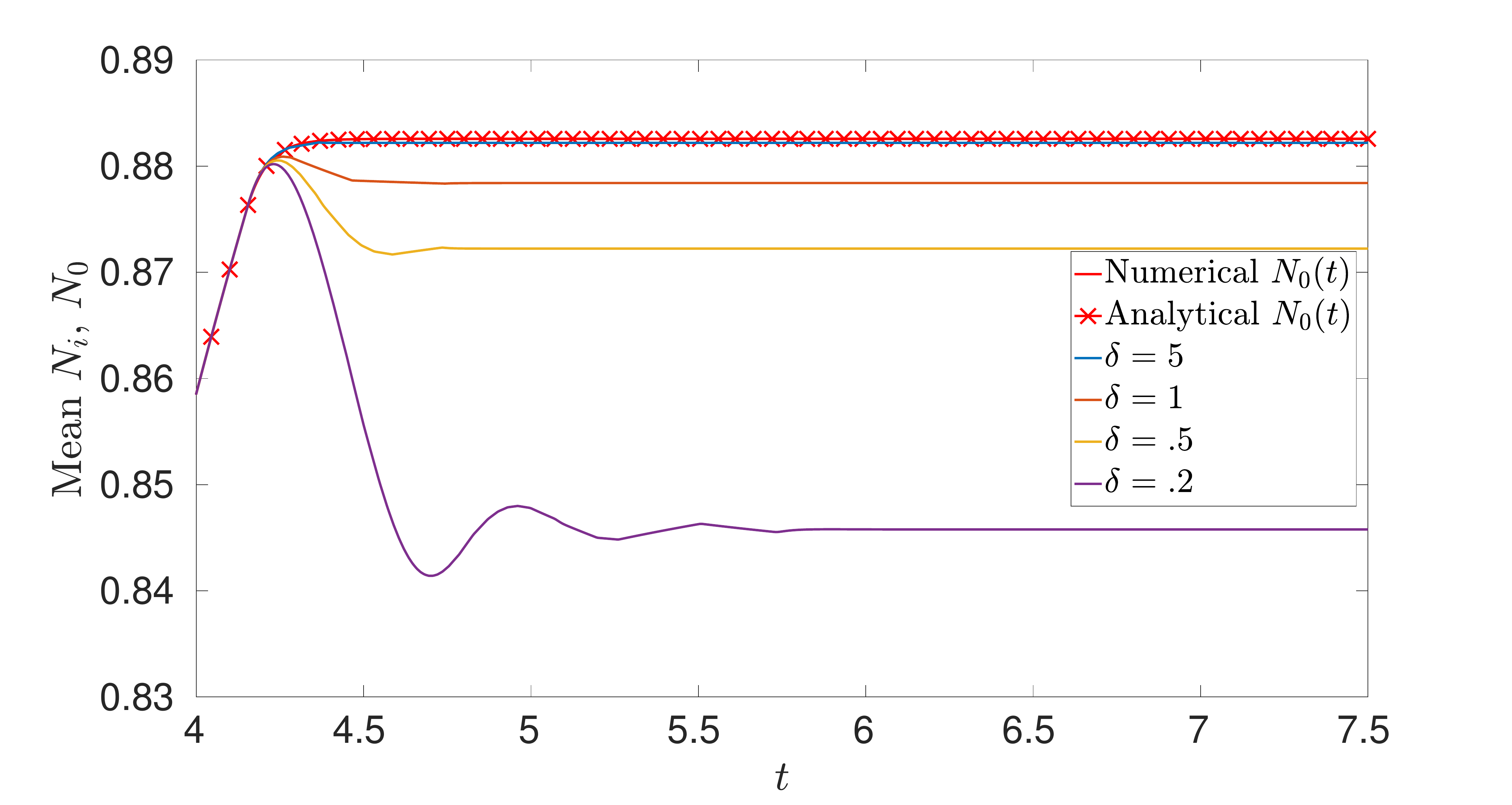}
\caption{Plots of the asymptotic cell density, $N_0$, computed from numerical solutions of equations \eqref{Cell_Expansion_2_unsteady}-\eqref{m_eq}, analytical solutions given by \eqref{Asym_Sol}, and simulations of the full two-dimensional model \eqref{lattice_fluid_eqns}-\eqref{f2} for $n=100$, $p_l=500$, and $\delta=0.1,0.5,1,$ and $5$.}\label{temporal_asymptotics2}
\end{figure}

We compare the asymptotic solution \eqref{Asym_Sol} for the cell density, $N_0$, against the (spatial) mean cell density for the full two-dimensional problem \eqref{lattice_fluid_eqns}-\eqref{f2} for various values of $\delta$ in Figure \ref{temporal_asymptotics2}. For small values of $\delta$, the oscillations in nodal cell densities have an overall effect on the spatial mean cell density such that our asymptotic solution over-predicts the mean cell density of oscillatory solutions. Increasing $\delta$, we see a relatively fast convergence to the asymptotic solution for the spatial mean cell density $N_0$. Comparable results hold for different values of $p_l$ and $n$, and for the solution given by equation \eqref{Asym_Sol2}.

\begin{figure}
\centering
\includegraphics[width=0.7\textwidth]{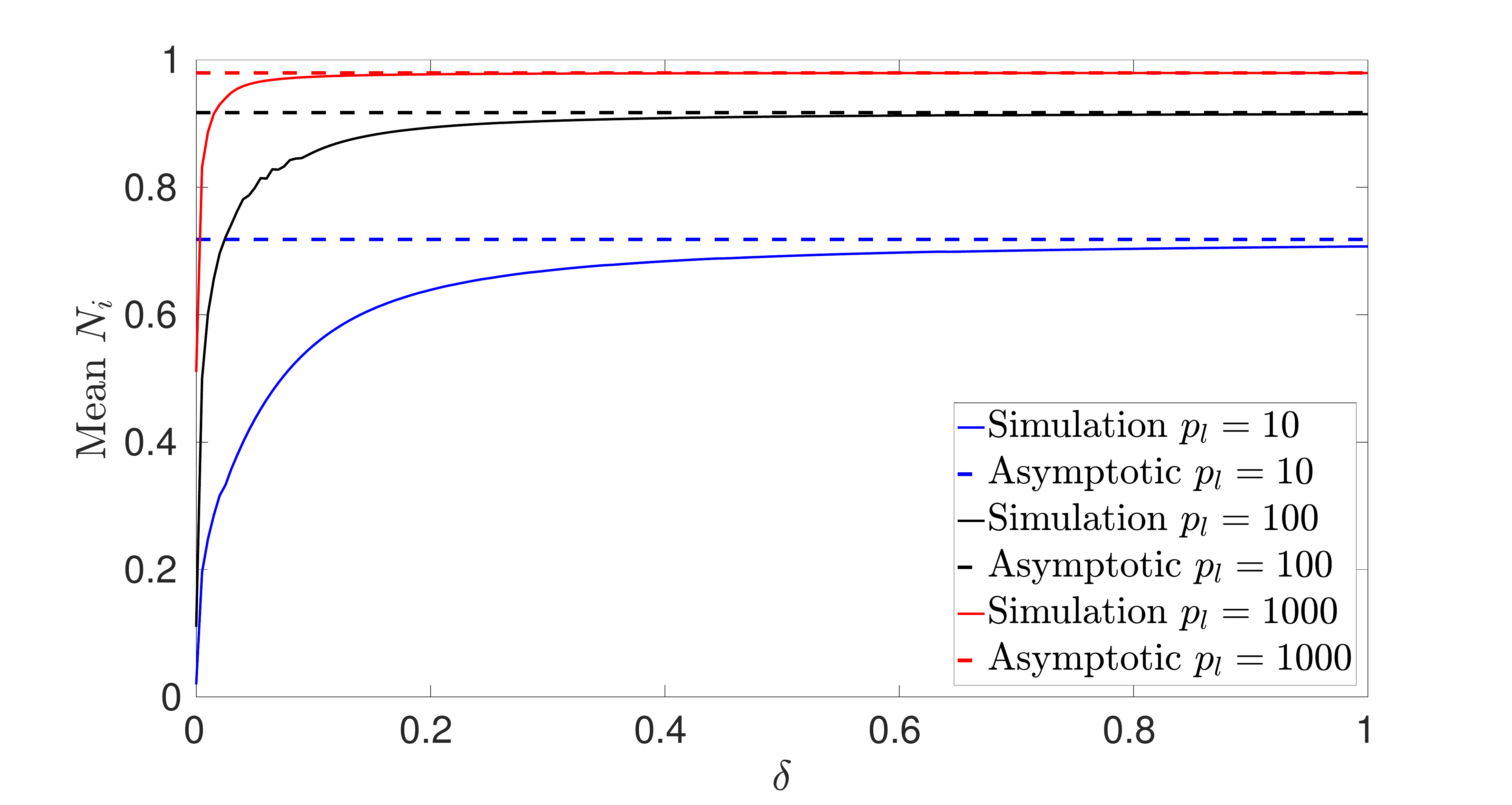}
\caption{Plots of steady state spatial mean cell density for lattice simulations with $n=10$ at three different values of the threshold $p_l$ against the predicted steady state solution from equation \eqref{lattice_d_ss}.}\label{lattice_large_delta}
\end{figure}

In Figure \ref{lattice_large_delta} we plot values of (spatial) mean cell densities for simulated solutions of equations \eqref{lattice_fluid_eqns}-\eqref{f2} for a lattice with $n=10$, along with corresponding $N_0^*$ computed from Equation \eqref{lattice_d_ss}. In both cases we plot the steady state behaviour, so if the simulations of the full problem were oscillatory, we averaged over $t=10$ units of time. The large $\delta$ steady state $N_0^*$ accurately predicts the long-time spatial mean cell density for $\delta \approx 1$ for all values of $p_l$, and for smaller values of $\delta$ for larger values of $p_l$. For smaller values of $\delta$, the variation from the asymptotic solution is partly due to significant spatial variation in the cell density $N_i$. The jagged regions in the plots are due to changes in the number of grid points that are growing or dying, rather than coarse plotting. The convergence in $\delta$ being faster for larger values of $p_l$ explains why in Figure \ref{pressure1figa} the cell density is very close to uniform, whereas in Figure \ref{pressure1figb} there is a substantial deviation from uniformity.

\subsection{Continuum asymptotics}\label{PDEAsym}
Analytical solutions for the PDE model can be obtained in exactly the same way as in Section \ref{lattice_d_asymptotics}, so we just mention key steps here. We expand our variables as $N = N_0 + \delta^{-1}N_1 + O(\delta^{-2})$ and $p = p_0 + \delta^{-1}p_1 + O(\delta^{-2})$. Substituting these into Equation \eqref{pde_cell_equation}, we find the leading-order equation for cell density,
\begin{equation}
\nabla^2 N_0 = 0,
\end{equation}
which, along with the boundary condition in equations \eqref{pde_cell_inits}, shows that $N_0$ is constant throughout the domain (e.g.~by the Maximum Principle). To find $N_0$ we consider the next order problem,
\begin{equation}\label{pde_d_second_order}
\frac{d N_0}{d t} = H(p_c-p_{0})N_0(1-N_0) - H(p_{0}-p_c)N_0+ \nabla^2 N_1.
\end{equation}

We now eliminate the dependence on $N_1$ using the divergence theorem. Integrating equation \eqref{pde_d_second_order} over the domain, along with the boundary conditions in \eqref{pde_cell_inits}, we find
\begin{equation}\label{pde_d_second_order_integrated}
\frac{d N_0}{d t} =  \int_0^1 H(p_c-p_{0}(x))N_0(1-N_0) - H(p_{0}(x)-p_c)N_0 dx,
\end{equation}
where $p_0(x)$ is a function of $x$ only due to the leading-order cell density $N_0$ being constant.

From equation \eqref{pde_sym} we have that $p_{0}(x)=(1-x)/k(N_0)$ which is monotonically decreasing in $x$. Given the bound from \eqref{pde_range}, which implies that we will see neither uniform logistic growth or uniform cell death, there must exist a point $x^*(N_0)$ such that for all $x\leq x^*(N_0)$, $p_{0}(x) \geq p_c$, and for all $x>x^*(N_0)$, $p_{0}(x) <p_c$. From Equation \eqref{pde_d_second_order_integrated} we then have,
\begin{equation}\label{pde_d_second_order_simplified}
\frac{d N_0}{d t} =  (1-x^*(N_0))N_0(1-N_0) - x^*(N_0) N_0,
\end{equation}
where $x^*(N_0) = [1-p_c k(N_0)]^+= [1-p_c (1-\rho N_0)^3]^+$, where again the positive part must be taken to avoid a negative value of $x^*(N_0)$ for small $N_0$. We have reduced a system of two partial differential equations in equations \eqref{pde_fluid1}-\eqref{pde_cell_equation} to equation \eqref{pde_d_second_order_simplified} which is an ordinary differential equation, although $x^*(N_0)$ is not everywhere-differentiable. At steady state, the solution $N_0^*$ must satisfy,
\begin{equation}\label{pde_d_ss}
0 =  p_c k(N_0^*)(2-N_0^*) - 1 = p_c (1-\rho N_0^*)^3(2-N_0^*) - 1,
\end{equation}
where we are neglecting the unstable steady state $N_0^*=0$. This is a quartic equation for $N_0^*$ which can be solved numerically. Up to constitutive differences giving different exponents between $k(N_0)$ and $R^{-4}(N_0)$, it can also be seen as the limit of equation \eqref{lattice_d_ss} as $n \to \infty$. As in equation \eqref{lattice_d_ss}, equation \eqref{pde_d_ss} can be shown to have at most one solution, $\tilde{N}$, in the interval $(0,1)$ which is globally attractive.  

We can follow the same procedure that was employed in Section \ref{lattice_d_asymptotics} to derive Equations \eqref{Asym_Sol} and \eqref{Asym_Sol2} to solve equation \eqref{pde_d_second_order_simplified}. We again consider a period of logistic growth (assuming the initial cell density is not too large), wherein the solution can be found analytically. As the cell density becomes large enough to induce non-uniform death (e.g. $x^*(N_0 > 0)$), we exploit the fact that the difference between $N_0(t)$ and the equilibrium solution given by Equation \eqref{pde_d_ss} is small to find the solution for all time $t$. We define $\hat{N_0}$ to be the value of cell density such that $x^*(\hat{N_0})=0$ and for all $N_0 > \hat{N_0}$, $x^*(N_0)>0$. If we assume that $N_0(0) < \hat{N_0}$, then the solution takes the form
\be\label{Asym_Sol_PDE}
N_0(t) = \begin{cases}\displaystyle
\frac{N_0(0)e^t}{N_0(0)(e^t-1)+1}, \quad &t \leq t_c,\\
\displaystyle N_0^*+\left(\frac{N_0(0)e^{t_c}}{N_0(0)(e^{t_c}-1)+1}-N_0^*\right)e^{-r_c(t-t_c)}, \quad &t > t_c,
\end{cases}
\ee
where $N_0(0)$ is the initial mean cell density, $N_0^*$ is the unique root of equation \eqref{pde_d_ss}, the switching time $t_c$ is given by
\be
t_c = \ln\left(\frac{(1-N_0(0))\left(1-\left(p_c^\frac{1}{3}-1\right)\right)}{N_0(0)(1-\left(1-\rho\right)p_c^\frac{1}{3}}\right),
\ee
and $r_c$ is given by
\be
r_c = p_c(3(1-\rho N_0^*)^2N_0^*(2-N_0^*)\rho - 2(1-\rho N_0^*)^3(1-N_0^*))+1.
\ee
Similarly, if we assume that the initial cell density satisfies $N_0(0) > \hat{N_0}$, then the solution is
\be\label{Asym_Sol_PDE2}
N_0(t) = N_0^* + (N_0(0)-N_0^*)e^{-r_c t}.
\ee

\begin{figure}
\centering
\includegraphics[width=0.7\textwidth]{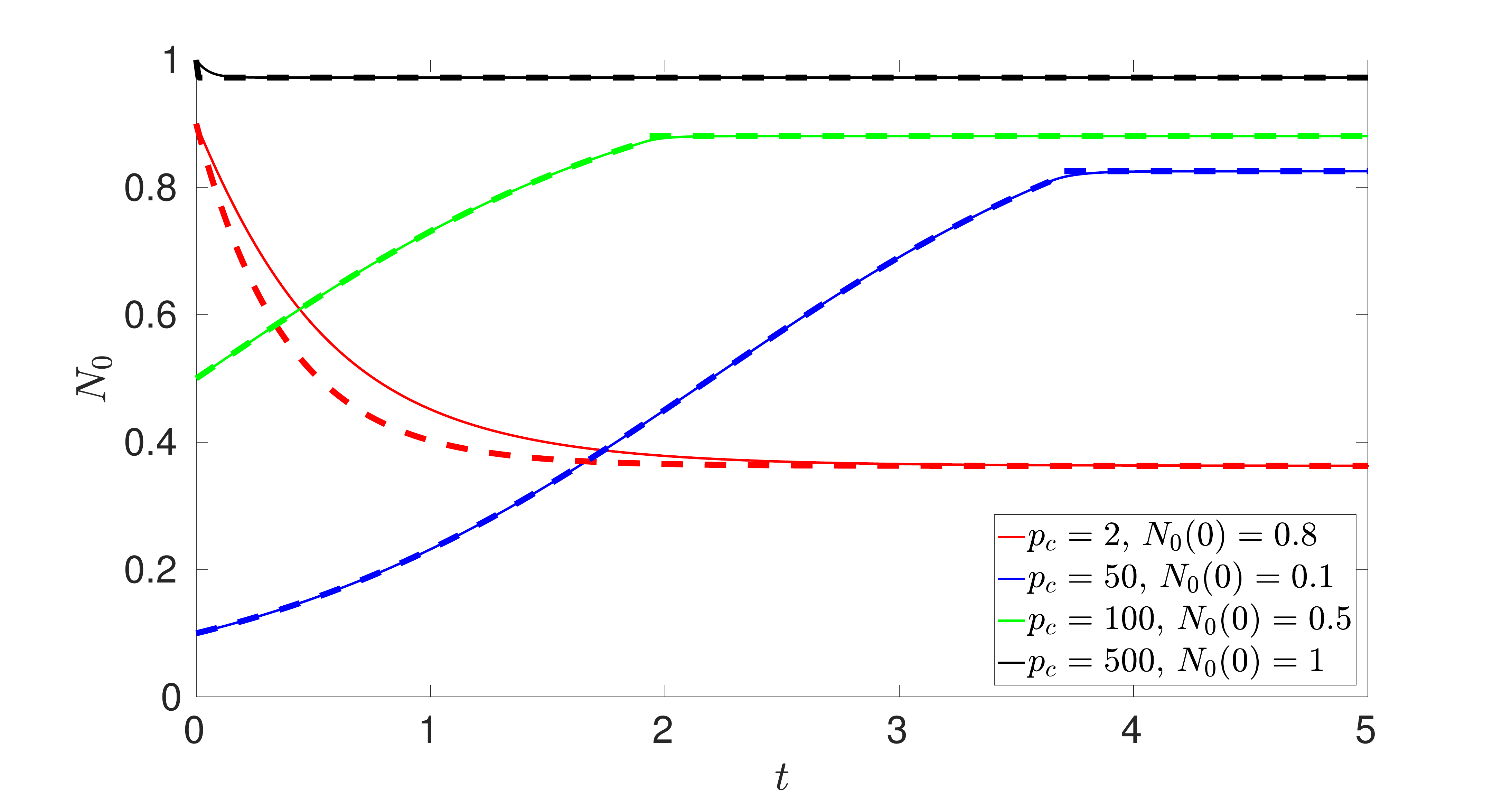}
\caption{Plots of $N_0(t)$ computed from simulations of equation \eqref{pde_d_second_order_simplified} as solid lines, as well as the analytical solutions, \eqref{Asym_Sol_PDE} for increasing curves and \eqref{Asym_Sol_PDE2} for decreasing curves, as dashed lines for different initial mean cell densities $N_0(0)$ and thresholds $p_c$.}\label{temporal_asymptoticsPDE}
\end{figure}

Solutions \eqref{Asym_Sol_PDE} and \eqref{Asym_Sol_PDE2} are analogous to the lattice solutions \eqref{Asym_Sol} and \eqref{Asym_Sol2}. In Figure \ref{temporal_asymptoticsPDE} we compare the analytical and numerical solutions to equation \eqref{pde_d_second_order_simplified}. The worst case behaviour, shown in red, corresponds to the asymptotic parameter $N_0^*-N_0(t_c) = 0.5$, and it is not surprising that there is a transient discrepancy between the numerical and analytical solutions in this case. 

\begin{figure}
\centering
\includegraphics[width=0.7\textwidth]{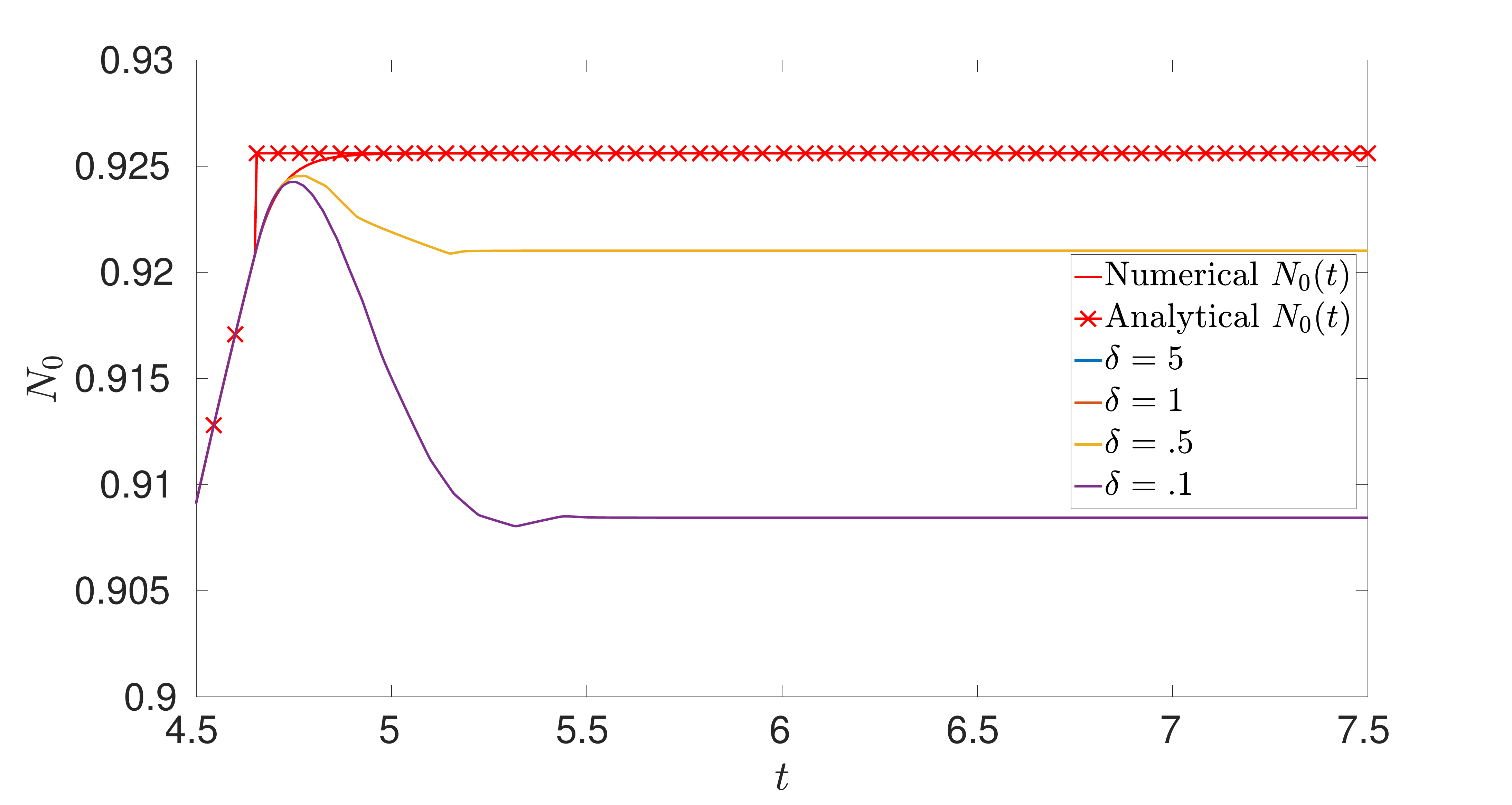}
\caption{Plots of $N_0(t)$ computed from numerical solutions of equations \eqref{pde_d_second_order_simplified}, analytical solutions given by \eqref{Asym_Sol_PDE}, and simulations of the full two-dimensional model \eqref{pde_fluid1}-\eqref{pde_cell_inits} for $p_c=200$, and $\delta=0.1,0.5,1,$ and $5$.}\label{temporal_asymptotics2PDE}
\end{figure}

We compare this solution for the cell density $N_0$ against the spatial mean cell density for the full 2-D problem \eqref{pde_fluid1}-\eqref{pde_cell_inits} for various values of $\delta$ in Figure \ref{temporal_asymptotics2PDE}. For small values of $\delta$, as in the lattice case, there is some discrepancy between the numerical and asymptotic solutions due to spatial structure. As we increase $\delta$, however, we see a relatively fast convergence to the asymptotic solution for $N_0$. Comparable results hold for different values of $p_c$. We note that the asymptotic solution converges more quickly to the equilibrium value $N_0^*$ than the numerical solution of Equation \eqref{pde_d_second_order_simplified}.

In this section, we have shown that there exist parameter regimes with relatively simple dynamics for large values of $\delta$, or moderate $\delta$ and large values of $p_l$ or $p_c$ (see Figure \ref{lattice_large_delta}; the PDE behaviour is qualitatively similar). In these regimes, the transient and long-time dynamics of both the lattice and PDE models can be captured by simple scalar ODEs each admitting a unique attracting steady state (Equations \eqref{Cell_Expansion_3} and \eqref{pde_d_second_order_simplified} respectively). We anticipate that the oscillatory solutions observed in Section \ref{Numerical_Overview} can be confined to regions of the parameter space where either $\delta$ is small, or the threshold $p_l$ or $p_c$ is small. We also note that other simulations show that the dimensionality (e.g. 1-D or 2-D) or the nature of the boundary conditions of the model (e.g. periodic) does not play a role in solution behaviour within this asymptotic regime (large $\delta$), whereas these model details do play a role in the nature of oscillations and steady states observed outside of this asymptotic regime (see Figure \ref{defect_oscillation2}).

\section{Bifurcations in small and large lattices}\label{Bifurcations}
We now consider the behaviours observed in Section \ref{Numerical_Overview} from the perspective of how solution behaviours change as parameters are varied. We begin by demonstrating the existence of Hopf and pitchfork bifurcations from steady states in an $n=4$ lattice as $\delta$ varies, and describe how the size of the lattice affects the existence of these bifurcations. Finally we classify behaviours observed in simulations throughout a bounded subset of the parameter space of $\delta$ and $p_l$ (or $p_c$) for larger lattices and the PDE.

\subsection{Existence of symmetry-breaking bifurcations}
We examine a low-dimensional ($n=4$) lattice model, and numerically continue steady state solutions over a range of $\delta$ with $p_l=15$. We track the local stability of these steady state solutions to small amplitude perturbations by evaluating the Jacobian of the system and its eigenvalues. We use natural parameter continuation and manually switch between solution branches by continuing solutions in $\delta$ from any stable equilibria found by solving Equation \eqref{general_lattice_cell_eqn}. That is, we take our equilibrium solution for a specific value of $\delta$, $N_i^*(\delta)$, then solve the algebraic steady state Equation \eqref{general_lattice_cell_eqn} using the \textsc{Matlab} function `fsolve' with $N_i^*(\delta)$ as an initial guess for the solution at a new value of the parameter $\delta' = \delta \pm \epsilon$ for some small $\epsilon$. We choose $\epsilon = 10^{-4}$, and continue solutions forward and backward for all $\delta \in [0,2]$. 

Equation \eqref{lattice_fluid_eqns} can be viewed as a weighted graph Laplace equation, and can be inverted numerically to find the pressures as functions of the cell density. We consider the $n^2 \times n^2$ matrix corresponding to the Jacobian of equation \eqref{general_lattice_cell_eqn} with $p_i$ a function of $N_1,\dots,N_{n^2}$ for all $i$. We denote the $ith$ eigenvalue of the Jacobian evaluated along a branch of steady states by $\sigma_i$. See \cite{seydel_practical_2009} for discussion of continuation procedures in general.
 
\begin{figure}
\centering
\subfloat[]{
\includegraphics[width=0.5\textwidth]{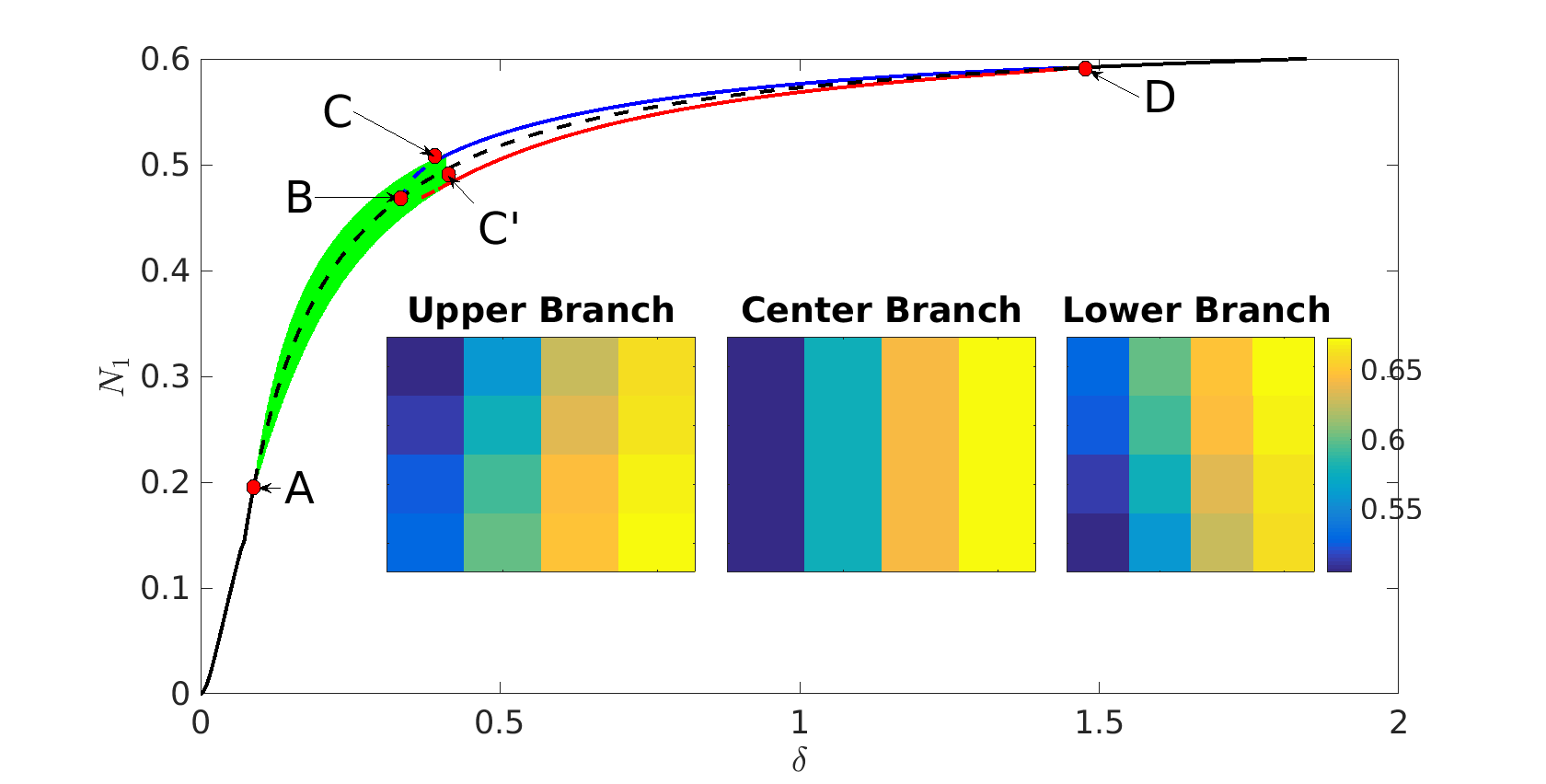}
    }
    \subfloat[]{
\includegraphics[width=0.5\textwidth]{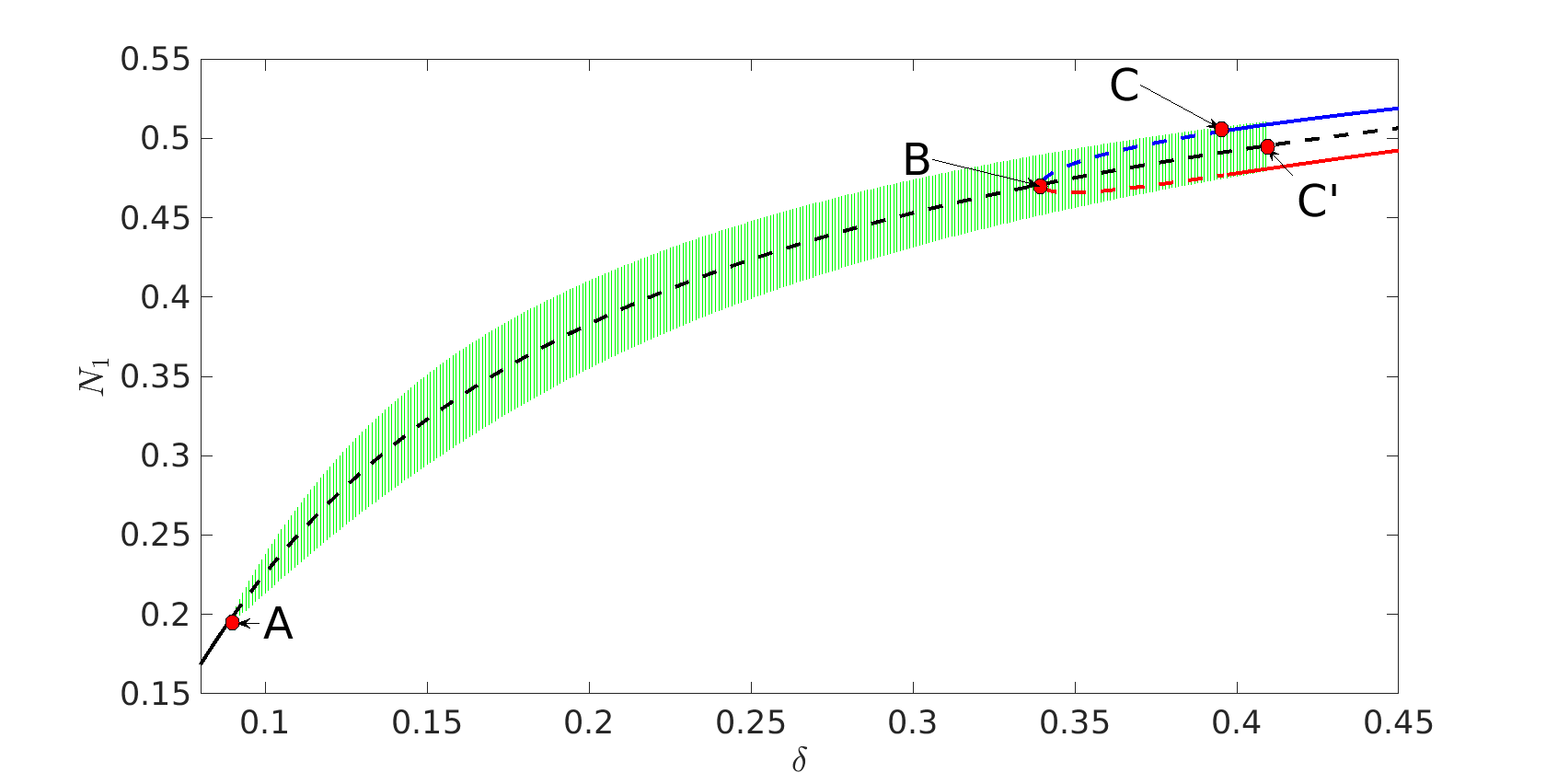}
    }
\caption{In (a) we plot the cell density $N_1$ at the bottom-left of the lattice corresponding to three steady state branches found via numerical continuation. Solid lines correspond to locally stable steady states, dashed lines to unstable steady states, and the green lines represent an envelope of oscillations about the steady state solutions. The insets show cell density plots across the lattice at $\delta=0.5$. Note that only the Center Branch (black line) solution is vertically symmetric, whereas the other two are not. The letters correspond to different bifurcations described in the text. For this case, $n=4$, $p_l = 15$ and $\rho = 0.9$. The plot in (b) is a closer look at the oscillatory regime.}\label{continuation}
\end{figure}

\begin{figure}
\centering
\includegraphics[width=0.6\textwidth]{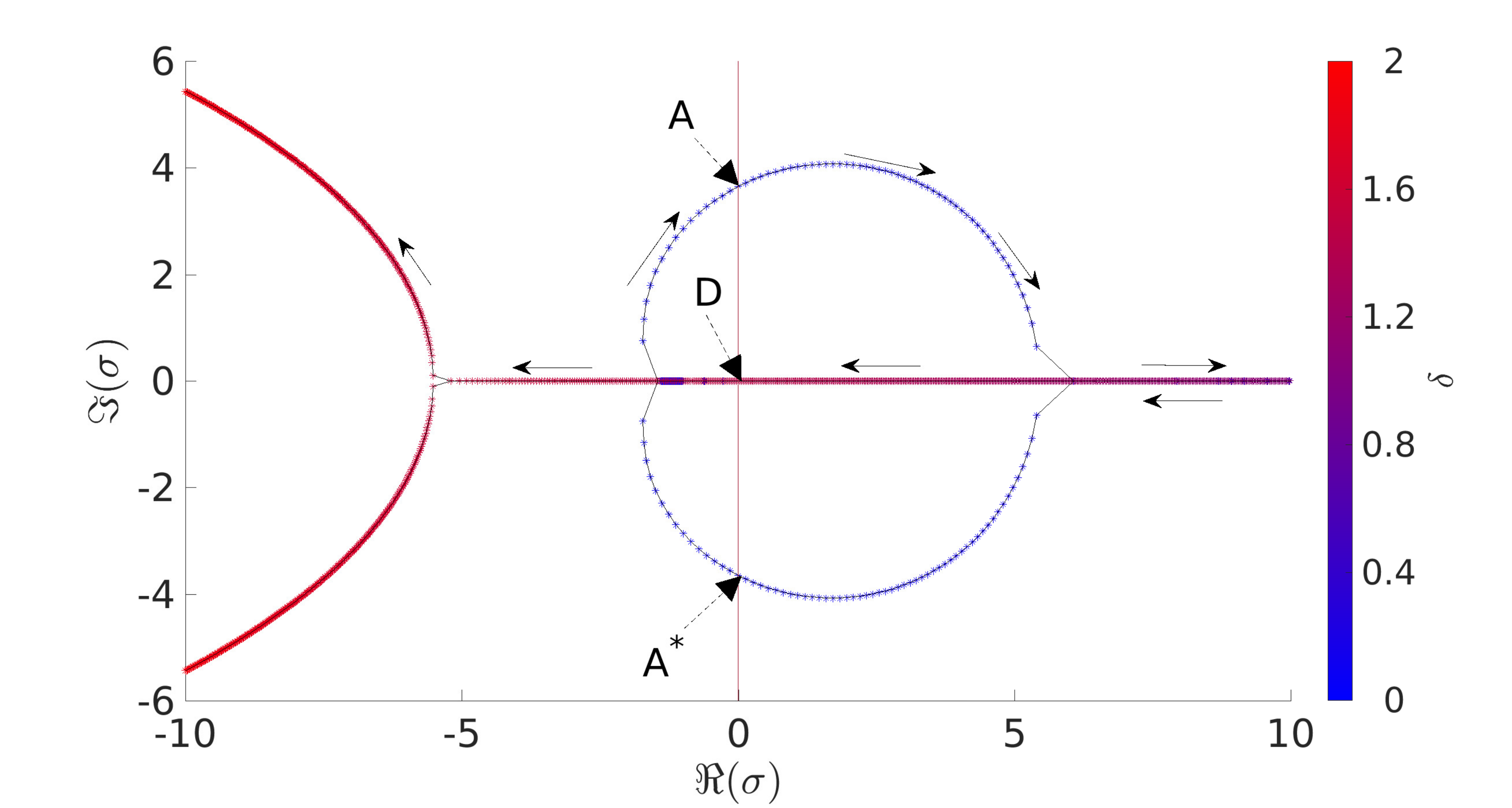}
\caption{Plots of the two eigenvalues $\sigma_i$ with largest real part of the Jacobian evaluated at the vertically symmetric branch, coloured according to the value of $\delta$. The arrows denote the direction of increasing $\delta$. The first crossing of the imaginary axis $\Re(\sigma_i)=0$, at A and its complex conjugate A$^*$, corresponds to a Hopf bifurcation. The second crossing of the imaginary axis, at point D, corresponds to a pitchfork bifurcation where the center branch again gains stability.}\label{continuation2}
\end{figure}

\begin{figure}
\centering
\includegraphics[width=0.6\textwidth]{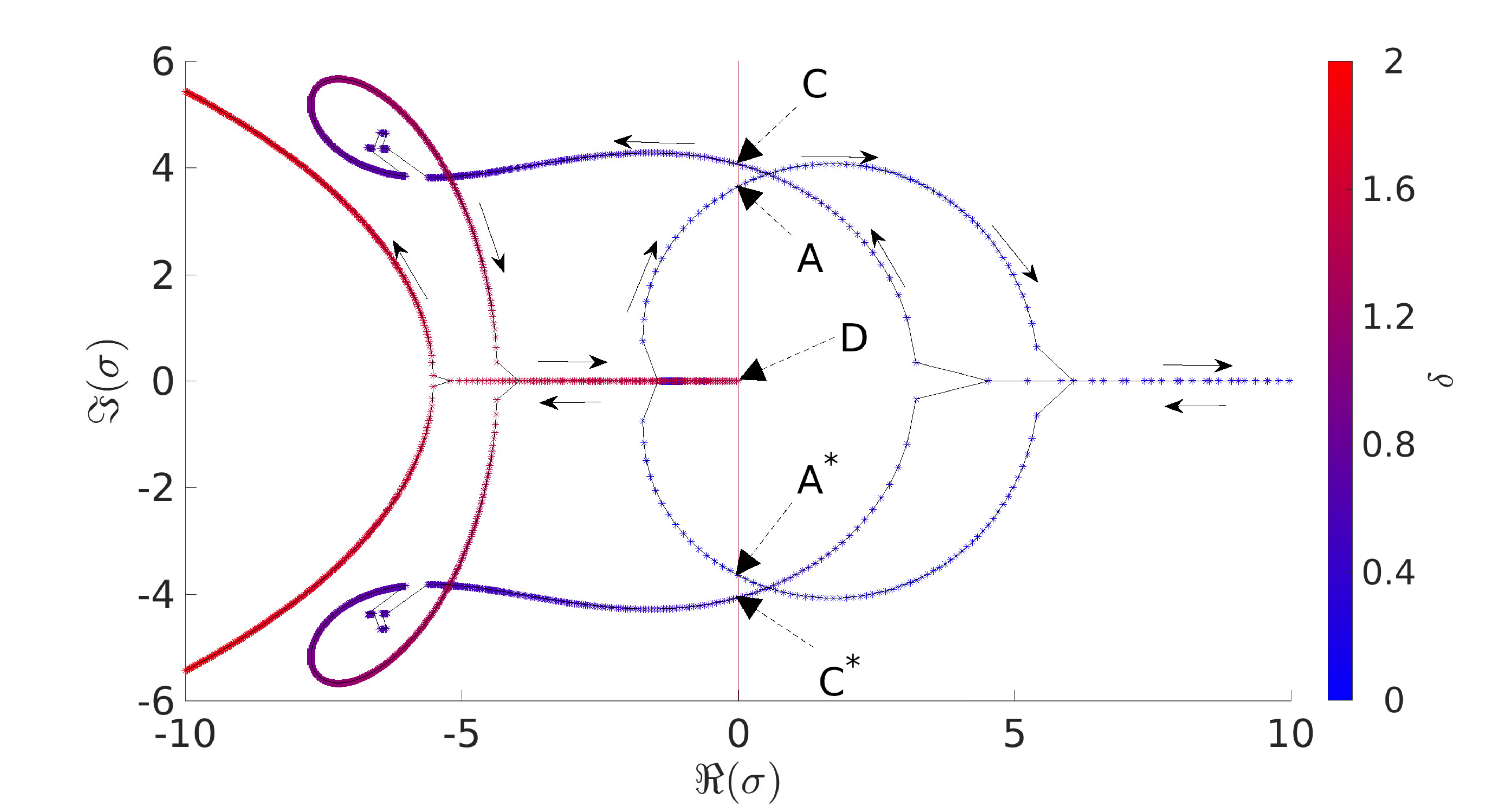}
\caption{Plots of the two eigenvalues $\sigma_i$ with largest real part of the Jacobian evaluated at the upper branch, coloured according to the value of $\delta$. The arrows denote the direction of increasing $\delta$. Note that the eigenvalues along the lower branch are precisely the same for small $\delta$. For $\delta< 0.4$ the eigenvalues follow the same trajectory as in Figure \ref{continuation2}, undergoing a Hopf Bifurcation at A (with complex conjugate A$^*$), but rather than returning along the real axis they develop a complex conjugate pair, and cross the imaginary axis in a pair at C and C$^*$. They approach the origin along the real axis at D, but remain in the left hand side of the plane, as this solution branch merges with the vertically symmetric branch which gains stability at this point.} \label{continuation3}
\end{figure}

In Figure \ref{continuation}, we plot the value of the cell density at the bottom-left node, $N_1$, corresponding to different steady state solution branches, for varying $\delta$ found from the above procedure. The solid lines denote steady states with all eigenvalues having negative real part, $\max_i \{\Re(\sigma_i)\} < 0$, and hence are locally linearly stable. The dashed lines have at least one eigenvalue with positive real part, and so are linearly unstable. The black line corresponds to a vertically symmetric equilibrium (which remains vertically symmetric for all $\delta$), and the red and blue lines to vertically asymmetric equilibria (see the insets of Figure \ref{continuation} which show cell density distributions for these equilibria at $\delta = 0.5$).

Around $\delta \approx 0.08$ (point A), the vertically symmetric steady state loses stability. Beyond point A there is an oscillating solution that attracts solutions perturbed from the now unstable vertically symmetric solution. This oscillating solution is of the vertically asymmetric kind plotted in Figure \ref{defect_oscillation}. We plot the envelope of the oscillation that $N_1$ undergoes during a period. We plot the two eigenvalues with largest real part in Figure \ref{continuation2}, and show that at the point where the vertically symmetric steady state loses stability, the eigenvalues cross the imaginary axis in a conjugate pair (e.g. $\Im(\sigma_i) \neq 0$ for the two complex conjugate $\sigma_i$ with $\Re(\sigma_i)=0$). Since the oscillatory solution immediately beyond this loss of stability is numerically stable and its amplitude is small for $\delta$ near the bifurcation point, this indicates a supercritical Hopf bifurcation giving rise to the stable limit cycle that we observe for $\delta > 0.08$ in Figure \ref{continuation}.

For $\delta \approx 1.5$ (point D), two solution branches emerge from the vertically symmetric branch as it loses stability (for decreasing $\delta$). These vertically asymmetric solutions are reflected copies of one another, as is typical in a pitchfork bifurcation. We plot the eigenvalues of these continued branches in Figure \ref{continuation3} showing evidence that this is a pitchfork bifurcation. The eigenvalues touch the imaginary axis with zero imaginary part at D in Figure \ref{continuation3} then return to the left hand side of the plane, whereas in Figure \ref{continuation2} the eigenvalues cross the imaginary axis as the vertically symmetric solution loses stability. The asymmetric branches both lose stability at $\delta \approx 0.46$ (point C), again satisfying the criteria for a Hopf bifurcation, as the eigenvalues with the largest real part again cross the imaginary axis in Figure \ref{continuation3} (but here for decreasing $\delta$). In this case, however, we do not find a small amplitude \emph{stable} limit cycle, and instead solutions are attracted to the limit cycle previously generated, which has a large amplitude at this point. This shows that there is a subcritical Hopf bifurcation for both of the asymmetric equilibria at C, which leads to the creation of an unstable limit cycle.

For larger values of diffusion the limit cycle is no longer observable and we suspect it becomes unstable around $\delta \approx 0.47$ (point C'). We know that an unstable limit cycle exists to the right of C, between C and C', and that this may lead to the loss of stability for the stable limit cycle at C', but we do not pursue these claims here. Lastly, we see that around $\delta \approx 0.4$ (point B) the three steady state solution branches merge, and the eigenvalues shown in Figure \ref{continuation3} have the same structure for the rest of the continued solution. Note that near B all branches have large real eigenvalues ($\sigma_i\approx 20$), and so there is no exchange of stability. While we have only shown plots for $N_1$, plots for all $N_i$ show that these behaviours are qualitatively consistent across all states, and hence these branches, their stability, and the merging of branches are faithfully captured by Figure \ref{continuation}.

The results in Figures \ref{continuation}-\ref{continuation3} suggest the existence of Hopf and pitchfork bifurcations, alongside multiple steady states, and oscillatory behaviour in the 2-D lattice with $n=4$. We did not observe any pulsing oscillations of the type shown in Figure \ref{cell1fig}. We briefly summarize results for other values of $n$. For $n=2$ and $n=3$ we only observed steady states for a large region of parameter space. Similar behaviours to $n=4$ (vertically asymmetric and symmetric steady states and asymmetric oscillations) were observed for $n=5,6$. For $n=7$, the 2-D lattice had both vertically symmetric pulsing oscillations, as in Figure \ref{cell1fig} and vertically asymmetric oscillations, in addition to steady state behaviour. For the 1-D model given by Equations \eqref{1dlattice}, we only observed steady state behaviour for $n \leq 6$, and oscillations and steady states for $n \geq 7$. 

For parameter values where a vertically symmetric oscillation exists but a vertically asymmetric oscillation does not, the vertically symmetric steady state remains locally stable. The 1-D model exhibits a locally stable steady state solution for every parameter combination we simulated. Together, this suggests that it is only these asymmetric oscillations or steady states that can induce instability in vertically symmetric 2-D steady states, as these vertically symmetric steady states corresponds to steady states in the 1-D model which are always (locally) stable. The multiple stability of steady states and oscillations (e.g. between point C and C' in Figure \ref{continuation}) observed also provides an explanation for the excitable behaviour in Figure \ref{excitable_behaviour}, as the perturbations need to be sufficiently large to move out of the basin of attraction of the locally stable steady state.

These results suggest that the vertically asymmetric oscillations in the smaller lattices are due to local symmetry-breaking bifurcations of Hopf type. For general differential-algebraic systems, codimension-1 bifurcations can also be due to a singularity in the algebraic subsystem, or due to saddle-node bifurcations \cite{venkatasubramanian_local_1995}. By the regularity of the graph Laplacian in \eqref{lattice_fluid_eqns}, we conjecture that no singularity-induced bifurcations can occur. While we are unable to rule out saddle-node bifurcations, we do not observe them in any of our continuation studies. The vertically symmetric pulsing oscillations are not, as far as we can detect numerically, created via a Hopf or other local bifurcation from a steady state.

\subsection{Classification of the parameter space}\label{Classification}
Even for small lattices, a complete classification of the parameter space is not tractable. Instead we now consider lattices of varying size and classify behaviours broadly as oscillatory or steady-state, and record properties of solutions. Namely, we record the amplitude and frequency of oscillations. The results from Section \ref{Asymptotics} suggest that for large values of diffusion the cell density is approximated accurately by a spatially constant solution. For this reason we choose to do parameter sweeps between $\delta=0.001$ and $\delta=0.1$, and variations in the threshold parameters within the bounds given by \eqref{lattice_range}-\eqref{pde_range}. We compute bifurcation plots in $(\delta, p_l)$ and $(\delta, p_c)$ in the following way. We first discretize the parameter space, and numerically simulate the governing equations at each discrete point for a sufficiently long period of time (we set this time to be $t=40$, and use identical initial conditions described at the start of Section \ref{Numerical_Overview}). We then truncate our simulation to analyze only the last $10$ units of time. For each node, or in the case of the PDE for each interpolated discretized element, we compute the largest and smallest values this node takes in the truncated time series to compute a nodal oscillation amplitude.

\begin{figure}\setlength{\tabcolsep}{5pt}
\centering
\begin{tabular}{r c c} 
 & 1D & 2D\\

$n=10$ &\raisebox{-.5\height}{\includegraphics[width=.30\textwidth]{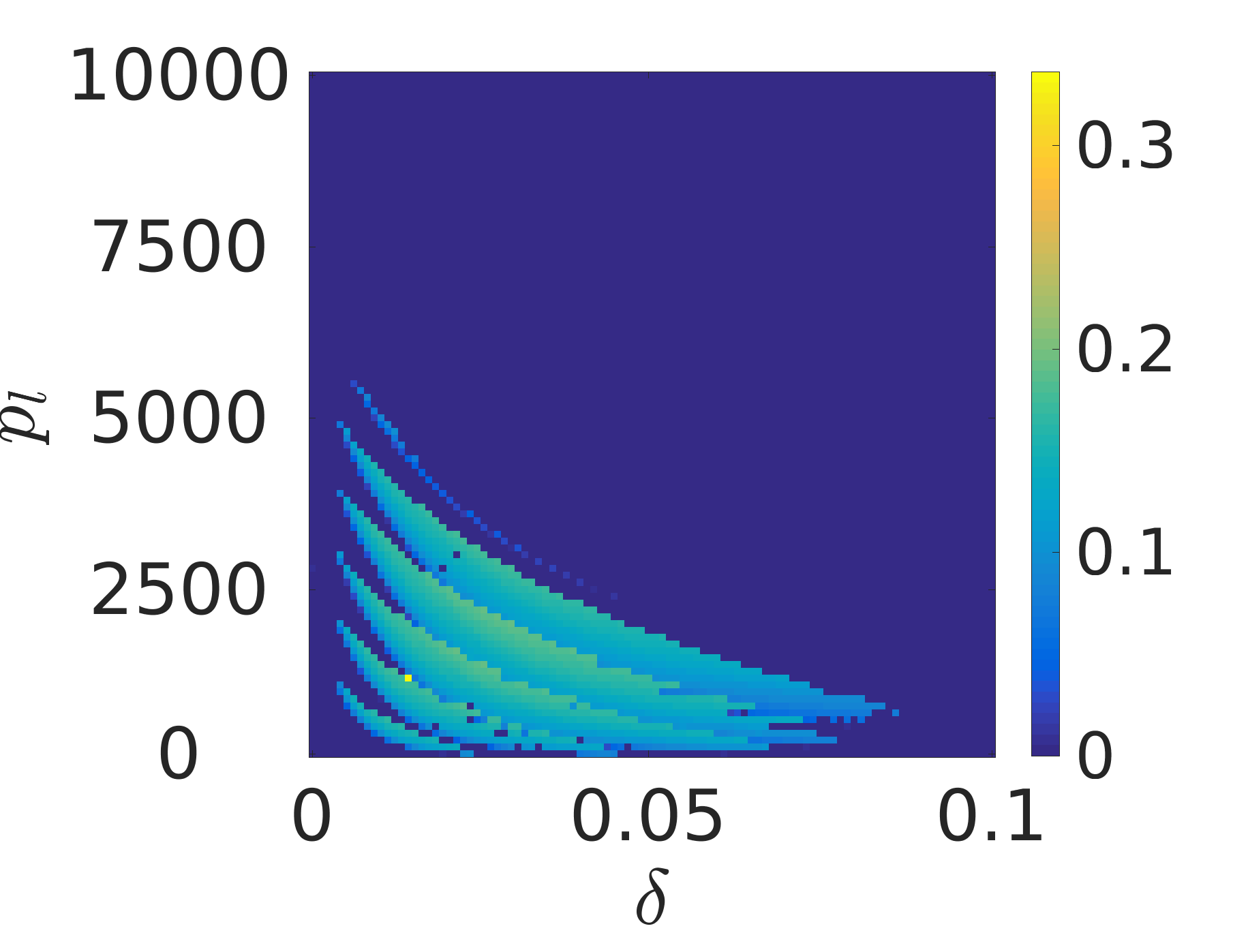}}&
\raisebox{-.5\height}{\includegraphics[width=.30\textwidth]{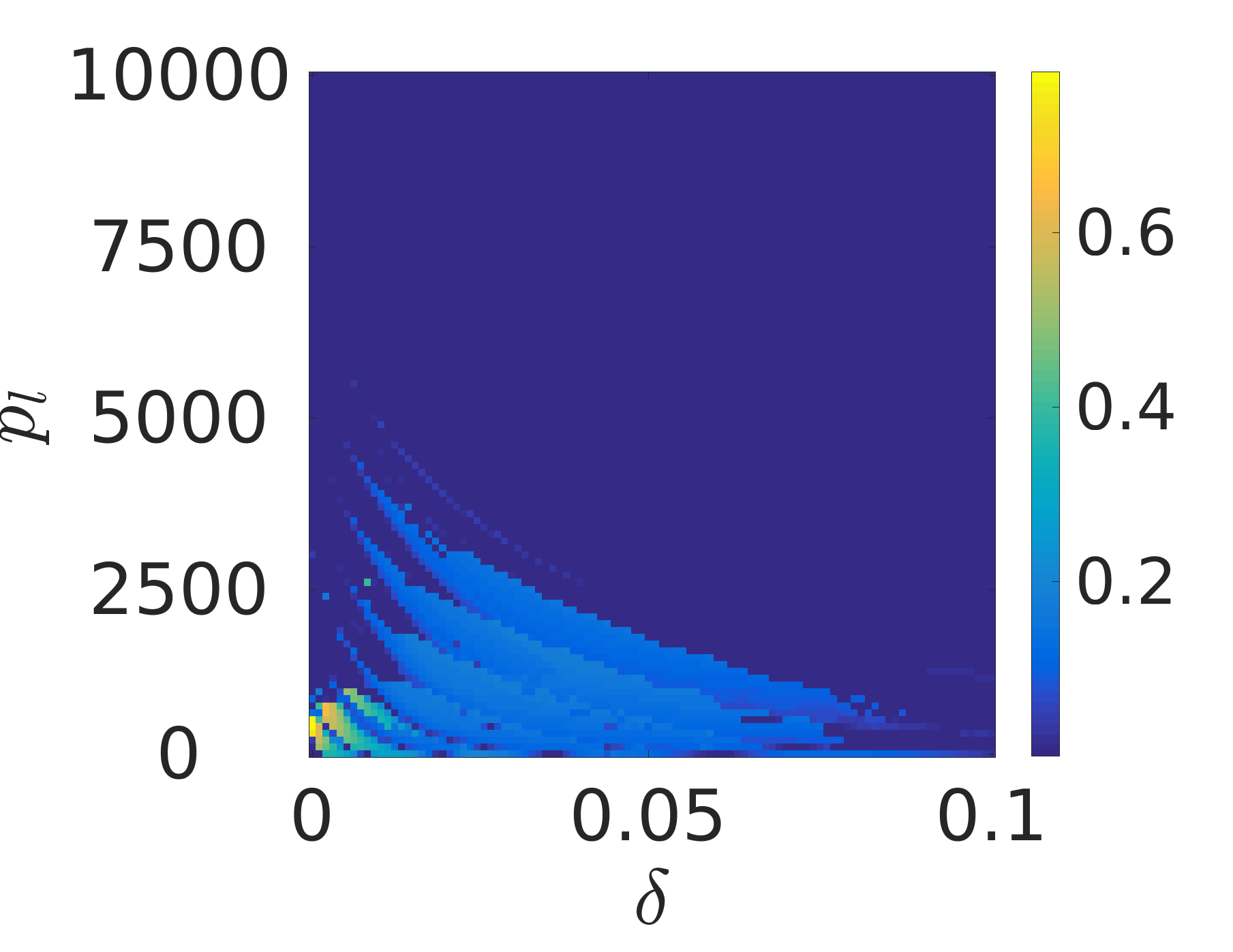}}\vspace{0mm}\\

$n=50$ & \raisebox{-.5\height}{\includegraphics[width=.30\textwidth]{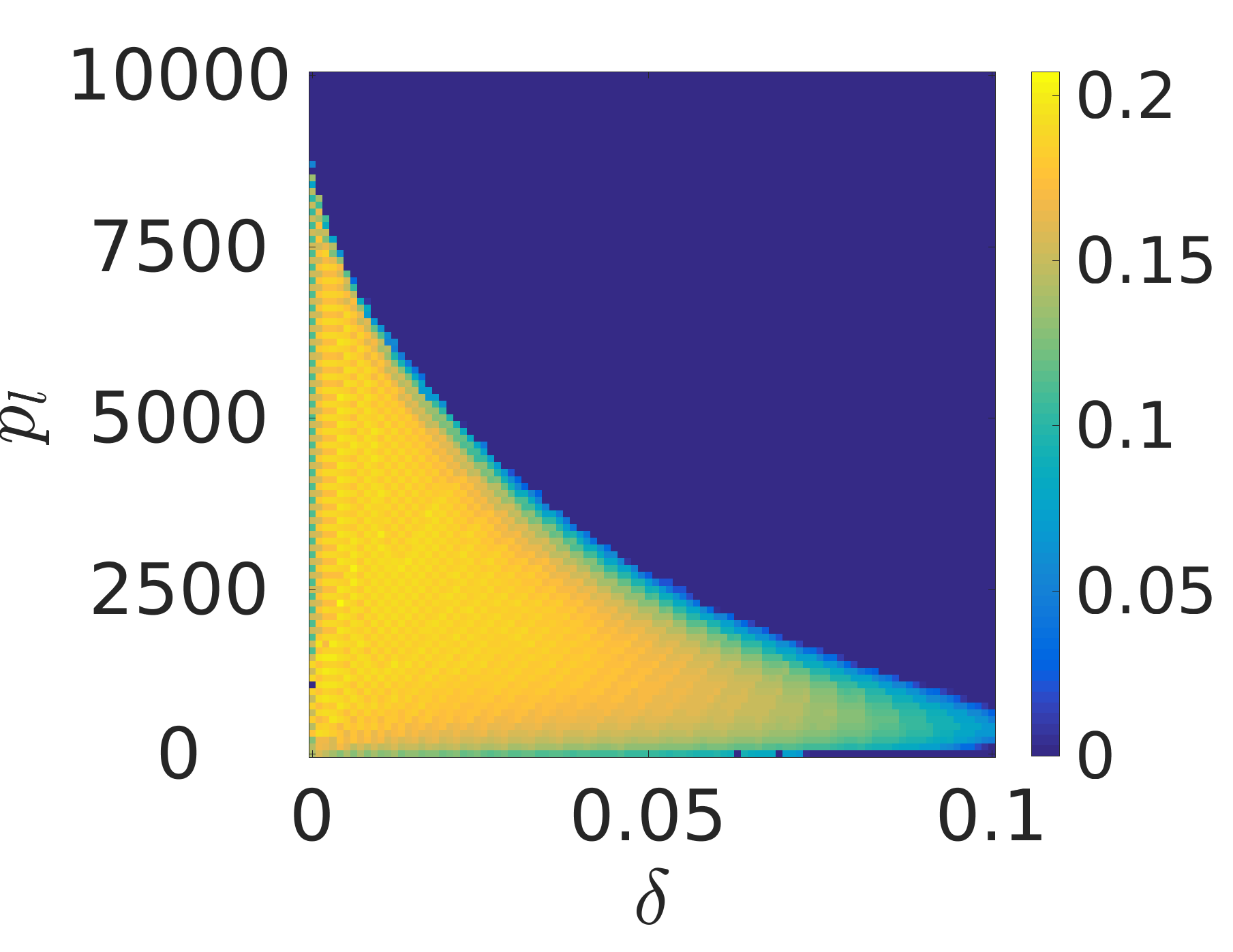}}&\raisebox{-.5\height}{\includegraphics[width=.30\textwidth]{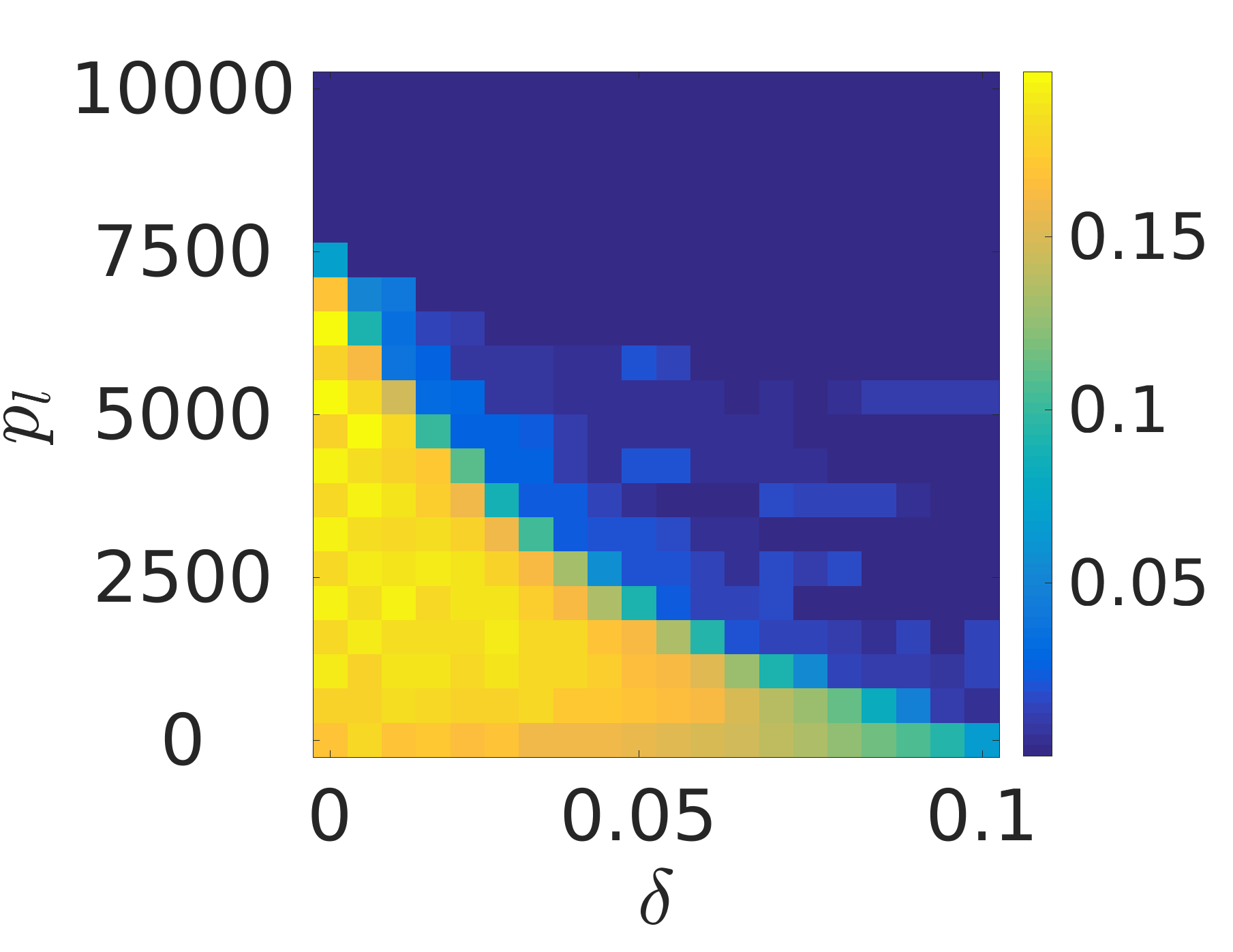}}
\vspace{0mm}\\

$n=100$ &\raisebox{-.5\height}{\includegraphics[width=.30\textwidth]{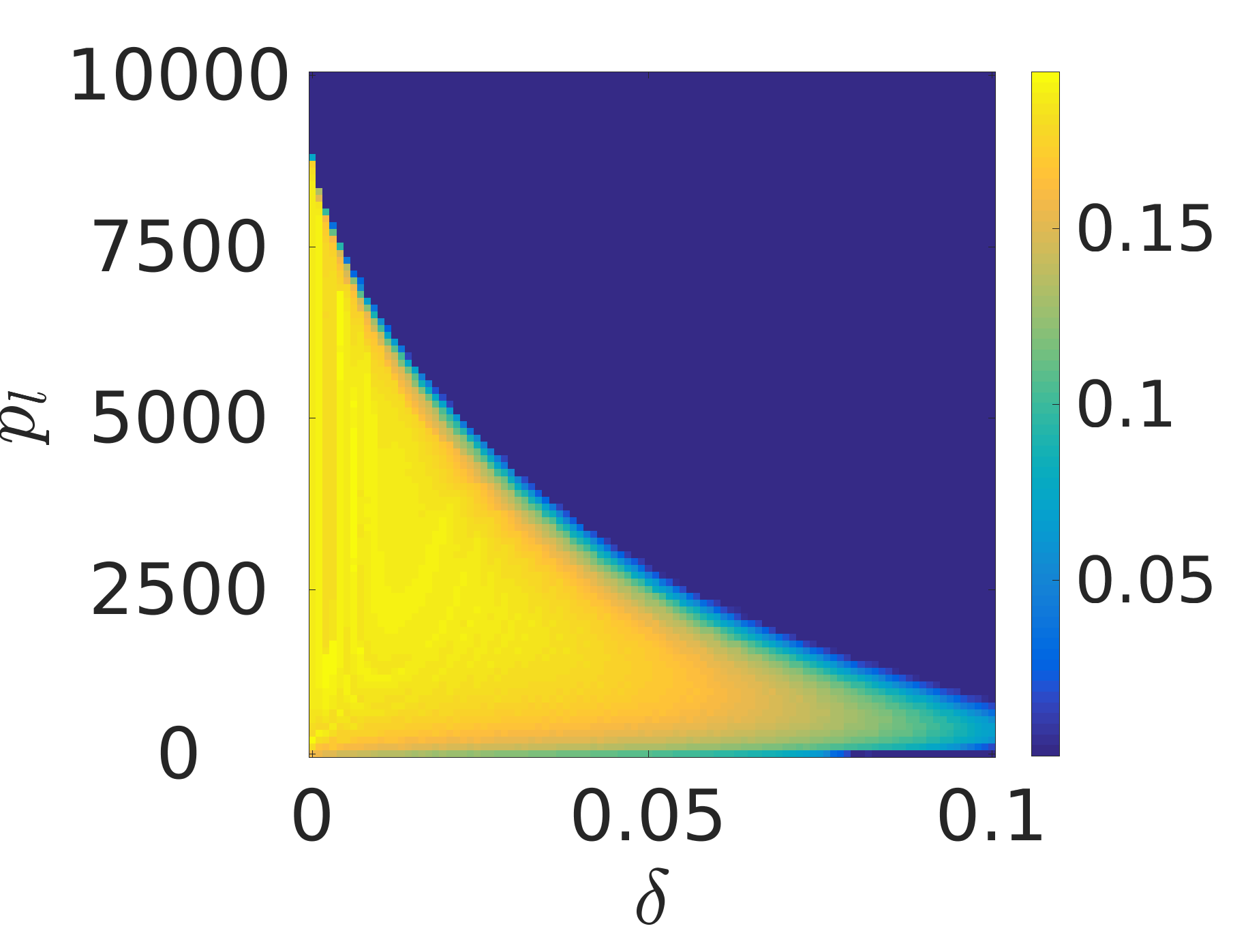}}&\raisebox{-.5\height}{\includegraphics[width=.30\textwidth]{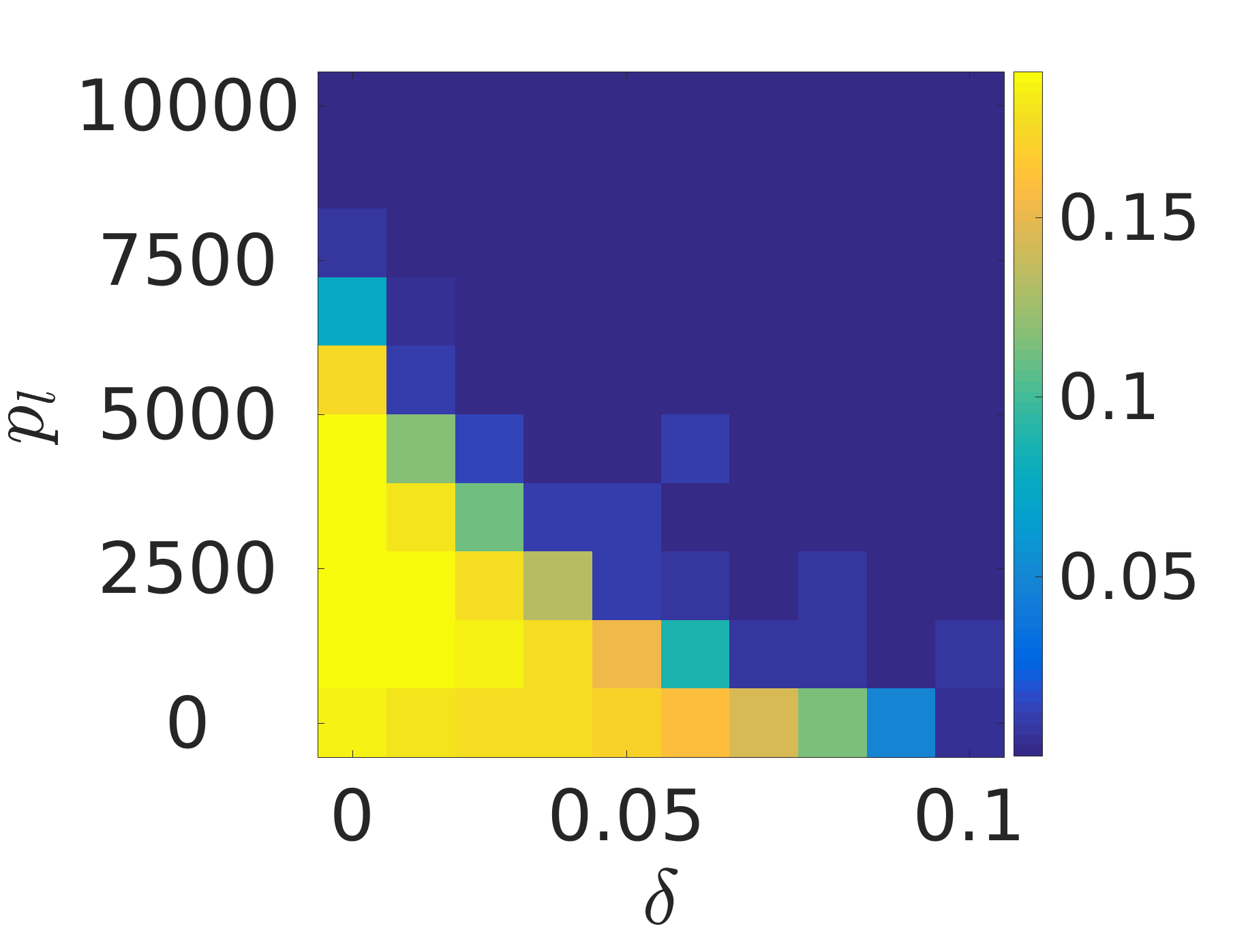}}
\end{tabular}
\caption{Amplitude of the maximal nodal oscillation after $t=40$ time units for different lattice sizes and dimensions over $\delta$ and $p_l$.}
\label{bifurcation_amplitude_plots}
\end{figure}

\begin{figure}\setlength{\tabcolsep}{5pt}
\begin{center}
\begin{tabular}{r c c} 
 & 1D & 2D\\
$n=10$ &\raisebox{-.5\height}{\includegraphics[width=.30\textwidth]{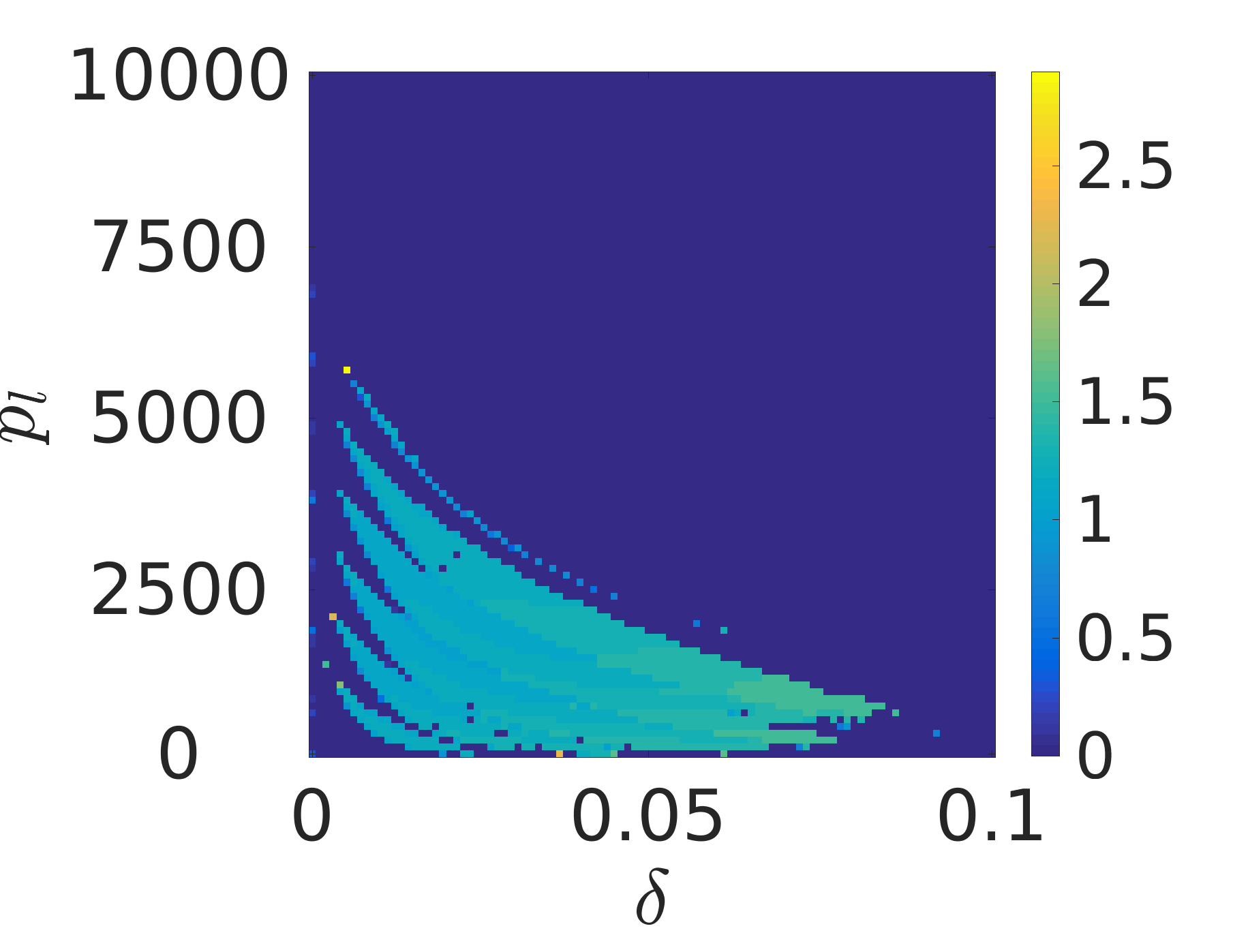}}&
\raisebox{-.5\height}{\includegraphics[width=.30\textwidth]{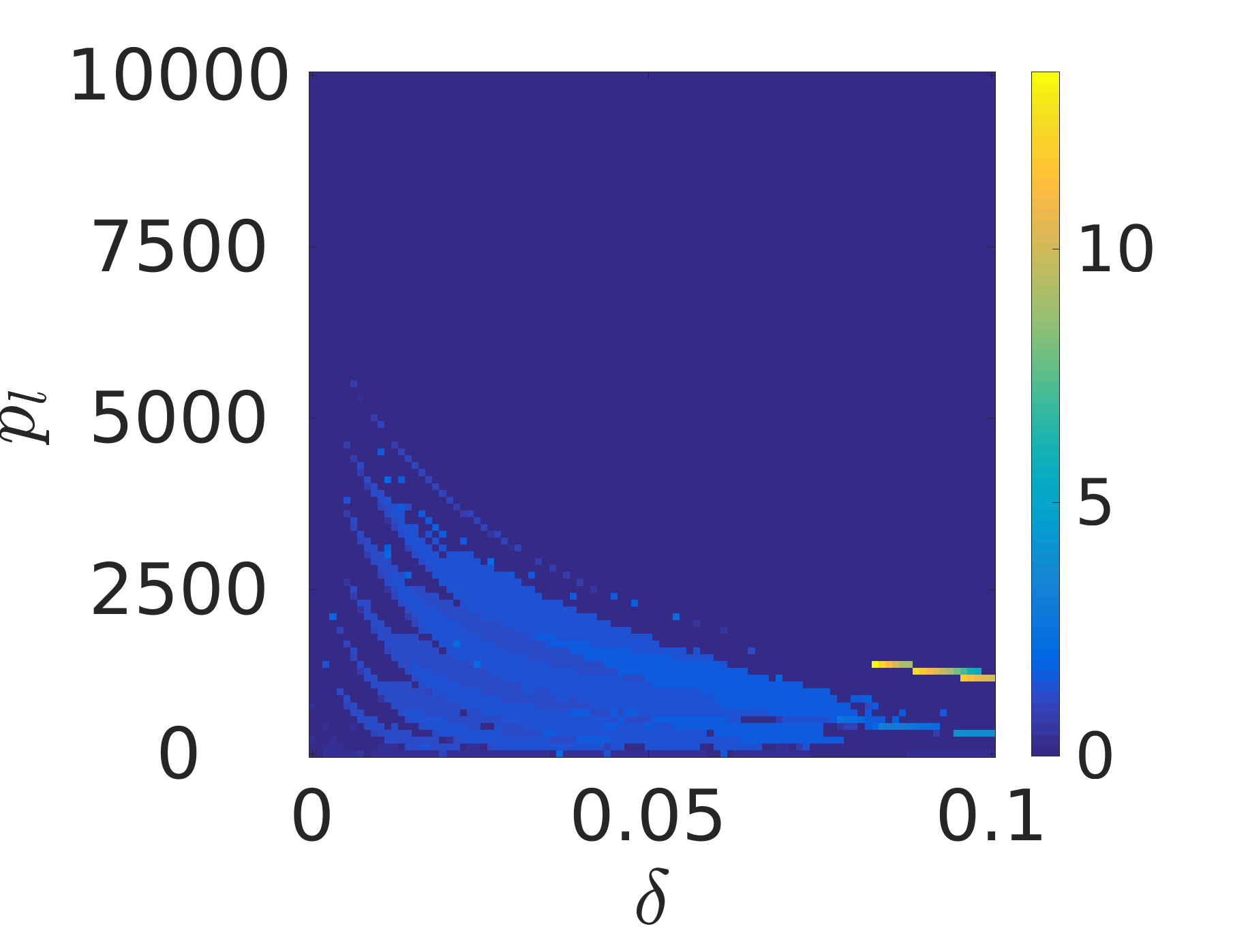}}\vspace{0mm}\\

$n=50$ & \raisebox{-.5\height}{\includegraphics[width=.30\textwidth]{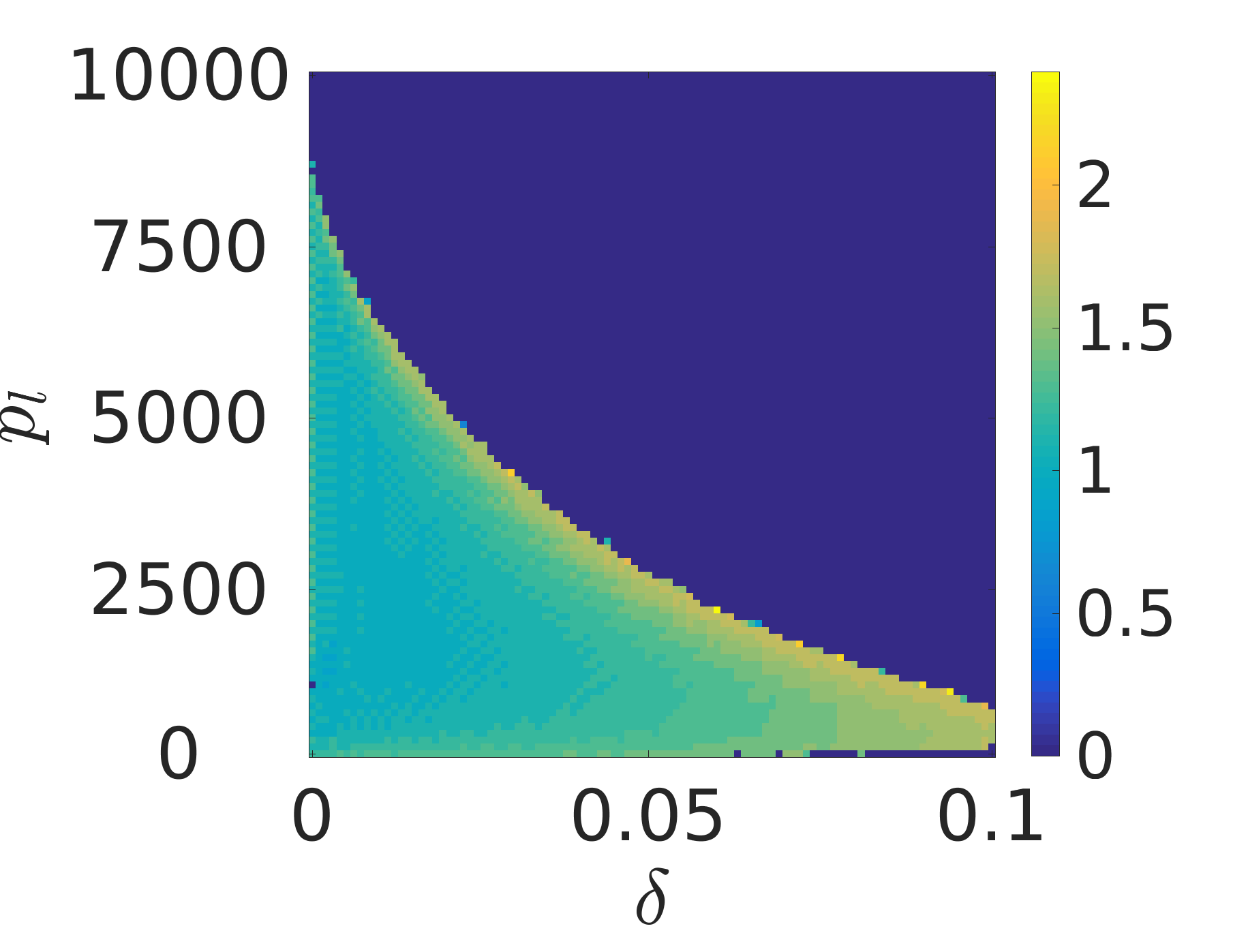}}&\raisebox{-.5\height}{\includegraphics[width=.30\textwidth]{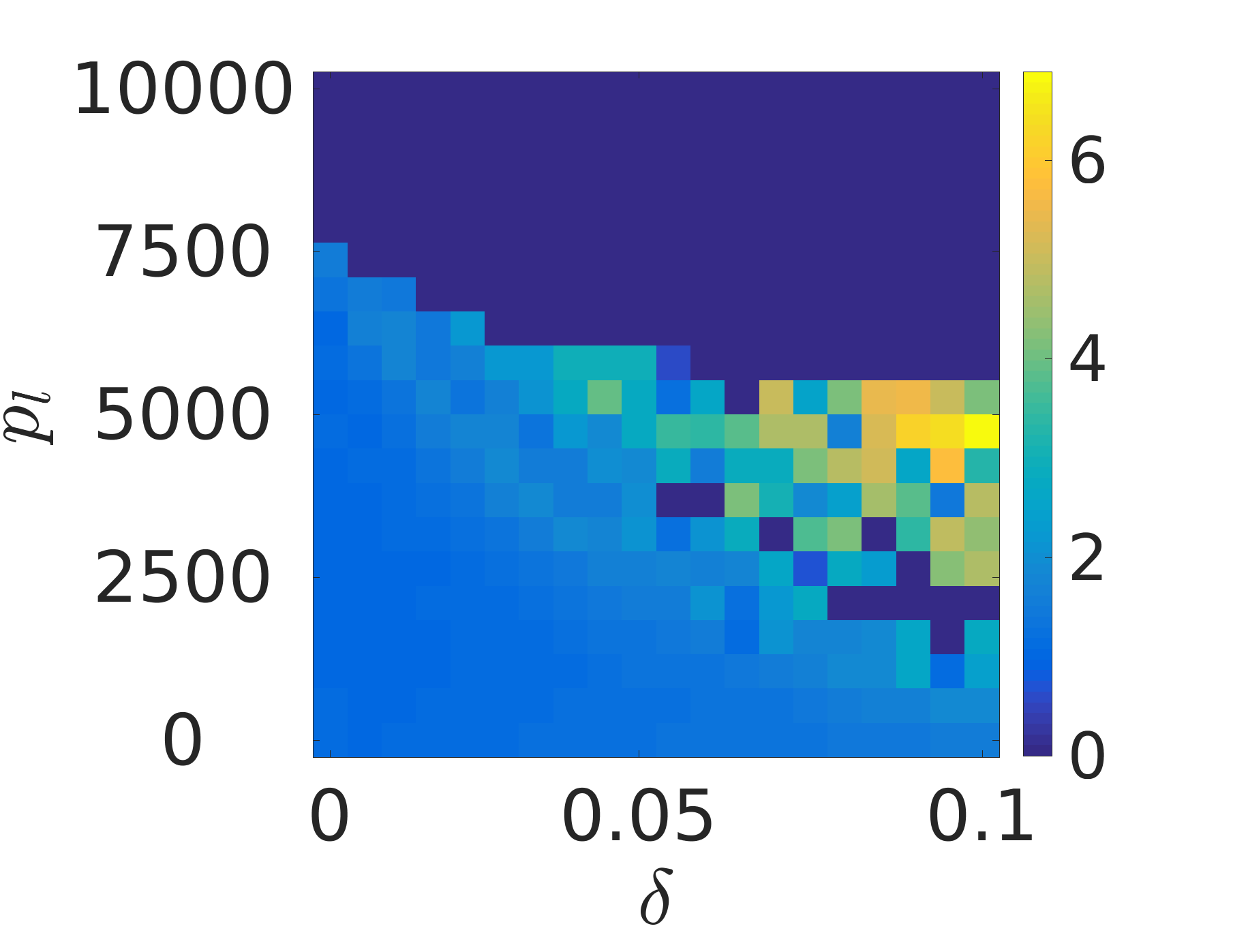}}
\vspace{0mm}\\

$n=100$ &\raisebox{-.5\height}{\includegraphics[width=.30\textwidth]{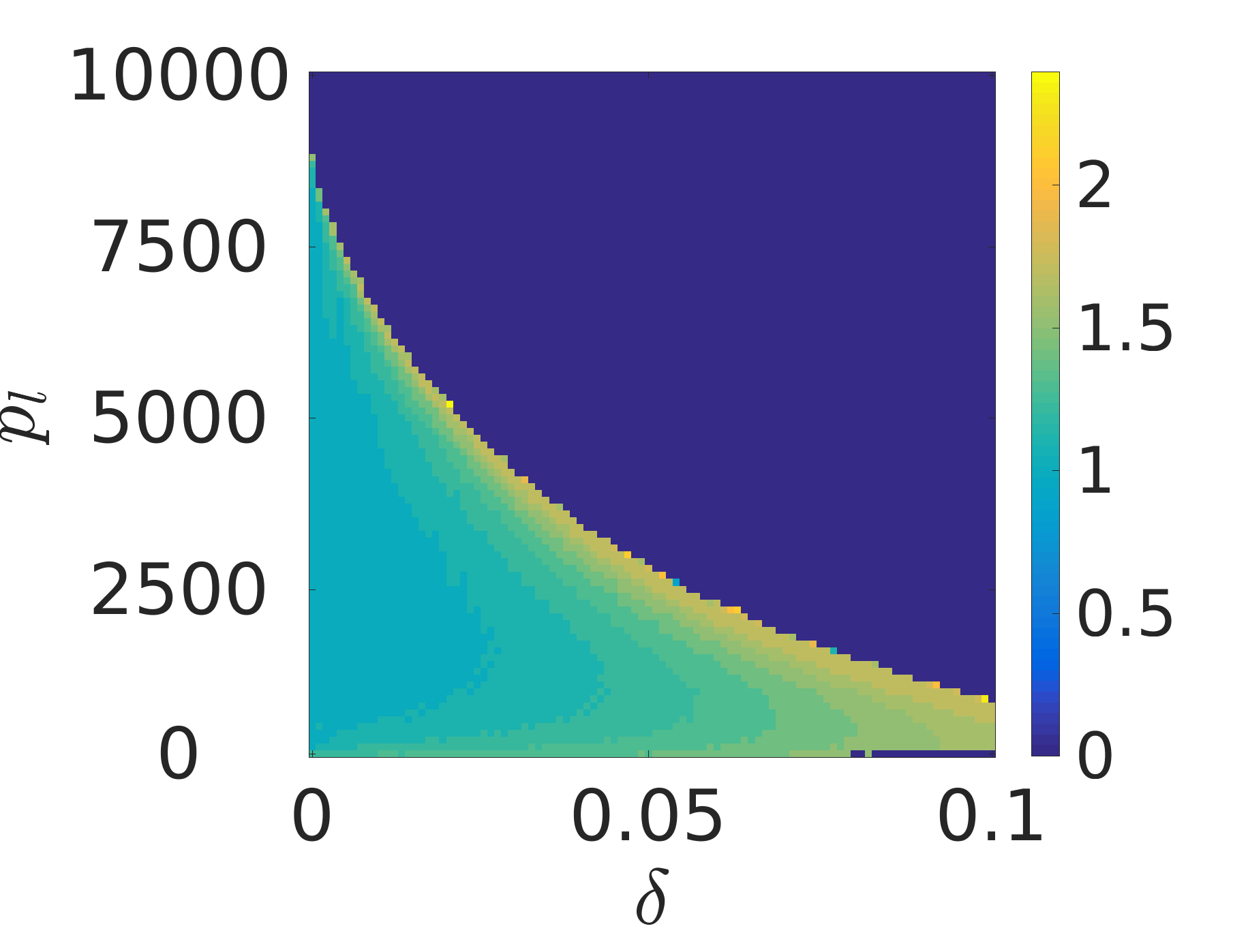}}&\raisebox{-.5\height}{\includegraphics[width=.30\textwidth]{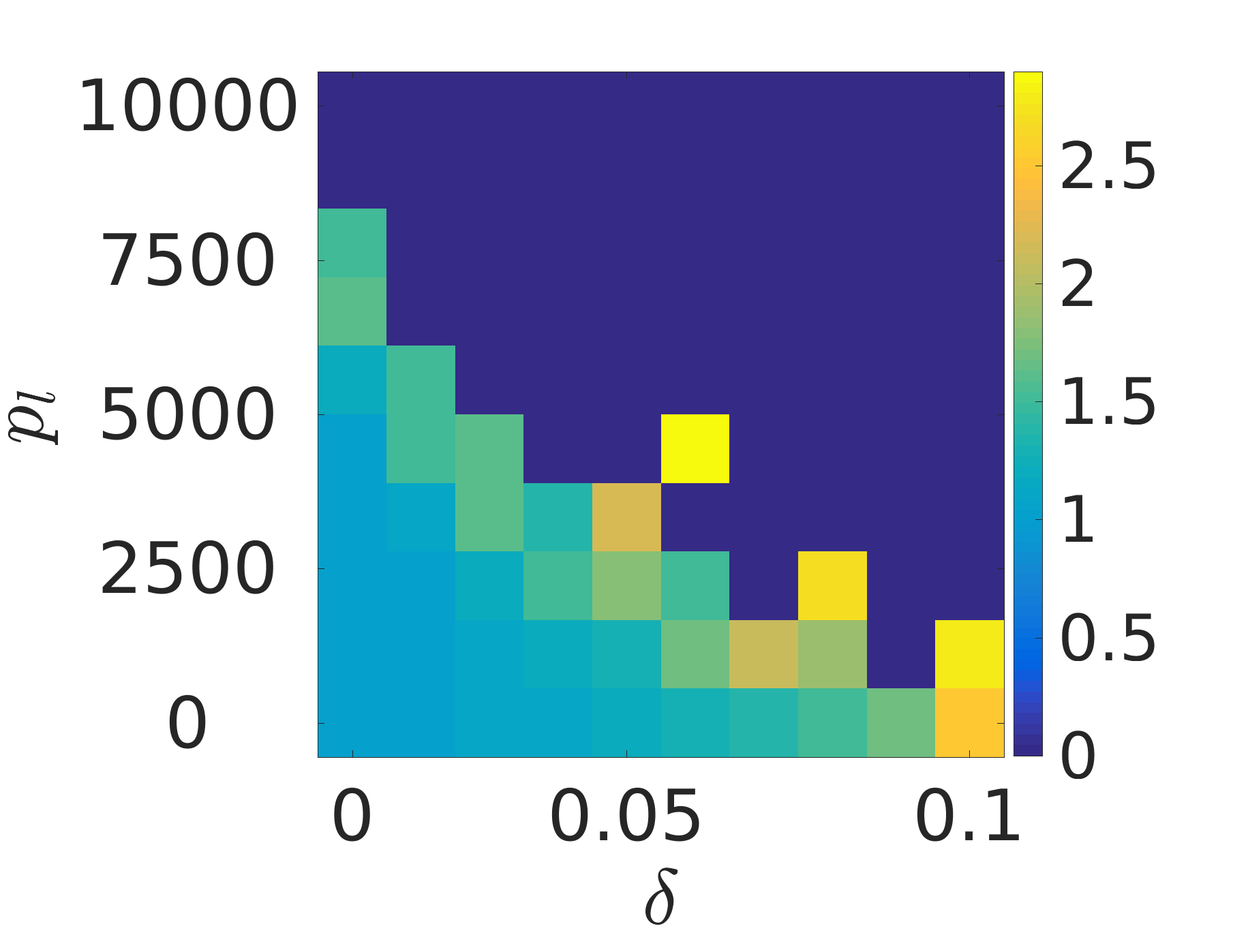}}
\end{tabular}
\caption{Frequency of the nodal oscillation after $t=40$ time units for different lattice sizes and dimensions over $\delta$ and $p_l$.}
\label{bifurcation_frequency_plots}
\end{center}
\end{figure}

\begin{figure}\setlength{\tabcolsep}{5pt}
\begin{center}
\begin{tabular}{r c c} 
 & 1D & 2D\\
Amplitude & \raisebox{-.5\height}{\includegraphics[width=.30\textwidth]{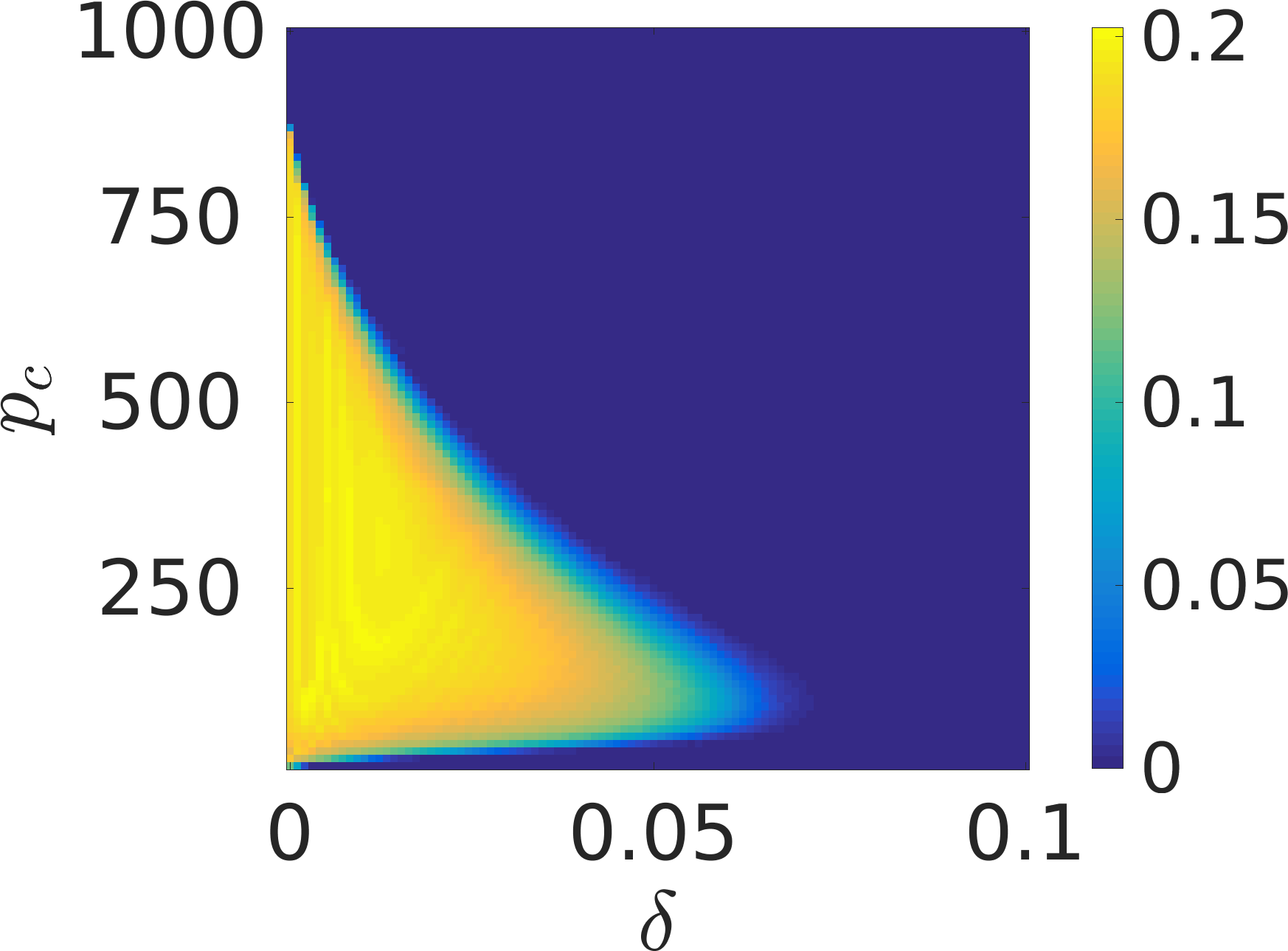}}&\raisebox{-.5\height}{\includegraphics[width=.30\textwidth]{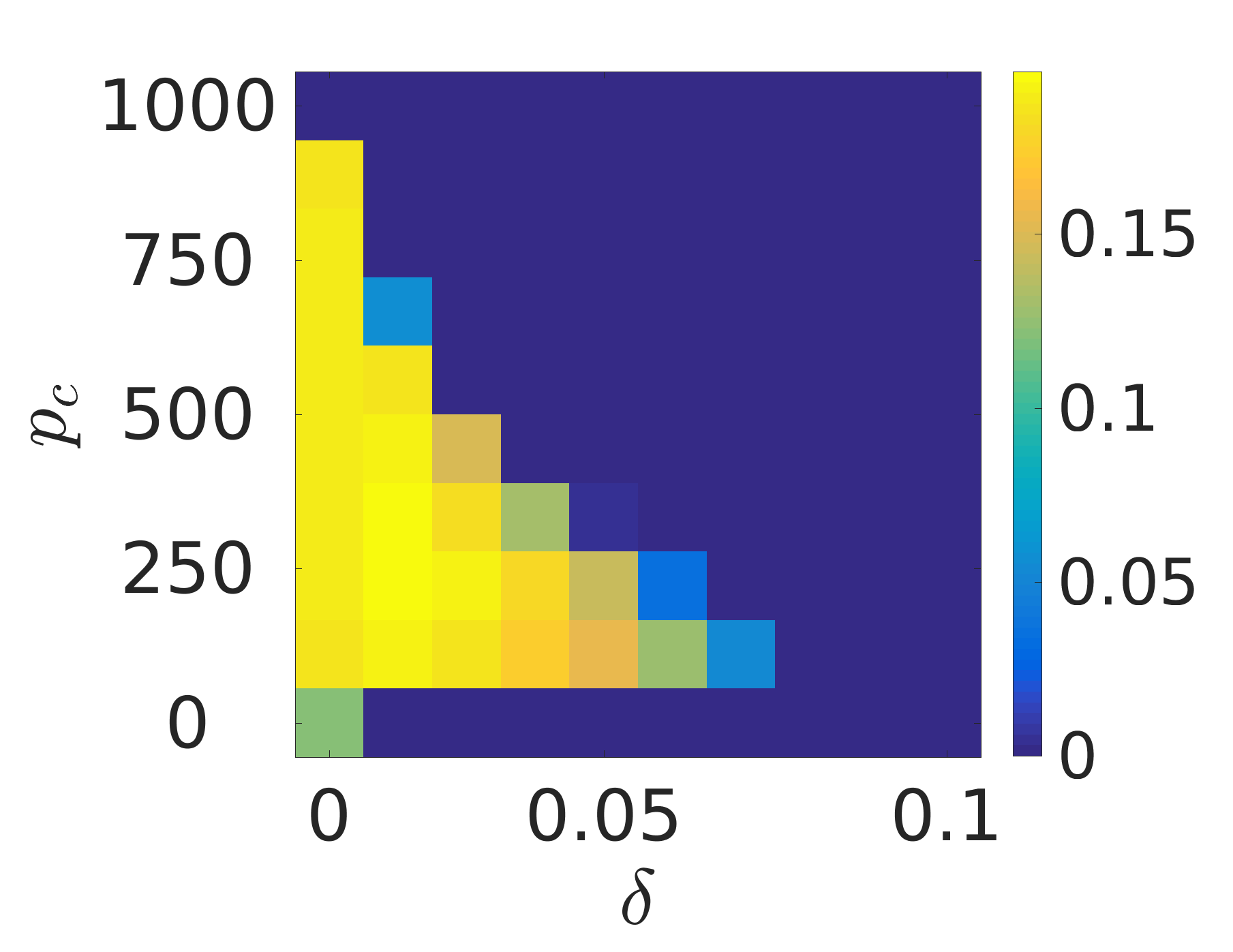}}
\vspace{0mm}\\

Frequency & \raisebox{-.5\height}{\includegraphics[width=.30\textwidth]{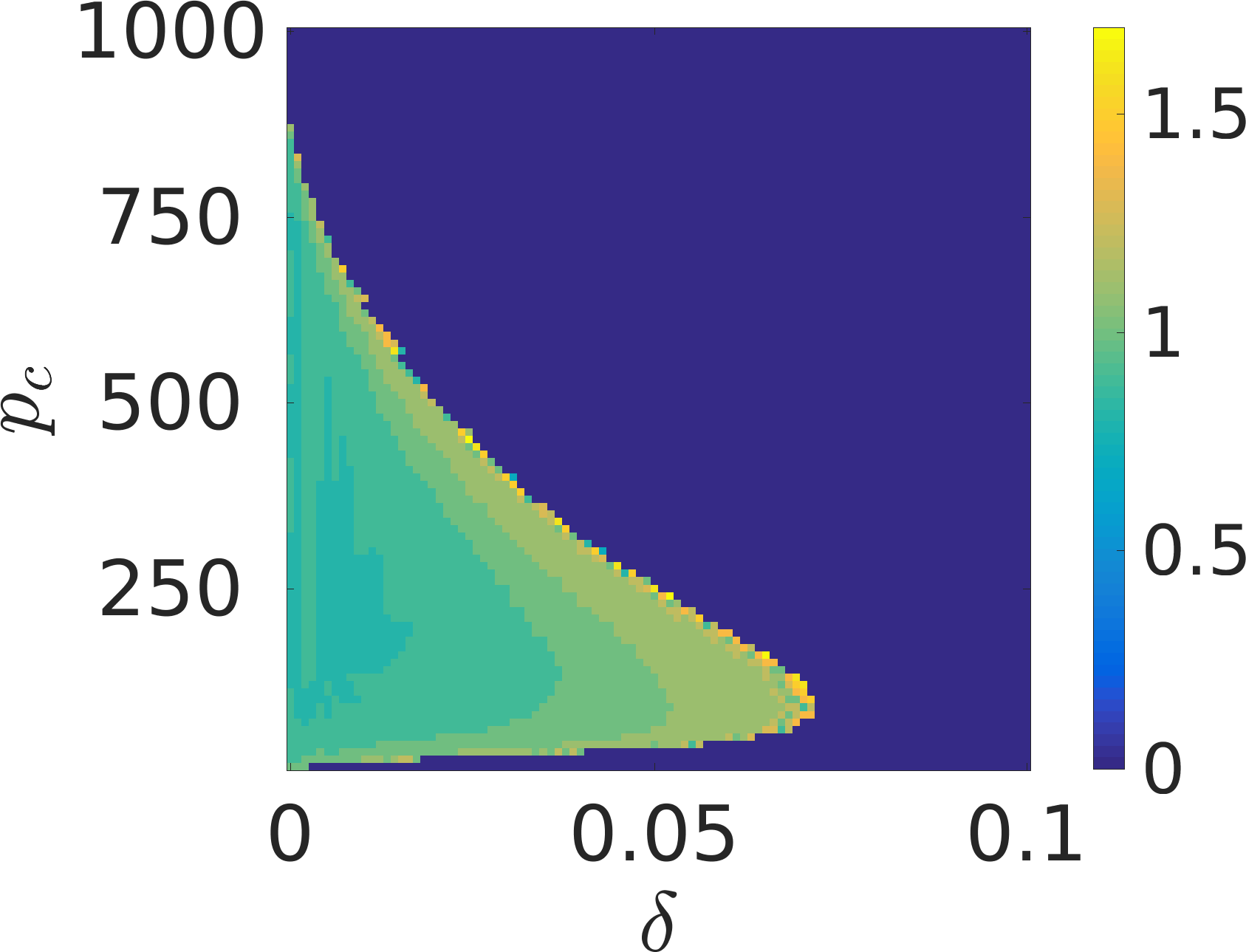}}&\raisebox{-.5\height}{\includegraphics[width=.30\textwidth]{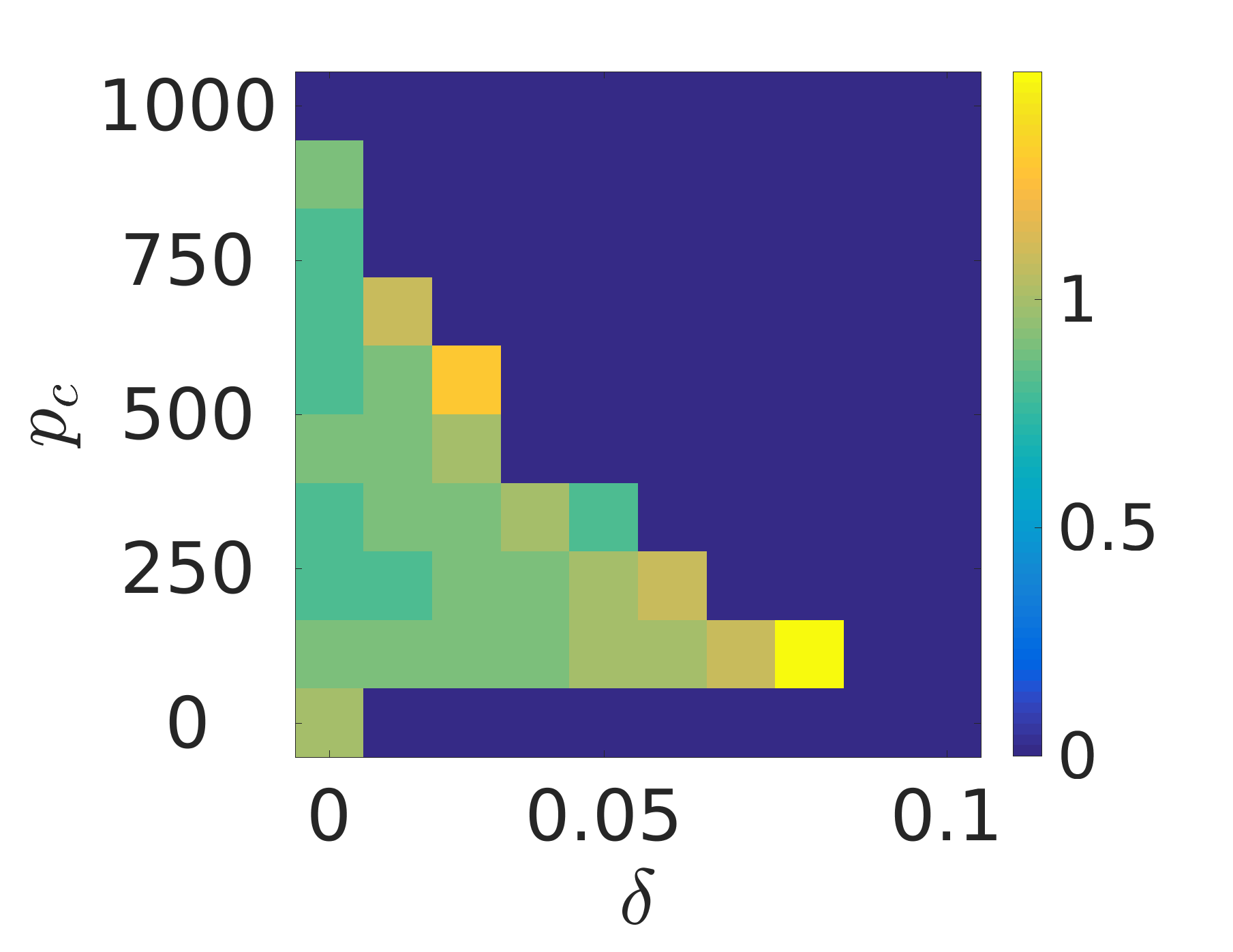}}
\vspace{0mm}
\end{tabular}
\caption{Bifurcation diagrams for the PDE in 1-D and 2-D over $\delta$ and $p_c$. Plotted in the first row is the amplitude of the largest nodal oscillation, and the second the frequency of the maximal nodal oscillation after $t=40$ time units.}
\label{bifurcation_PDE_plots}
\end{center}
\end{figure}

\begin{figure}\setlength{\tabcolsep}{5pt}
\begin{center}
\begin{tabular}{r c c} 
 & 1D & 2D\\
$n=50$ & \raisebox{-.5\height}{\includegraphics[width=.30\textwidth]{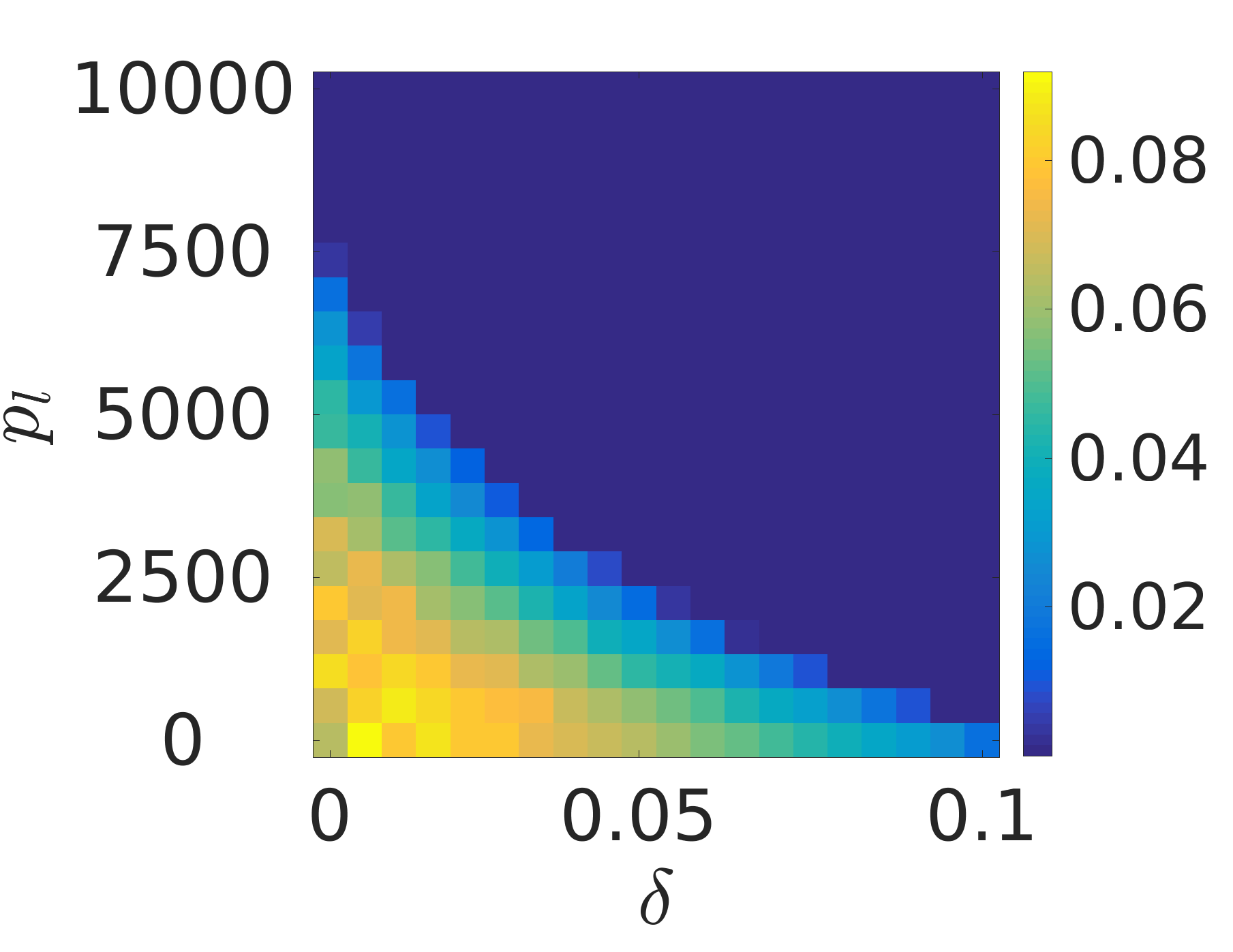}}&\raisebox{-.5\height}{\includegraphics[width=.30\textwidth]{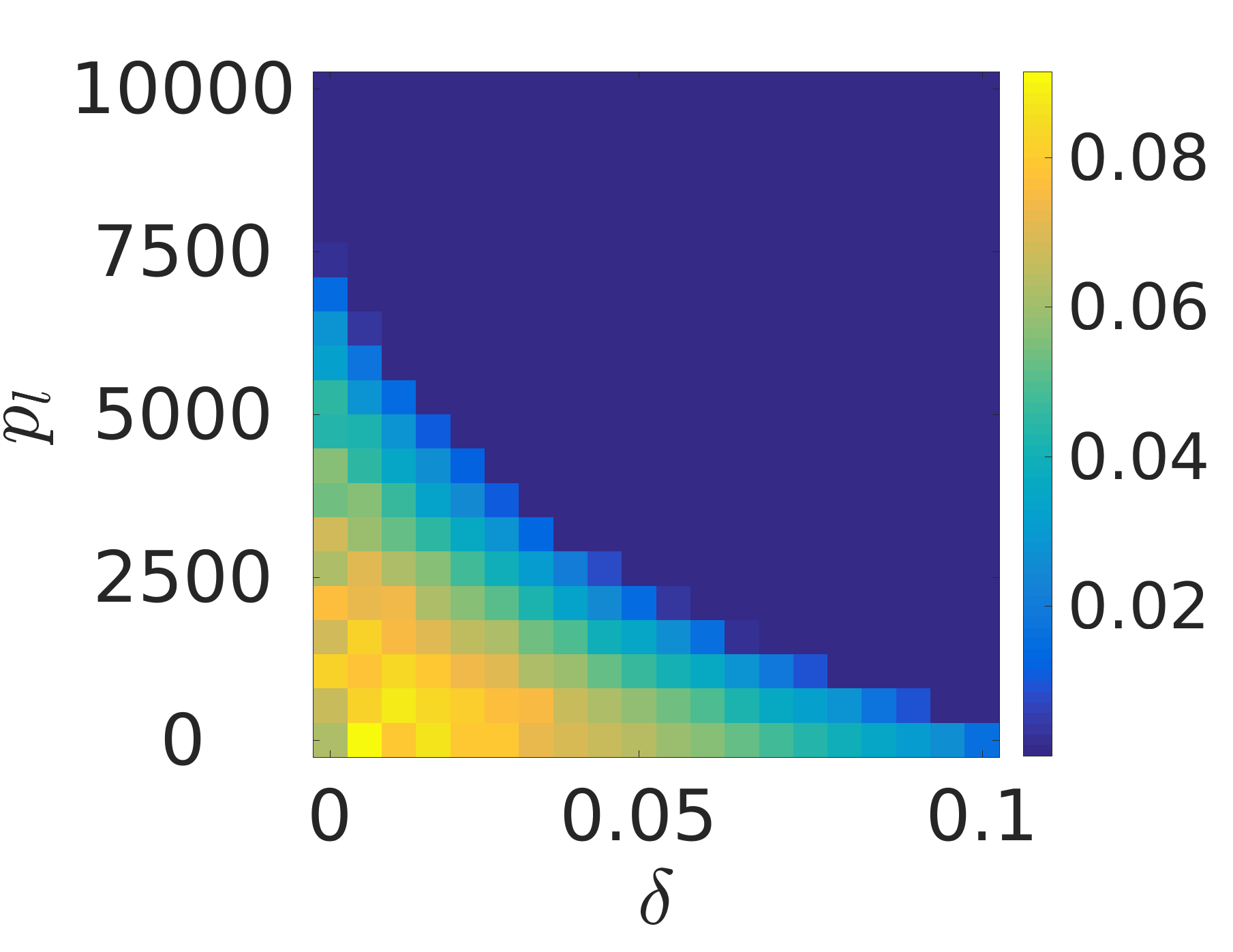}}
\vspace{0mm}\\

$n=100$ &\raisebox{-.5\height}{\includegraphics[width=.30\textwidth]{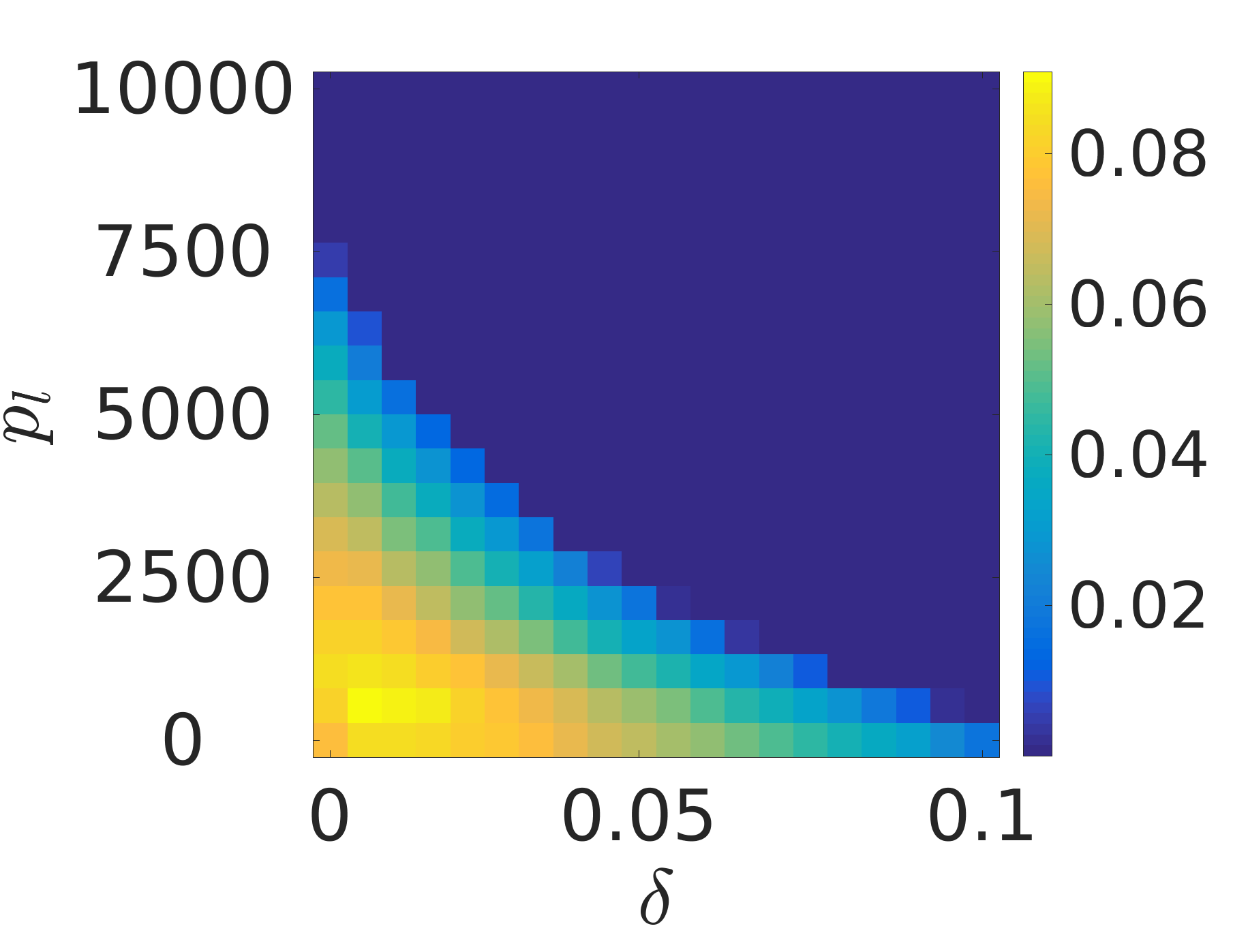}}&\raisebox{-.5\height}{\includegraphics[width=.30\textwidth]{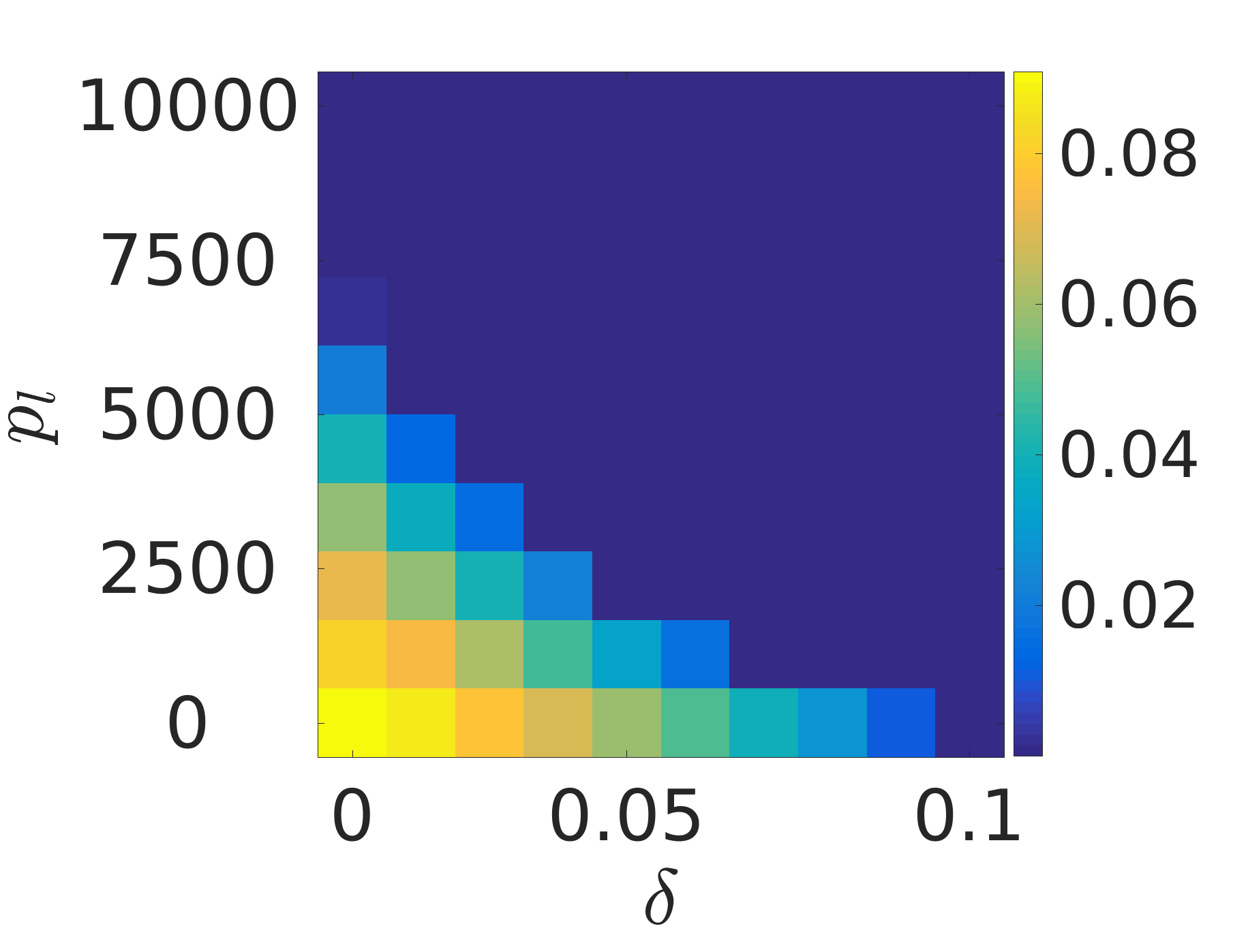}}
\vspace{0mm}\\

PDE & \raisebox{-.5\height}{\includegraphics[width=.30\textwidth]{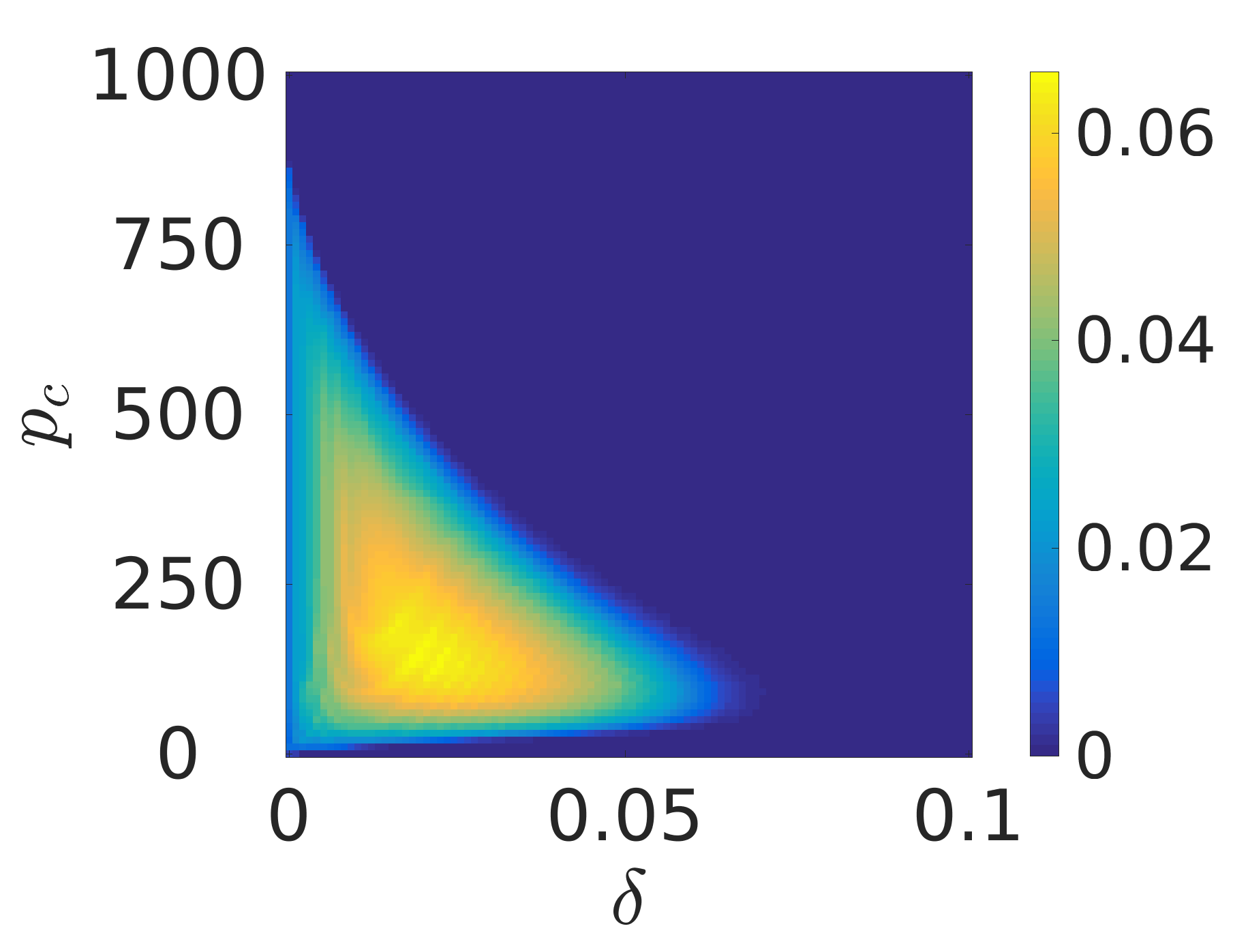}}&\raisebox{-.5\height}{\includegraphics[width=.30\textwidth]{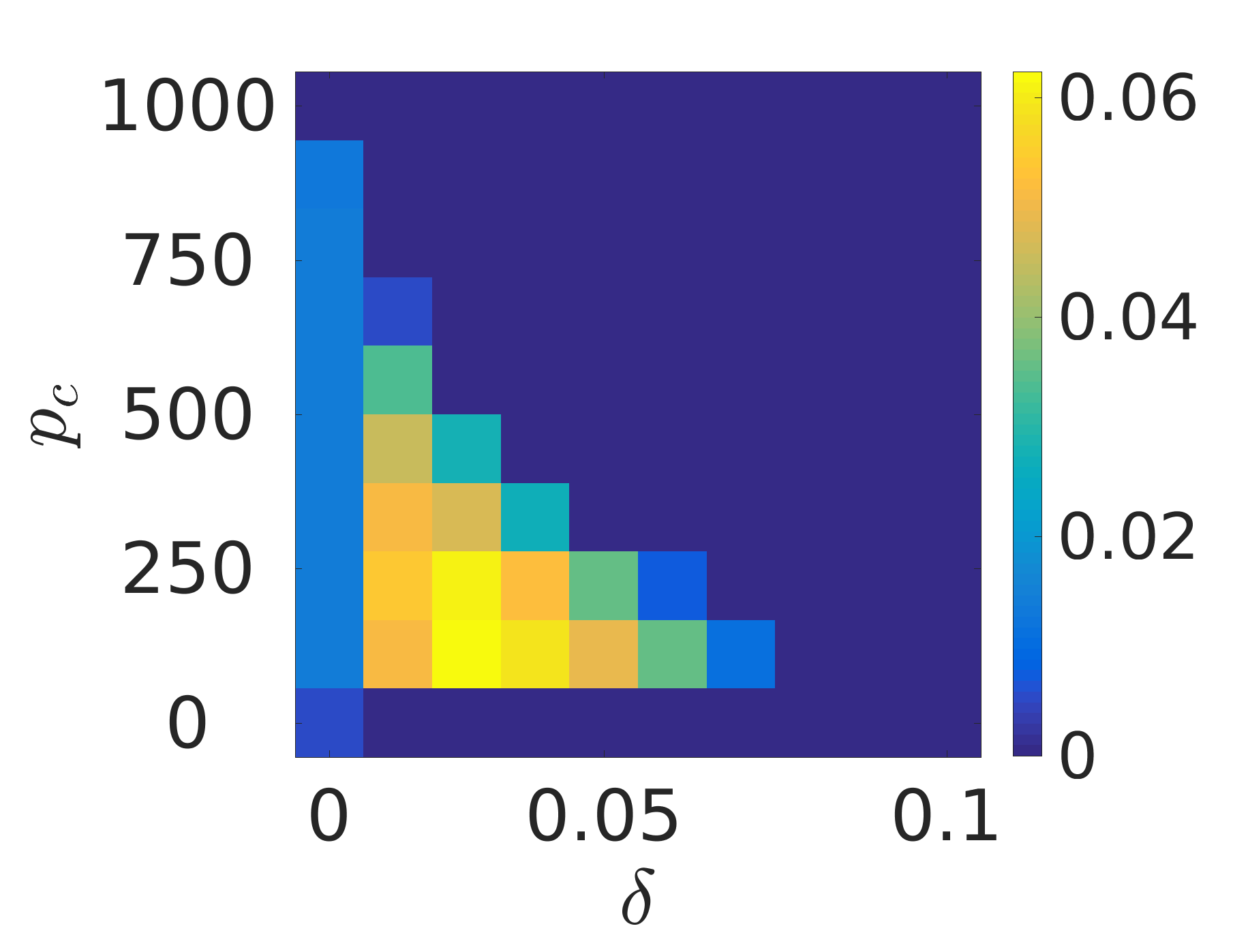}}
\end{tabular}
\caption{Amplitude of oscillations in the spatial mean cell density after $t=40$ time units for different lattice sizes and dimensions over $\delta$ and $p_l$.}
\label{bifurcation_density_amplitude_plots}
\end{center}
\end{figure}

In Figure \ref{bifurcation_amplitude_plots} we plot the maximal (over all nodes) amplitude of these nodal oscillations in the lattice, and in Figure \ref{bifurcation_frequency_plots} we plot the frequency of nodal oscillations computed using the Fast-Fourier Transform of these truncated time series. We plot corresponding PDE bifurcation diagrams in Figure \ref{bifurcation_PDE_plots}. The larger lattices in Figure \ref{bifurcation_amplitude_plots} and the PDE in Figure \ref{bifurcation_PDE_plots} appear to have an oscillatory region for small $\delta$ or small $p_l$. The 2-D lattices have more complicated regions not present in the 1-D models where vertically asymmetric oscillations are present (compare the left and right columns of Figure \ref{bifurcation_amplitude_plots}, especially for $n\geq 50$). The magnitude of the nodal oscillations away from the pulsing regime (bottom-right region of each plot in the right column of Figure \ref{bifurcation_amplitude_plots}) appears to decrease as a function of $n$, whereas the pulsing oscillations maintain a comparable amplitude for $n \geq 25$ (not shown). Regions with large (nodal) amplitudes in the 1-D model correspond to pulsing oscillations in the 2-D model. The mixed oscillation (vertically symmetric and asymmetric) behaviours shown in Figure \ref{interp} exist only when the 1-D model has smaller amplitude oscillations, and hence along the boundaries of the oscillatory regions in the plots along the left of Figure \ref{bifurcation_amplitude_plots}. Note too that for $n=10$ both the 1-D and 2-D oscillation regions are much more disconnected than the corresponding regions for larger lattices (this disconnected bifurcation structure also existed for smaller lattices). For $n \geq 25$ the region where oscillatory behaviour is observed appears to become simply connected in the 1-D models, but it is unclear that this happens in the 2-D case for large $n$.

The frequency of oscillations away from the small $\delta$ or $p_l$ regime increases with $\delta$ (see Figure \ref{bifurcation_frequency_plots}). The oscillation frequency is substantially larger for oscillations present in the 2-D lattices that are not present in their 1-D counterparts; compare the colour ranges in Figure \ref{bifurcation_frequency_plots}. The PDE behaviours shown in Figure \ref{bifurcation_PDE_plots} have no appreciable difference between the 1-D and 2-D model behaviours. This suggests that the symmetry-breaking oscillations of the smaller lattices may disappear in the limit of $n \to \infty$. 

We also consider the amplitudes of oscillations in the spatial mean cell density over the entire lattice in Figure \ref{bifurcation_density_amplitude_plots}, in comparison to the maximal nodal oscillations plotted in Figure \ref{bifurcation_amplitude_plots}. We see no qualitative difference between the solutions of the 1-D and 2-D models, suggesting that for large $n$, the asymmetric oscillations become less influential on the spatial mean cell density. This is also consistent with the maximal amplitudes of the nodal oscillations decaying with $n$.

\section{Discussion}\label{Conclusions}
In this paper we have presented and analyzed a pair of new models based on those proposed in \cite{krause_lattice_2017} that describe an active porous medium with interactions between cell growth and fluid flow. We have demonstrated several notable differences between the lattice models of various sizes, as well as between the lattice models and the spatially continuous model. 

In Section \ref{Model} we presented the lattice and continuum models, and demonstrated a reduction to a 1-D model available for vertically symmetric cell density distributions. In Section \ref{Numerical_Overview}, we solved the (1-D and 2-D) systems numerically to demonstrate the existence of steady state solutions and oscillations, either preserving or breaking vertical symmetry. Vertically symmetric solutions were robust to changing the boundary conditions and dimension of the models, as well as the size of the lattice above a minimum threshold. Vertically asymmetric solutions appeared in smaller lattices, and these were more sensitive to the nature of the boundary conditions (and did not exist in the 1-D models). In Section \ref{Asymptotics} we derived large $\delta$ asymptotic solutions for cell density in both the continuum and lattice models. In Section \ref{Bifurcations} we exhibited Hopf bifurcations from steady states to vertically asymmetric oscillations. The vertically symmetric oscillations, on the other hand, were not created from local bifurcations from steady states that we found numerically. The asymptotic solutions from Section \ref{Asymptotics} allowed us to confine parameter sweeps in Section \ref{Bifurcations} to ranges of the model parameters where variation in spatial and temporal structure may be found, and we characterized model behaviours across these parameter ranges. We found that there is a region of parameter space for small $\delta$ or small pressure thresholds where oscillations are observed. This region appears simply connected for large lattices, and for the continuum model, but for smaller lattices the parameter space consists of disconnected regions corresponding to oscillatory behaviour. Smaller lattices also exhibit both vertically symmetric and asymmetric oscillations. 

Results in Sections \ref{Numerical_Overview}-\ref{Bifurcations} demonstrate an important role played by vertically symmetric steady states and oscillatory solutions to these models. Our numerical explorations show that in various limits, notably the large lattice limit of $n \to \infty$, these two solutions are the main stable long-time behaviours of the models, at least within the parameter regimes we explored. While asymmetric oscillations do exist in large lattices, they become almost indistinguishable from the vertically symmetric oscillations. For large lattices, quantities such as mean cell density oscillation amplitudes, shown in Figure \ref{bifurcation_density_amplitude_plots}, are completely indistinguishable. For smaller lattices we have shown the existence of Hopf and pitchfork bifurcations that break the symmetry of the vertically symmetric steady state solutions and give rise to asymmetric steady states, and asymmetric oscillating solutions. The phase space and the parameter space for moderate lattice sizes (e.g.~$n \sim O(10)$) is difficult to analyze as it admits all of the various solutions, alongside what are complicated combinations of them (as demonstrated in Figure \ref{interp}).

We can interpret our results in terms of solution symmetry and nonlocal reaction-diffusion mechanisms. Due to results from \cite{casten_instability_1978, matano_convergence_1978}, and others (see \cite{smith_monotone_2008} for a more thorough review), it is known that \emph{local} scalar reaction-diffusion equations with Neumann data on a bounded domain have relatively simple asymptotic behaviour - namely, all bounded solutions converge to spatially uniform equilibria. Equation \eqref{pde_cell_equation_sym} is a simplification of our model that still admits spatially-structured and oscillatory solutions despite being a scalar reaction-diffusion equation, and hence this nonlocality is a necessary component for both the spatially structured steady states and oscillatory behaviours. Simpler models of this kind of nonlocal reaction-diffusion have been considered in \cite{billingham_dynamics_2004,gourley_spatio-temporal_2001, hamel_nonlocal_2014, levchenko_asymptotics_2016}, and the references therein, where oscillations and non-uniform equilibria are found and analyzed. We note that these models are qualitatively similar to Equation \eqref{pde_cell_equation_sym} with different nonlinearities, both in the reaction kinetics and the integral kernel. 

We conjecture that it is this inherently nonlocal structure, driven by the long-range quasi-static pressure forcing, that gives rise to the oscillatory behaviours we observe. For smaller lattices, the vertically asymmetric steady states and oscillations are related to the discrete structure of these finite-size lattices, as these are not present in the continuum analogue. We conjecture that this discrete structure leads to symmetry-breaking effects due, for example, to the anisotropic nature of discrete diffusion on a lattice \cite{cardy_finite-size_2012}. We conclude that the vertically asymmetric oscillations are then an interplay of this symmetry breaking with the nonlocal mechanism described before. 

Our observations about vertically asymmetric oscillations are consistent with the oscillations observed in \cite{krause_lattice_2017}, where a similar model captured cell death due to high values of shear stress. While the solution behaviours in \cite{krause_lattice_2017} were generally more complicated (e.g.~lacked the spatial symmetries present here), the models there satisfied the same fluid equations and hence also had a nonlocal coupling between cell growth and fluid pressure as demonstrated explicitly in Sections \ref{Vertically_Symmetric} and \ref{Vertically_Symmetric_PDE}. Oscillations in cell density were also observed in the lattice model of \cite{krause_lattice_2017}, with the amplitude of oscillations decreasing with increasing $n$, but no oscillations were not observed in the spatially continuous analogue. By comparison, the spatially continuous pressure model in Section \ref{Model} did give rise to oscillations. The difference between these models may be due to a localization of the nonlocal pressure in the fluid shear stress, which is a function of gradients in pressure. This is in contrast to the models presented in Section \ref{Model}, as the pressure itself is forcing the cell growth problem, and hence oscillations are observed in both lattice and continuum models. Overall, these results suggest that oscillations observed in \cite{krause_lattice_2017} are analogous to the vertically asymmetric oscillations in the pressure-forced model, as these are also a finite-size effect observed in smaller lattices. 

In the context of bioactive porous media, these models demonstrate several interesting effects due to the finite pore network captured in our lattice model. In the limit of very large lattices, we suspect that spatially continuous models are good approximations to these kinds of pore networks, at least for the simple topology considered here \cite{verwer_convergence_1984}. For smaller networks, we have demonstrated nontrivial effects due to the discrete symmetries inherent in the finite-scale lattice that cannot, as far as we are aware, be captured in continuous analogues \cite{cardy_finite-size_2012}. We suspect that more realistic (and hence more complicated) pore networks may exhibit other behaviours due to the presence or lack of symmetries in the topology of the pore network. The tissue engineering literature, for instance, emphasizes macroscopic and microscopic properties of porous scaffolds, such as permeability and pore size respectively, but does not seem to consider as much the importance of the discrete structure of pore networks \cite{hutmacher_scaffolds_2000, jones_quantifying_2008, wu_biomimetic_2014}. Further explorations of network models of bioactive porous media, such as those studied here, will be both quantitatively useful for applications, and provide a number of interesting mathematical questions to pursue.

\subsection*{Acknowledgments}
D.B. was partially funded by Engineering \& Physical Sciences Research Council (EPSRC) Fellowship ref. EP/M002896/1. S.L.W. gratefully acknowledges funding in the form of a Royal Society Leverhulme Trust Senior Research Fellowship. In compliance with EPSRC's open access initiative, the data in this paper is available from http://dx.doi.org/xxx/xxx.

\bibliographystyle{ws-ijbc}
\bibliography{PressureBib}

\begin{thebibliography}{61}
\newcommand{\enquote}[1]{``#1''}
\providecommand{\natexlab}[1]{#1}
\providecommand{\url}[1]{\texttt{#1}}
\providecommand{\urlprefix}{URL }
\expandafter\ifx\csname urlstyle\endcsname\relax
  \providecommand{\doi}[1]{doi:\discretionary{}{}{}#1}\else
  \providecommand{\doi}{doi:\discretionary{}{}{}\begingroup
  \urlstyle{rm}\Url}\fi

\bibitem[{Anderson \& Chaplain(1998)}]{anderson1998continuous}
Anderson, A. R.~A. \& Chaplain, M. A.~J. [1998] \enquote{Continuous and
  discrete mathematical models of tumor-induced angiogenesis,} \emph{Bulletin
  of Mathematical Biology} \textbf{60},  857--899.

\bibitem[{Barbotteau \emph{et~al.}(2003)Barbotteau, Irigaray \&
  Mathiot}]{barbotteau_modelling_2003}
Barbotteau, Y., Irigaray, J.~L. \& Mathiot, J.~F. [2003] \enquote{Modelling by
  percolation theory of the behaviour of natural coral used as bone
  substitute,} \emph{Physics in Medicine and Biology} \textbf{48},  3611--3623.

\bibitem[{Bear(1972)}]{bear_dynamics_1972}
Bear, J. [1972] \emph{Dynamics of {Fluids} in {Porous} {Media}} (American
  Elsevier, New York).

\bibitem[{Bendix \emph{et~al.}(2009)Bendix, Fleischmann, Kottos \&
  Shapiro}]{bendix2009exponentially}
Bendix, O., Fleischmann, R., Kottos, T. \& Shapiro, B. [2009]
  \enquote{Exponentially fragile $\mathcal{PT}$ symmetry in lattices with
  localized eigenmodes,} \emph{Physical Review Letters} \textbf{103},  030402.

\bibitem[{Billingham(2004)}]{billingham_dynamics_2004}
Billingham, J. [2004] \enquote{Dynamics of a strongly nonlocal
  reaction-diffusion population model,} \emph{Nonlinearity} \textbf{17},
  313--346.

\bibitem[{Blunt(2001)}]{blunt_flow_2001}
Blunt, M.~J. [2001] \enquote{Flow in porous media — pore-network models and
  multiphase flow,} \emph{Current Opinion in Colloid \& Interface Science}
  \textbf{6},  197--207.

\bibitem[{Brazhnyi \& Malomed(2011)}]{brazhnyi2011spontaneous}
Brazhnyi, V.~A. \& Malomed, B.~A. [2011] \enquote{Spontaneous symmetry breaking
  in schr{\"o}dinger lattices with two nonlinear sites,} \emph{Physical Review
  A} \textbf{83},  053844.

\bibitem[{Buzano \& Golubitsky(1983)}]{buzano1983bifurcation}
Buzano, E. \& Golubitsky, M. [1983] \enquote{Bifurcation on the hexagonal
  lattice and the planar b{\'e}nard problem,} \emph{Phil. Trans. R. Soc. Lond.
  A} \textbf{308},  617--667.

\bibitem[{Byrne \& Drasdo(2009)}]{byrne2009individual}
Byrne, H. \& Drasdo, D. [2009] \enquote{Individual-based and continuum models
  of growing cell populations: a comparison,} \emph{Journal of Mathematical
  Biology} \textbf{58},  657--687.

\bibitem[{Byrne \& Preziosi(2003)}]{byrne2003modelling}
Byrne, H. \& Preziosi, L. [2003] \enquote{Modelling solid tumour growth using
  the theory of mixtures,} \emph{Mathematical Medicine and Biology}
  \textbf{20},  341--366.

\bibitem[{Cardy(2012)}]{cardy_finite-size_2012}
Cardy, J. [2012] \emph{Finite-{Size} {Scaling}} (Elsevier).

\bibitem[{Casten \& Holland(1978)}]{casten_instability_1978}
Casten, R.~G. \& Holland, C.~J. [1978] \enquote{Instability results for
  reaction diffusion equations with {Neumann} boundary conditions,}
  \emph{Journal of Differential Equations} \textbf{27},  266--273.

\bibitem[{Chow \emph{et~al.}(1996)Chow, Mallet-Paret \&
  Van~Vleck}]{chow1996dynamics}
Chow, S.-N., Mallet-Paret, J. \& Van~Vleck, E.~S. [1996] \enquote{Dynamics of
  lattice differential equations,} \emph{International Journal of Bifurcation
  and Chaos} \textbf{6},  1605--1621.

\bibitem[{Christodoulou(2008)}]{christodoulou_discrete_2008}
Christodoulou, N. [2008] \enquote{Discrete {Hopf} bifurcation for
  {Runge}-{Kutta} methods,} \emph{Applied Mathematics and Computation}
  \textbf{206},  346--356.

\bibitem[{Coletti \emph{et~al.}(2006)Coletti, Macchietto \&
  Elvassore}]{coletti_mathematical_2006}
Coletti, F., Macchietto, S. \& Elvassore, N. [2006] \enquote{Mathematical
  {Modeling} of {Three}-{Dimensional} {Cell} {Cultures} in {Perfusion}
  {Bioreactors},} \emph{Industrial \& Engineering Chemistry Research}
  \textbf{45},  8158--8169.

\bibitem[{Cox \emph{et~al.}(2015)Cox, Thornby, Gibbons, Williams \&
  Mallick}]{cox_3d_2015}
Cox, S.~C., Thornby, J.~A., Gibbons, G.~J., Williams, M.~A. \& Mallick, K.~K.
  [2015] \enquote{3d printing of porous hydroxyapatite scaffolds intended for
  use in bone tissue engineering applications,} \emph{Materials Science and
  Engineering: C} \textbf{47},  237--247.

\bibitem[{Crawford \& Knobloch(1991)}]{crawford1991symmetry}
Crawford, J.~D. \& Knobloch, E. [1991] \enquote{Symmetry and symmetry-breaking
  bifurcations in fluid dynamics,} \emph{Annual Review of Fluid Mechanics}
  \textbf{23},  341--387.

\bibitem[{Duncan \emph{et~al.}(2015)Duncan, Liao, Vejchodsk{\`y}, Erban \&
  Grima}]{duncan2015noise}
Duncan, A., Liao, S., Vejchodsk{\`y}, T., Erban, R. \& Grima, R. [2015]
  \enquote{Noise-induced multistability in chemical systems: Discrete versus
  continuum modeling,} \emph{Physical Review E} \textbf{91},  042111.

\bibitem[{Fraternali \emph{et~al.}(2014)Fraternali, Farina \&
  Carpentieri}]{fraternali2014discrete}
Fraternali, F., Farina, I. \& Carpentieri, G. [2014] \enquote{A
  discrete-to-continuum approach to the curvatures of membrane networks and
  parametric surfaces,} \emph{Mechanics Research Communications} \textbf{56},
  18--25.

\bibitem[{Geris(2013)}]{geris_computational_2013}
Geris, L. (ed.) [2013] \emph{Computational {Modeling} in {Tissue}
  {Engineering}}, Studies in {Mechanobiology}, {Tissue} {Engineering} and
  {Biomaterials}, Vol.~10 (Springer Berlin Heidelberg, Berlin, Heidelberg).

\bibitem[{German \& Madihally(2016)}]{german_applications_2016}
German, C.~L. \& Madihally, S.~V. [2016] \enquote{Applications of
  {Computational} {Modelling} and {Simulation} of {Porous} {Medium} in {Tissue}
  {Engineering},} \emph{Computation} \textbf{4},  7.

\bibitem[{Gillis \& Golubitsky(1997)}]{gillis1997patterns}
Gillis, D. \& Golubitsky, M. [1997] \enquote{Patterns in square arrays of
  coupled cells,} \emph{Journal of Mathematical Analysis and Applications}
  \textbf{208},  487--509.

\bibitem[{Golubitsky \emph{et~al.}(2004)Golubitsky, Nicol \&
  Stewart}]{golubitsky2004some}
Golubitsky, M., Nicol, M. \& Stewart, I. [2004] \enquote{Some curious phenomena
  in coupled cell networks,} \emph{Journal of Nonlinear Science} \textbf{14},
  207--236.

\bibitem[{Gourley \emph{et~al.}(2001)Gourley, Chaplain \&
  Davidson}]{gourley_spatio-temporal_2001}
Gourley, S.~A., Chaplain, M. A.~J. \& Davidson, F.~A. [2001]
  \enquote{Spatio-temporal pattern formation in a nonlocal reaction-diffusion
  equation,} \emph{Dynamical Systems} \textbf{16},  173--192.

\bibitem[{Hamel \& Ryzhik(2014)}]{hamel_nonlocal_2014}
Hamel, F. \& Ryzhik, L. [2014] \enquote{On the nonlocal {Fisher}-{KPP}
  equation: steady states, spreading speed and global bounds,}
  \emph{Nonlinearity} \textbf{27},  2735.

\bibitem[{Hutmacher(2000)}]{hutmacher_scaffolds_2000}
Hutmacher, D. [2000] \enquote{Scaffolds in tissue engineering bone and
  cartilage,} \emph{Biomaterials} \textbf{21},  2529--2543.

\bibitem[{Janssen \& Tjon(1983)}]{janssen1983bifurcations}
Janssen, T. \& Tjon, J. [1983] \enquote{Bifurcations of lattice structures,}
  \emph{Journal of Physics A: Mathematical and General} \textbf{16},  673.

\bibitem[{Jones \emph{et~al.}(2008)Jones, Atwood, Poologasundarampillai, Yue \&
  Lee}]{jones_quantifying_2008}
Jones, J.~R., Atwood, R.~C., Poologasundarampillai, G., Yue, S. \& Lee, P.~D.
  [2008] \enquote{Quantifying the 3d macrostructure of tissue scaffolds,}
  \emph{Journal of Materials Science: Materials in Medicine} \textbf{20},
  463--471.

\bibitem[{Kamei(2009)}]{kamei2009existence}
Kamei, H. [2009] \enquote{The existence and classification of
  synchrony-breaking bifurcations in regular homogeneous networks using lattice
  structures,} \emph{International Journal of Bifurcation and Chaos}
  \textbf{19},  3707--3732.

\bibitem[{Kevrekidis \emph{et~al.}(2005)Kevrekidis, Chen, Malomed,
  Frantzeskakis \& Weinstein}]{kevrekidis2005spontaneous}
Kevrekidis, P., Chen, Z., Malomed, B., Frantzeskakis, D. \& Weinstein, M.
  [2005] \enquote{Spontaneous symmetry breaking in photonic lattices: Theory
  and experiment,} \emph{Physics Letters A} \textbf{340},  275--280.

\bibitem[{Krause \emph{et~al.}(2017)Krause, Beliaev, Van~Gorder \&
  Waters}]{krause_lattice_2017}
Krause, A.~L., Beliaev, D., Van~Gorder, R.~A. \& Waters, S.~L. [2017]
  \enquote{Lattice and continuum modelling of a bioactive porous tissue
  scaffold,} \emph{arXiv preprint arXiv:1702.07711} .

\bibitem[{Levchenko \emph{et~al.}(2016)Levchenko, Shapovalov \&
  Trifonov}]{levchenko_asymptotics_2016}
Levchenko, E.~A., Shapovalov, A.~V. \& Trifonov, A.~Y. [2016]
  \enquote{Asymptotics semiclassically concentrated on curves for the nonlocal
  {Fisher}-{Kolmogorov}-{Petrovskii}-{Piskunov} equation,} \emph{Journal of
  Physics A: Mathematical and Theoretical} \textbf{49},  305203.

\bibitem[{Loh \& Choong(2013)}]{loh_three-dimensional_2013}
Loh, Q.~L. \& Choong, C. [2013] \enquote{Three-{Dimensional} {Scaffolds} for
  {Tissue} {Engineering} {Applications}: {Role} of {Porosity} and {Pore}
  {Size},} \emph{Tissue Engineering Part B: Reviews} \textbf{19},  485--502.

\bibitem[{Ma{\~n}as \& Mackey(2004)}]{manas_morphological_2004}
Ma{\~n}as, P. \& Mackey, B.~M. [2004] \enquote{Morphological and physiological
  changes induced by high hydrostatic pressure in exponential-and
  stationary-phase cells of escherichia coli: relationship with cell death,}
  \emph{Applied and Environmental Microbiology} \textbf{70},  1545--1554.

\bibitem[{Matano \emph{et~al.}(1978)}]{matano_convergence_1978}
Matano, H. \emph{et~al.} [1978] \enquote{Convergence of solutions of
  one-dimensional semilinear parabolic equations,} \emph{Journal of Mathematics
  of Kyoto University} \textbf{18},  221--227.

\bibitem[{Matsiaka \emph{et~al.}(2018)Matsiaka, Penington, Baker \&
  Simpson}]{matsiaka2018discrete}
Matsiaka, O.~M., Penington, C.~J., Baker, R.~E. \& Simpson, M.~J. [2018]
  \enquote{Discrete and continuum approximations for collective cell migration
  in a scratch assay with cell size dynamics,} \emph{Bulletin of Mathematical
  Biology} \textbf{80},  738--757.

\bibitem[{McDougall(2002)}]{mcdougall_mathematical_2002}
McDougall, S. [2002] \enquote{Mathematical {Modelling} of {Flow} {Through}
  {Vascular} {Networks}: {Implications} for {Tumour}-induced {Angiogenesis} and
  {Chemotherapy} {Strategies},} \emph{Bulletin of Mathematical Biology}
  \textbf{64},  673--702.

\bibitem[{McDougall \& Sorbie(1997)}]{mcdougall1997application}
McDougall, S. \& Sorbie, K. [1997] \enquote{The application of network
  modelling techniques to multiphase flow in porous media,} \emph{Petroleum
  Geoscience} \textbf{3},  161--169.

\bibitem[{Murisic \emph{et~al.}(2015)Murisic, Hakim, Kevrekidis, Shvartsman \&
  Audoly}]{murisic2015discrete}
Murisic, N., Hakim, V., Kevrekidis, I.~G., Shvartsman, S.~Y. \& Audoly, B.
  [2015] \enquote{From discrete to continuum models of three-dimensional
  deformations in epithelial sheets,} \emph{Biophysical Journal} \textbf{109},
  154--163.

\bibitem[{Nessler \emph{et~al.}(2016)Nessler, Henstock, El~Haj, Waters,
  Whiteley \& Osborne}]{nessler_influence_2016}
Nessler, K. H.~L., Henstock, J.~R., El~Haj, A.~J., Waters, S.~L., Whiteley,
  J.~P. \& Osborne, J.~M. [2016] \enquote{The influence of hydrostatic pressure
  on tissue engineered bone development,} \emph{Journal of Theoretical Biology}
  \textbf{394},  149--159.

\bibitem[{Newman(2010)}]{newman_networks:_2010}
Newman, M. E.~J. [2010] \emph{Networks: an introduction} (Oxford University
  Press, Oxford; New York).

\bibitem[{O'Dea \emph{et~al.}(2012)O'Dea, Byrne \&
  Waters}]{odea_continuum_2012}
O'Dea, R.~D., Byrne, H.~M. \& Waters, S.~L. [2012] \enquote{Continuum
  {Modelling} of {In} {Vitro} {Tissue} {Engineering}: {A} {Review},}
  \emph{Computational {Modeling} in {Tissue} {Engineering}}, ed. Geris, L., no.
  10  Studies in {Mechanobiology}, {Tissue} {Engineering} and {Biomaterials}
  (Springer Berlin Heidelberg), pp. 229--266.

\bibitem[{Oh \emph{et~al.}(2012)Oh, Lee, Ahn \& Furlani}]{oh2012design}
Oh, K.~W., Lee, K., Ahn, B. \& Furlani, E.~P. [2012] \enquote{Design of
  pressure-driven microfluidic networks using electric circuit analogy,}
  \emph{Lab on a Chip} \textbf{12},  515--545.

\bibitem[{Pikovsky \& Grassberger(1991)}]{pikovsky1991symmetry}
Pikovsky, A.~S. \& Grassberger, P. [1991] \enquote{Symmetry breaking
  bifurcation for coupled chaotic attractors,} \emph{Journal of Physics A:
  Mathematical and General} \textbf{24},  4587.

\bibitem[{Rothos \emph{et~al.}(2002)Rothos, Antonopoulos \&
  Drossos}]{rothos2002chaos}
Rothos, V.~M., Antonopoulos, C. \& Drossos, L. [2002] \enquote{Chaos in a
  near-integrable hamiltonian lattice,} \emph{International Journal of
  Bifurcation and Chaos} \textbf{12},  1743--1754.

\bibitem[{Sattinger(1980)}]{sattinger1980bifurcation}
Sattinger, D. [1980] \enquote{Bifurcation and symmetry breaking in applied
  mathematics,} \emph{Bulletin of the American Mathematical Society}
  \textbf{3},  779--819.

\bibitem[{Scianna \emph{et~al.}(2013)Scianna, Bell \&
  Preziosi}]{scianna_review_2013}
Scianna, M., Bell, C. \& Preziosi, L. [2013] \enquote{A review of mathematical
  models for the formation of vascular networks,} \emph{Journal of Theoretical
  Biology} \textbf{333},  174--209.

\bibitem[{Seydel(2009)}]{seydel_practical_2009}
Seydel, R. [2009] \emph{Practical {Bifurcation} and {Stability} {Analysis}}
  (Springer Science \& Business Media, New York).

\bibitem[{Shakeel \emph{et~al.}(2013)Shakeel, Matthews, Graham \&
  Waters}]{shakeel_continuum_2013}
Shakeel, M., Matthews, P.~C., Graham, R.~S. \& Waters, S.~L. [2013] \enquote{A
  continuum model of cell proliferation and nutrient transport in a perfusion
  bioreactor,} \emph{Mathematical Medicine and Biology: A Journal of the IMA}
  \textbf{30},  21--44.

\bibitem[{Silber \& Knobloch(1991)}]{silber1991hopf}
Silber, M. \& Knobloch, E. [1991] \enquote{Hopf bifurcation on a square
  lattice,} \emph{Nonlinearity} \textbf{4},  1063.

\bibitem[{Silber \emph{et~al.}(1992)Silber, Riecke \&
  Kramer}]{silber1992symmetry}
Silber, M., Riecke, H. \& Kramer, L. [1992] \enquote{Symmetry-breaking hopf
  bifurcation in anisotropic systems,} \emph{Physica D: Nonlinear Phenomena}
  \textbf{61},  260--278.

\bibitem[{Smith(2008)}]{smith_monotone_2008}
Smith, H.~L. [2008] \emph{Monotone {Dynamical} {Systems}: {An} {Introduction}
  to the {Theory} of {Competitive} and {Cooperative} {Systems}} (American
  Mathematical Soc.).

\bibitem[{Thullner \& Baveye(2008)}]{thullner_computational_2008}
Thullner, M. \& Baveye, P. [2008] \enquote{Computational pore network modeling
  of the influence of biofilm permeability on bioclogging in porous media,}
  \emph{Biotechnology and Bioengineering} \textbf{99},  1337--1351.

\bibitem[{Vafai(2010)}]{vafai_porous_2010}
Vafai, K. [2010] \emph{Porous {Media}: {Applications} in {Biological} {Systems}
  and {Biotechnology}} (CRC Press, London).

\bibitem[{Van~Blitterswijk \& Thomsen(2008)}]{van_blitterswijk_tissue_2008}
Van~Blitterswijk, C.~A. \& Thomsen, P. (eds.) [2008] \emph{Tissue engineering},
  Academic {Press} series in biomedical engineering (Elsevier, Acad. Press,
  Amsterdam).

\bibitem[{Venkatasubramanian \emph{et~al.}(1995)Venkatasubramanian, Schattler
  \& Zaborszky}]{venkatasubramanian_local_1995}
Venkatasubramanian, V., Schattler, H. \& Zaborszky, J. [1995] \enquote{Local
  bifurcations and feasibility regions in differential-algebraic systems,}
  \emph{IEEE Transactions on Automatic Control} \textbf{40},  1992--2013.

\bibitem[{Verwer \& Sanz-Serna(1984)}]{verwer_convergence_1984}
Verwer, J.~G. \& Sanz-Serna, J.~M. [1984] \enquote{Convergence of method of
  lines approximations to partial differential equations,} \emph{Computing}
  \textbf{33},  297--313.

\bibitem[{Wang \emph{et~al.}(2006)Wang, Lu \& Chen}]{wang2006spatio}
Wang, Q., Lu, Q. \& Chen, G. [2006] \enquote{Spatio-temporal patterns in a
  square-lattice hodgkin-huxley neural network,} \emph{The European Physical
  Journal B-Condensed Matter and Complex Systems} \textbf{54},  255--261.

\bibitem[{Winterbottom(2004)}]{winterbottom2004mode}
Winterbottom, D.~M. [2004] \enquote{Mode interactions on a square lattice,}
  \emph{International Journal of Bifurcation and Chaos} \textbf{14},
  3883--3897.

\bibitem[{Wolfrum(2012)}]{wolfrum2012turing}
Wolfrum, M. [2012] \enquote{The turing bifurcation in network systems:
  Collective patterns and single differentiated nodes,} \emph{Physica D:
  Nonlinear Phenomena} \textbf{241},  1351--1357.

\bibitem[{Wu \emph{et~al.}(2014)Wu, Liu, Yeung, Liu \&
  Yang}]{wu_biomimetic_2014}
Wu, S., Liu, X., Yeung, K. W.~K., Liu, C. \& Yang, X. [2014]
  \enquote{Biomimetic porous scaffolds for bone tissue engineering,}
  \emph{Materials Science and Engineering: R: Reports} \textbf{80},  1--36.

\end{thebibliography}
\end{document}